\documentclass[12pt,twoside,a4paper]{article}
\usepackage[dvips]{epsfig}
\usepackage{pstricks}
\renewcommand{\theequation}{\arabic{section}.\arabic{equation}}

\newcommand{\lb}{\rm \scriptstyle label}
\newcommand{\ee}{{$e^+e^-$\ }}
\newcommand{\epem}{e^+e^-}

\newcommand{\dstar}{$D^{*\pm}$ }
\newcommand{\msbar}{$\overline{\rm MS}$}
\newcommand{\n}{\hspace*{-2.5mm}}
\newcommand{\PS}{{\rm PS}}
\newcommand{\li}{{\rm Li}_2}
\newcommand{\lithree}{{\rm Li}_3}
\newcommand\COFm{{\rm\kern.24em
    \vrule width.02em height1.4ex depth-.05ex
    \kern-.26em C}}
\newcommand{\aqed}{\alpha_{\rm \scriptstyle QED}}
\newcommand\N{{\rm I\kern-.18em N}}
\newcommand\R{{\rm I\kern-.21em R}}
\newcommand{\ve}{\varepsilon}
\newcommand{\COMMA}{\hspace{0.1cm},}
\newcommand{\STOP}{\hspace{0.1cm}.}
\newcommand{\SPACE}{\hspace{0.4cm}}

\catcode`@=11
\newcount\@tempcntc
\def\@citex[#1]#2{\if@filesw\immediate\write\@auxout{\string\citation{#2}}\fi
  \@tempcnta\z@\@tempcntb\m@ne\def\@citea{}\@cite{\@for\@citeb:=#2\do
    {\@ifundefined
       {b@\@citeb}{\@citeo\@tempcntb\m@ne\@citea\def\@citea{,}{\bf ?}\@warning
       {Citation `\@citeb' on page \thepage \space undefined}}%
    {\setbox\z@\hbox{\global\@tempcntc0\csname b@\@citeb\endcsname\relax}%
     \ifnum\@tempcntc=\z@ \@citeo\@tempcntb\m@ne
       \@citea\def\@citea{,}\hbox{\csname b@\@citeb\endcsname}%
     \else
      \advance\@tempcntb\@ne
      \ifnum\@tempcntb=\@tempcntc
      \else\advance\@tempcntb\m@ne\@citeo
      \@tempcnta\@tempcntc\@tempcntb\@tempcntc\fi\fi}}\@citeo}{#1}}
\def\@citeo{\ifnum\@tempcnta>\@tempcntb\else\@citea\def\@citea{,}%
  \ifnum\@tempcnta=\@tempcntb\the\@tempcnta\else
   {\advance\@tempcnta\@ne\ifnum\@tempcnta=\@tempcntb \else \def\@citea{--}\fi
    \advance\@tempcnta\m@ne\the\@tempcnta\@citea\the\@tempcntb}\fi\fi}
\catcode`@=12
\begin{document}
\title{\vskip-3cm{\baselineskip14pt
\centerline{\normalsize DESY 97--128\hfill}
\centerline{\normalsize hep--ph/9707???\hfill}
\centerline{\normalsize July 1997\hfill}}
\vskip1.5cm
Fragmentation Functions in Next--To--Leading Order QCD}
\author{J. Binnewies\\
II. Institut f\"ur Theoretische Physik\thanks{Supported
by Bundesministerium f\"ur Forschung und Technologie, Bonn, Germany,
under Contract 05~6~HH~93P~(5),
and by EEC Program {\it Human Capital and Mobility} through Network
{\it Physics at High Energy Colliders} under Contract
CHRX--CT93--0357 (DG12 COMA).},
Universit\"at Hamburg,\\
Luruper Chaussee 149, 22761 Hamburg, Germany}
\date{}
\maketitle
\begin{abstract}
We present new
sets of fragmentation functions in next--to--leading
order QCD that are determined from \ee annihilation data
of inclusive particle production.
In addition to the ${\cal O}(\alpha_s)$ 
unpolarized cross section the longitudinal
cross section is also used 
to extract the gluon fragmentation function from \ee annihilation data.
As the ${\cal O}(\alpha_s^0)$ vanishes for longitudinal
polarized photons (or $Z$ bosons), the  ${\cal O}(\alpha_s^2)$ 
corrections are required to reduce the scale ambiguities.
Recently, P.J.~Rijken and W.L.~van~Neerven presented the 
longitudinal coefficient functions to next--to--leading order.
We confirm part of their results
in this thesis and complete the calculation by the results
for the color class $C_F T_R$ that must be included
for a consistent comparison with LEP1 data.
The complete set of coefficient functions is 
then used together with novel data from ALEPH 
to determine the fragmentation functions for charged hadrons.
This set, and also sets 
for charged pions, kaons, and $D^*$ mesons as well as neutral kaons
published previously, can then be employed 
to test QCD in \ee annihilation, photoproduction, $\gamma \gamma$ 
collisions,  $p\overline{p}$ scattering and DIS.
Finally, we suggest how the improved knowledge on the fragmentation
in particular of the gluon could be used to
determine the gluon and charm content of the photon.

%%\noindent
%%PACS numbers: 
\end{abstract}
\newpage

%%%%%%%%%%%%%%%%%%%%%%%%%%%%%%%%%%%%%%%%%%%%%%%%%%%%%%%%%%%%%%%%%%%%%%%%%%

\section{Introduction}
\setcounter{equation}{0}
\bigskip

At the heart of high--energy physics lies the desire to find the
fundamental building blocks of matter and to understand how they make up
our world.
This is a very old quest that has made substantial progress only in
the past 100 years with the advent of high--energy experiments,
as the spatial resolution achieved in experiments is inversely proportional
to the energy transfer.
The picture that has emerged contains leptons (like the electron),
quarks (that build up hadrons like the proton \cite{partonm}), 
and gauge bosons 
(like the photon) that
mediate the four fundamental forces gravitation, electromagnetism,
weak, and strong interactions. 
Due to its weakness, gravitation can be disregarded in high--energy
experiments.
\smallskip

The first complete theory was formulated for electromagnetism.
That theory, Quantum Electrodynamics (QED), succeeded in describing
all experimental data on electromagnetic interactions with a
precision rivaled by no other theory to date.
However, QED is not exactly solvable, it is solved in a perturbative
expansion of its coupling constant $\aqed$.
As that is of the order of 10$^{-2}$, neglecting terms of 
 ${\cal O}(\aqed^2)$ or even ${\cal O}(\aqed^3)$
introduces but a small uncertainty.
Because of its tremendous success, the concept of a perturbatively
solvable field theory was adopted to formulate theories for the other
forces as well.
In this process the weak interactions turned out to be closely
related to the electromagnetic ones, leading to a unification
of the two theories to the theory of electroweak interactions
\cite{electroweak}.
\smallskip

The field theory for the strong interactions is Quantum Chromodynamics
(QCD) \cite{qcd}. 
Like QED it can be solved only perturbatively
(the alternative lattice techniques are not applicable to the subject
under study, at present, and are not discussed).
In contrast to QED, however, the coupling 
constant $\alpha_s$ is much larger,
and the fundamental particles of the 
theory, the quarks, are only free in the
limit of asymptotically high energies \cite{afree}.
This corresponds to small distances, i.e.\ the quarks are not free on
macroscopic scales but are confined inside hadrons made up of
two or three quarks in the case of mesons or baryons, respectively.
\smallskip

Therefore, only part of the strong interactions can be 
calculated perturbatively, the confinement can not be computed from QCD (yet),
and its properties must be measured.
For a number of high--energy reactions, the hard and soft regions can be
disentangled and treated separately. 
This is the content of the factorization theorem of QCD.
The hard cross sections for parton scattering can then be calculated
to fixed order in perturbation theory.
Beyond leading order, virtual corrections to the process
under study must be taken into account, too.
These give rise to singularities in certain regions of the
phase space, as do the real corrections.
For inclusive processes, the infrared and collinear
singularities cancel when the real and virtual contributions are added
\cite{kln}.
The remaining ultraviolet singularities are removed by
wave function and coupling constant renormalization.
\smallskip

For semi--inclusive processes 
this does not hold true anymore.
In the case of single hadron production, e.g., the transition
of the parton from a hard, quasi--free particle to confinement
within the observed final--state hadron can not be calculated from
perturbative QCD.
It must be modeled by a phenomenological fragmentation function (FF).
Because this function applies to a single parton, some
(collinear) final--state singularities remain.
These residual singularities are factored off into the
bare, infinite FF, leading to finite partonic cross sections and
finite fragmentation functions
(an analogous procedure applies to initial--state singularities and
parton density functions, PDF's).
\smallskip

Inclusive particle production (IPP) is of particular interest for the 
investigation of QCD.
It has been measured in a variety of processes and
over a wide range of energies.
Due to the well defined final--state, the measurements
have small systematic errors and the observables are also unambiguously
defined theoretically.
This enables a test of QCD via two of its central properties:
scaling violation and universality of factorization.
As pointed out, the FF's required for the calculation of
hadron production cross sections
are not calculable, but must be fitted to data.
Once this has been accomplished with the help of data
of one specific process at one particular center--of--mass (CM) energy,
the FF's so obtained can be applied to any process due to
the universality of factorization.
Their scale dependence
is determined by the Altarelli--Parisi splitting functions.
The determination of these FF's from data, and their application in a
number of phenomenological studies are
the subjects of this thesis.
\smallskip

The choice process for the extraction of FF's from data 
is \ee annihilation.
From the experimental viewpoint it is distinguished
by an abundance of data that have been collected over the past
fifteen years at energies from 3.6 to 91.2~GeV \cite{eecomp}.
In addition, the background is fully under control so that
not only statistical but also systematical errors are very small.
From the theoretical viewpoint,
the greatest advantage of \ee annihilation is that no additional
non--perturbative input is needed,
avoiding possible bias in the determination of the FF's. 
Moreover, the Born approximation is of purely electroweak
origin, i.e.\ it is precisely known, and any deviation from it
is due to the strong interaction. 
Last but not least, the phase space integrations required for
the calculation of single particle production are particularly
easy for this process, owing to the simplicity of the initial state.
This permits to perform the integration over phase space
analytically (which is not possible, e.g., 
for $p\overline{p}$ scattering),
speeding up the numerical evaluation of the cross section
tremendously, which is crucial for fitting.
\smallskip

To salvage the most of the predictive power of QCD it is desirable
to use data from one specific process only in extracting the FF's.
One disadvantage of \ee annihilation in this context is that gluons are not
produced at LO, leading to a small contribution to the
unpolarized cross section and consequently to little information
on their fragmentation.
However, the gluons do contribute a major part of 
the longitudinal cross section.
From a comparison of this measurement with theory one can
determine the gluon as well.
Until recently, the longitudinal cross section has been computed to LO
accuracy only -- 
which is ${\cal O}(\alpha_s)$ -- introducing large scale dependences.
The NLO
(${\cal O}(\alpha_s^2)$) corrections are required to reduce
this uncertainty.
The calculation of the relevant coefficient functions has been
presented by P.J.~Rijken and W.L.~van Neerven \cite{rij1}
for the color classes $C_F^2$, $C_F ( C_F - 1/2 N_C)$, $C_F N_C$, 
and $N_F T_R C_F$.
In this work we confirm the results for the class $N_F T_R C_F$ 
and augment the results by the color class $C_F T_R$.
The contribution from this class turns out to be numerically small.
\smallskip

Subsequently, we use the complete
set of NLO longitudinal coefficient functions 
together with the NLO transverse coefficient functions to determine
a set of charged hadron FF's 
to NLO that feature a better constrained gluon.
Although the transverse cross section has been calculated to 
NNLO \cite{rij2}, which
is of order ${\cal O}(\alpha_s^2)$ and hence formally of the same order
as the longitudinal cross section to NLO, these higher orders
are not included in the present analysis, as they would lead to
inconsistency with the Altarelli--Parisi evolution kernels that
are only known to NLO.
For later use in consistent $\alpha_s$ fits to 
deep inelastic scattering (DIS) and
photoproduction data, the new fit is performed for
distinct values of $\Lambda$.
The range from $\Lambda^{(4)}= 150$ to 400 MeV is
covered in 50 MeV steps,
our central choice is $\Lambda^{(4)}=350$~MeV.
\smallskip

In addition, a number of studies undertaken in the past years
in collaboration with B.A.~Kniehl and G.~Kramer
will be reviewed in order to demonstrate the potential
of inclusive particle production as a tool to test QCD in
a variety of processes.
\smallskip

Specifically, we fitted charged pion and charged kaon fragmentation
functions to data from TPC \cite{tpc1} and to data
from ALEPH \cite{ale0,ale4}.
The novel data in \cite{ale0} for the sum over
charged hadrons distinguished between different quark flavors.
Together with data on the gluon FF from OPAL \cite{opa2},
this enabled us to determine the FF's for individual
flavors for the first time \cite{binpk}.
We also presented sets of neutral kaon \cite{bink0}
and $D^{*\pm}$ meson FF's \cite{binds}.
Owing to the universality of factorization,
these FF's are applicable to any process.
We employed them in photoproduction, $p\overline{p}$,
and $\gamma\gamma$ scattering and found good agreement with data.
The main emphasis was put on photoproduction, where very good
data from HERA are available for study 
\cite{h11,h13,h14,h16,zeus2,zeus3}.
In particular the rapidity spectra have the potential to improve
our knowledge of the gluon and charm content
of the photon \cite{binds,binps}.
In $\gamma\gamma$ scattering, NLO predictions for inclusive
particle production at LEP2 and at the NLC were presented
for the first time \cite{bingg}.
\smallskip

The thesis is organized as follows.
Section~2 contains a short review of the theoretical framework
for using perturbative QCD and a discussion of the implications of the
factorization theorem.
In section~3 the process of \ee annihilation is discussed with emphasis
on the longitudinal cross section. 
Sets of fragmentation functions for various mesons,
extracted from \ee annihilation data, are presented in section~4, and 
with the help of the ${\cal O}(\alpha_s^2)$ longitudinal 
coefficient functions
a new set of particularly well constrained charged hadron FF's
is obtained.
Subsequently, these sets are applied to test QCD, via scaling violation 
and the property of universality, in section~5. 
A new approach to constraining the (incalculable) soft regime
of QCD in the case of the photonic gluon and charm densities 
is also presented.
The results are summarized in section~6 together with an outlook
on the potential of future analyses.

\vfill
\clearpage
\section{Factorization of Hard and Soft Regions}
\setcounter{equation}{0}
\bigskip

The strong coupling constant
that usually serves as the expansion parameter in perturbative QCD
becomes small only for short distances.
This corresponds to high momenta so that 
the bound states of colored objects can not be 
calculated and the problem of confinement remains
unsolved.
To put perturbation theory to any use in QCD, the short--distance and
long--distance processes therefore have to be separated.
Only for processes and experimental observables where this is feasible 
a comparison between perturbative QCD calculations and experimental data
is possible.
\smallskip

For this reason the factorization theorem of QCD that ensures
the factorizability of hard and soft regions (corresponding to
short and long distances, respectively) for a number of processes is
crucial to any calculation in perturbative QCD.
Moreover it is of particular interest for this analysis where the
non--perturbative fragmentation of
partons into hadrons is treated expressly by the parametrization
of the ignorance about the confinement into phenomenological functions.
\smallskip

\begin{figure}[hht!]
\begin{center}
\epsfig{file=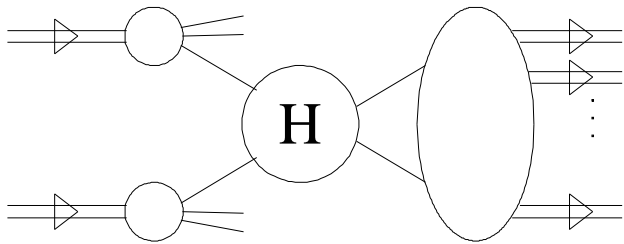 , clip=,width=14cm }
\end{center}
\vspace{-0cm}
\caption[]{\small 
General structure of factorization of a high--energy
cross section. 
\protect\label{factorization}}
\end{figure}

In Fig.~\ref{factorization}, the fragmentation process is represented
by the large blob to the right. The double lines with the arrowheads
represent hadrons in the initial or final state, the blob labeled 
{\bf H} represents the calculable hard scattering and the small blobs
in the initial state represent another non--perturbative
ingredient required for some processes, namely parton density 
functions of hadrons\footnote{
   Here as in all Feynman diagrams in this work, time 
   flow is from the left towards the right.}.
\smallskip

This schematic diagram shows the general features of factorization
in any factorizable process.
The large blob to the right may, however, also 
be taken to represent jet
fragmentation instead of inclusive particle production.
The reader is referred to
\cite{fab1,kla1} and references therein for details on the
subject of jet production.
\smallskip

The purpose of this section is to give a short account of
the content of the factorization theorem and to discuss its consequences
in some detail.
The theorem will be employed in section~4 to fit FF's to data from
\ee annihilation and some of its implications will be used as
rigorous tests of QCD when applying them in section~5.
A more detailed intuitive picture of factorization 
can be found in \cite{col1}.

\vspace{2cm}
\subsection{Factorization Theorem}
\smallskip

The factorization theorem 
states that the soft and the hard parts
of the cross section factorize (in fourier space) so that the
cross section can be written as the convolution of a non--perturbative
part with a perturbatively calculable hard scattering cross section.
It holds in every renormalizable field theory
for all leading twist two contributions, in particular, it has
been proven in \cite{ippfac} that the inclusive cross section
for any hard process factorizes.
To leading order, both the partonic, hard cross section and the
FF's are finite.
In higher orders, however, singularities are generated through the
real corrections.
These are in part canceled by singularities in the virtual corrections
that have to be added.
Due to the Kinoshita--Lee--Nauenberg theorem,
the cancellation works perfectly for inclusive quantities \cite{kln}.
For semi--inclusive observables, like inclusive particle production,
residual singularities are factored off into the bare FF's.
This is achieved with universal transition functions $\Gamma$.
In the dimensional regularization scheme \cite{tho1,dimreg} the space--time
dimension $n$ is taken slightly off the physical 4 dimensions in intermediate 
steps of the calculation.
This procedure isolates the singularities as poles in $4-n\equiv 2\ve$. 
The transition functions have the general structure
\begin{eqnarray}
\label{transf}
\Gamma_{a\to b}^{T} &\n=\n& \delta_{ab} {\rm \bf 1} 
                        -\frac{1}{\ve}\frac{\overline{\alpha}_s}{2\pi}
                         S_{\ve}\left(\frac{\mu^2}{M_f^2}\right)^{\ve}
                         \left[ P_{a\to b}^{T(0)} +d_{a\to b} \right]
                         + {\cal O}\left(\overline{\alpha}_s^2\right)
\SPACE {\rm and}
\\
\Gamma_{a\to b}^{S} &\n=\n& \delta_{ab} {\rm \bf 1} 
                        -\frac{1}{\ve}\frac{\overline{\alpha}_s}{2\pi}
                         S_{\ve}\left(\frac{\mu^2}{M_f^2}\right)^{\ve}
                         \left[ P_{a\to b}^{S(0)} +f_{a\to b} \right]
                         + {\cal O}\left(\overline{\alpha}_s^2\right)
\COMMA
\end{eqnarray}
where the factors $\mu^{2\ve}$ and
\begin{equation}
\label{Sepsdef} 
S_{\ve}\equiv \Gamma(1+\ve) (4\pi)^{\ve}
\end{equation}
are artefacts of dimensional regularization.
$M_f$ defines the scale at which the divergences are subtracted.
The unrenormalized strong coupling constant is denoted 
by $\overline{\alpha}_s$.
Dependence of the functions on variables is suppressed to simplify notation;
 ${\rm \bf 1}$ is shorthand for the $\delta$~function.
The superscripts $T$, $S$ stand for the timelike and
the spacelike processes, respectively.
The formul{\ae} are fully analogous and as \ee annihilation
is timelike ($Q^2>0$), only the timelike functions will
be considered in the following.
The superscript will be dropped for 
ease of notation.
The $d_{a\to b}$ functions specify the factorization scheme,
the {\msbar}--scheme \cite{msbar} adopted in this work
is defined by $d_{a\to b}=0$ for all $a$, $b$.
\smallskip

Up to order $\overline{\alpha}_s$, the timelike $\Gamma$~functions 
in the {\msbar}--scheme 
are given by: 
%__nocit__%\cite{rij2,str1}:
\begin{eqnarray}
\label{defGNS}
\Gamma_{q\to q}^{NS} &\n=\n& {\rm \bf 1} -\frac{1}{\ve}
   \frac{\overline{\alpha}_s}{2\pi}S_{\ve}
   \left(\frac{\mu^2}{M_f^2}\right)^{\ve}
   P_{q\to q}^{(0)} \\
\label{defGPS}
\Gamma_{q\to q}^{PS} &\n=\n& 0 \\
\label{defGSIG}
\Gamma_{q\to q}^{\Sigma} &\n=\n& \Gamma_{q\to q}^{NS} + \Gamma_{q\to q}^{PS} \\
\label{defGqG}
\Gamma_{q\to G} &\n=\n&  -\frac{1}{\ve}   
\frac{\overline{\alpha}_s}{2\pi}S_{\ve}\left(\frac{\mu^2}{M_f^2}\right)^{\ve}
   P_{q\to G}^{(0)}\\
\label{defGGq}
\Gamma_{G\to q} &\n=\n&  -\frac{1}{2\ve}   
\frac{\overline{\alpha}_s}{2\pi}S_{\ve}\left(\frac{\mu^2}{M_f^2}\right)^{\ve}
   P_{G\to q}^{(0)}\\
\label{defGGG}
\Gamma_{G\to G} &\n=\n& {\rm \bf 1} -\frac{1}{\ve}   
\frac{\overline{\alpha}_s}{2\pi}S_{\ve}\left(\frac{\mu^2}{M_f^2}\right)^{\ve}
   P_{G\to G}^{(0)}
\STOP
\end{eqnarray}
The singularities in the bare cross section $d\overline{\sigma}$
are factored into the bare fragmentation functions $\overline{D}$
according to
\begin{eqnarray}
\label{factNS}
d\overline{\sigma}_{L,q}^{NS} &\n=\n& \Gamma_{q\to q}^{NS} 
     \otimes d\sigma_{L,q}^{NS} \\ 
\label{factSig}
d\overline{\sigma}_{L,q}^{\Sigma} &\n=\n& \Gamma_{q\to q}^{\Sigma} \otimes 
        d\sigma_{L,q}^{\Sigma} + \Gamma_{G\to q} \otimes d\sigma_{L,G} \\ 
\label{factG}
d\overline{\sigma}_{L,G} &\n=\n& 2\Gamma_{q\to G} 
       \otimes d\sigma_{L,q}^{\Sigma} 
       +\Gamma_{G\to G} \otimes d\sigma_{L,G} 
\end{eqnarray}
and analogous equations
\begin{eqnarray}
 D_{NS}^h &\n=\n& \Gamma_{q\to q}^{NS} \otimes \overline{D}_{NS}^h
\label{DfactNS}
\\
 D_{\Sigma}^h &\n=\n& \Gamma_{q\to q}^{\Sigma} \otimes 
      \overline{D}_{\Sigma}^h+2N_F\Gamma_{q\to G}
       \otimes \overline{D}_G^h
\label{DfactSig}
\\
 D_{G}^h &\n=\n& \Gamma_{G\to G} \otimes \overline{D}_{G}^h
      +\frac{1}{N_F}\Gamma_{G\to q} \otimes \overline{D}_{\Sigma}^h
\label{DfactG}
\end{eqnarray}
for the FF's.
The definition of the combinations $\Sigma$ and $NS$
of FF's in the above
equations will be given in the next section.
The physical hadronic cross section remains
unchanged by this operation,
\begin{equation}
d\sigma^{h} = \sum_l d\sigma_l \otimes D_l^h 
            = \sum_l d\overline{\sigma}_l \otimes \overline{D}_l^h
\nonumber
\STOP
\end{equation}
The details of the summation are given in (\ref{sigxpol}).
The shorthand ${\otimes} $ denotes convolution,
\begin{eqnarray}
\left[f \otimes g\right](x) &\n \equiv \n& 
     \int_0^1dx_1\int_0^1dx_2f(x_1)g(x_2)\delta(x-x_1x_2)
\nonumber \\
&\n=\n&
 \int_x^1 \frac{dz}{z} f\left(\frac{x}{z}\right) g(z)
\label{convolution}
\STOP
\end{eqnarray}
The actual factorization for color class $N_F T_R C_F$ will be performed in
subsection~3.3.3.
\smallskip

One of the main results of the factorization theorem is that the
functions parametrizing the non--perturbative part are process independent.
This feature, known as universality of the FF's and the PDF's,
is instrumental to the predictive power of QCD analyses.
It is one of the most directly testable properties of QCD and one of the
reasons for the phenomenological significance of inclusive
particle production and FF's.
\smallskip

Albeit, there are some caveats to 
the predictive power of the 
factorization theorem as applied in this work. 

(i)
As stated at the outset, the factorization theorem holds for
leading twist two contributions, i.e.\ for terms of ${\cal O}(1/Q^2)$.
Thus, the results of
present day perturbative calculations are correct at least in the 
limit $Q^2 \to \infty$.
For any finite energy, higher twist
of the order $1/Q^k$, $k\ge3$ may be important, too.
For deep inelastic scattering
it has been proven that the subleading twist is twist four,
hence those contributions are suppressed by a relative factor of
 $1/Q^2$ and may safely be neglected.
In \ee annihilation, the situation is less clear as the subleading
terms may be of ${\cal O}(1/Q^3)$ for some observables.
We demonstrated some time ago \cite{bin1} that present data on the
unpolarized inclusive particle production do not necessitate the
introduction of such terms
as even the scaling violation, where they would show up
most prominently, is in good agreement with data.
This is demonstrated, for example, by Fig.~\ref{scaling} in section~5.2. 
Very recently, M.~Dasgupta and B.R.~Webber \cite{das1} calculated the
corrections with the help of renormalon techniques and found that the
${\cal O}(1/Q^3)$ is absent in unpolarized IPP cross sections
also in \ee annihilation.
In the same work they also maintain that the longitudinal and
transverse contributions separately do have $1/Q^3$ contributions.
The uncalculable coefficient of those 
contributions, however, is small 
and omitting such terms altogether is consistent with the
data on $R_{ee,L}$ \cite{opa3} at the 1.5~$\sigma$ level.
Power corrections will thus be neglected
throughout this work.
Some phenomenology of higher twist and power corrections in \ee annihilation
can be found in \cite{nas1,web1,dok1}.

(ii)
How much of the cross section is factored into the non--perturbative
functions is unequivocal only when the calculation is performed to
all orders\footnote{
   Strictly speaking, this is not feasible even in principle, as
   the perturbative expansion in $\alpha_s$ is an asymptotic series, i.e.\
   does not converge. For the sake of the argument one may pretend it is
   a convergent series as only the first few orders will likely ever be
   calculable in practice, anyway.}.
For a finite order calculation the scale $M_f$ sets the point at which
the infrared singularities are factored off, introducing an
additional unphysical parameter in the results of the calculations.
The main reason why one needs to calculate at least the subleading order
for most practical purposes is to reduce the dependence of the result on this
scale (and on related scales for the renormalization of the coupling,
 $\mu_R$, and for the factorization of 
the initial state singularities, $M_i$).
In subleading order, called NLO in the following (i.e.\ NLO to a
process that receives its leading contribution in ${\cal O}(\alpha_s^k)$
is of order ${\cal O}(\alpha_s^{k+1})$), the sensitivity to the
choice of scales is reduced compared to the LO.
For actual fits to data it is quite cumbersome to vary the scales as one
has to produce a separate fit for every choice of scales.
For this reason the discussion of scales in this work is limited
to the demonstration of the reduced scale dependence in NLO versus LO 
calculations.
In the actual fits
and applications the renormalization scales and fragmentation scales
will generally
all be identified with an appropriate hard scale of the process.
In \ee annihilation, there is only one hard scale, the virtuality of the
intermediate vector boson, which somewhat reduces the ambiguity of
choice of scales in this process.

(iii)
Last but not least, all particles, hadrons and partons alike, are treated
as massless in this work.
For the partons, the non--negligible masses of the $c$ and $b$ 
quarks are usually taken into account only via thresholds in the
evolution of the FF's, according to the 
variable flavor number scheme (VFNS) \cite{vfns}.
In the recent past, also some massive calculations of cross sections
have been performed \cite{mel1,mass}.
In this thesis, the VFNS philosophy is adopted.
Only for the heavy $D^{*\pm}$ mesons this topic is considered in more
detail and the factorization scheme is modified also to 
properly account for both the intermediate-- and high--energy regions.
As for the hadrons, there is no easy way to account for their masses.
The masses are thus taken to vanish throughout. 
For momenta that are large compared to all the masses involved,
this simplification is well justified. 
However, when the momentum of the final--state hadron is of the same
order as its mass, threshold effects arise and the massless results
become unreliable.
This is a general feature of all massless calculations and is also
observed, e.g.\ in jet production when approaching the boundary of
phase space.
In \ee annihilation the non--vanishing quark masses are important only
for small virtualities of the photon.
The finite hadron masses
always become important at some point for 
production of a hadron with small momentum fraction $x$,
notably for heavy
hadrons like $D^{*\pm}$ or protons/antiprotons.
On the other hand, the methods employed traditionally for
fitting and evolving FF's become unreliable for 
very small $x$ in any case.
This is due to the large logarithms of $1/x$ that have to be
resummed, eventually.
The Altarelli--Parisi evolution thus becomes unreliable at some point
and double--leading--logarithmic evolution equations 
are relevant instead \cite{web1}.
In an intermediate region of moderately small $x$, the description of the
fragmentation process can be improved by the modified leading log 
approximation (MLLA) in connection with local parton hadron duality
\cite{web1,lphd}.
This fundamental limitation of the traditional approach 
turns up as a singularity in differential NLO cross sections at $x=0$.
In this work, $x_{min}$ in the range 0.05 ... 0.20 will be taken as
lower limit for the validity of the massless approach.

\vspace{2cm}
\subsection{Altarelli--Parisi Equations}
\smallskip

The fragmentation functions that have been introduced in
the previous subsection are non--perturbative.
They can not be calculated, at present, but must be taken
from data at some scale $\mu$.
Their evolution to other scales, however, is then
determined by perturbative QCD.  
In this subsection, we will present the equations governing the
evolution of FF's.
For brevity, they will be referred to as the Altarelli--Parisi (AP)
equations, for a more complete bibliography of contributions
see \cite{dglap}.
\smallskip

Let $D_q^h(x,\mu^2)$, $D_{\overline{q}}^h(x,\mu^2)$, 
and $D_G^h(x,\mu^2)$ be the FF's of
 $N_F$ quarks $q$, antiquarks $\overline{q}$, and 
the gluon $G$, respectively, into
some hadron $h$ with momentum fraction $x$ at the scale $\mu$.
The $\mu^2$ evolution of these FF's is conveniently formulated for the
linear combinations
\begin{eqnarray}
D_{\Sigma}^h(x,\mu^2)&\n=\n&\sum_{i=1}^{N_F}\left( 
      D_{q_i}^h(x,\mu^2)+D_{\overline{q}_i}^h(x,\mu^2) \right)
\COMMA
\\
\nonumber\\
D_{(+),i}^h(x,\mu^2)&\n=\n&D_{q_i}^h(x,\mu^2) 
     + D_{{\overline{q}},i}^h(x,\mu^2)
   -\frac{1}{N_F} D_{\Sigma}^h(x, \mu^2)
\COMMA
\\
\nonumber\\
D_{(-),i}^h(x,\mu^2)&\n=\n&D_{q_i}^h(x,\mu^2)
     - D_{{\overline{q}},i}^h(x,\mu^2)
\COMMA
\end{eqnarray}
as for these the gluon decouples from the 
non--singlet $\scriptstyle (+)$ and the
asymmetric $\scriptstyle (-)$
combinations, leaving only the singlet and the gluon 
fragmentation functions coupled;
\begin{eqnarray}
\mu^2\frac{d}{d\mu^2}
D_{(+),i}^h(x,\mu^2)
&\n=\n&\left[P_{(+)}\left(\alpha_s(\mu^2)\right)\otimes
D_{(+),i}^h(\mu^2)\right](x) 
\COMMA \label{APplus}
\\
\nonumber\\
\mu^2\frac{d}{d\mu^2}D_{(-),i}^h(x,\mu^2)
&\n=\n&\left[P_{(-)}\left(\alpha_s(\mu^2)\right)\otimes
D_{(-),i}^h(\mu^2)\right](x)
\COMMA \label{APminus}
\\
\nonumber\\
\mu^2\frac{d}{d\mu^2}D_{\Sigma}^h(x,\mu^2)
&\n=\n&\left[
P_{\Sigma}\left(\alpha_s(\mu^2)\right)\otimes D_{\Sigma}^h(\mu^2)
\right](x)
\nonumber\\
& & \hspace{1cm}
+2N_F\left[P_{q\to G}\left(\alpha_s(\mu^2)\right)\otimes D_G^h(\mu^2)
\right](x)
\COMMA \label{APS1}
\\
\nonumber\\
\mu^2\frac{d}{d\mu^2}D_G^h(x,\mu^2)
&\n=\n&\frac{1}{2N_F}\left[
P_{G\to q}\left(\alpha_s(\mu^2)\right)\otimes D_{\Sigma}^h(\mu^2)
\right](x)
\nonumber\\
& & \hspace{1cm}
+\left[
P_{G\to G}\left(\alpha_s(\mu^2)\right)D_G^h(\mu^2)\right](x)
\STOP \label{APS2}
\end{eqnarray}
These equations can be derived from (\ref{DfactNS}) through (\ref{DfactG})
by taking the derivative.
The splitting functions 
that appear in above equations are defined as
\begin{eqnarray}
P_{(+)}\left(x,\alpha_s(\mu^2)\right)&\n=\n&
P_{q\to q}^{V}\left(x,\alpha_s(\mu^2)\right) +
P_{q\to \overline{q}}^{V}\left(x,\alpha_s(\mu^2)\right)
\COMMA
\\
\nonumber\\
P_{\Sigma}\left(x,\alpha_s(\mu^2)\right)&\n=\n&
P_{(+)}\left(x,\alpha_s(\mu^2)\right)+
2N_FP_{q\to q}^{S}\left(x,\alpha_s(\mu^2)\right)
\COMMA
\\
\nonumber\\
P_{(-)}\left(x,\alpha_s(\mu^2)\right)&\n=\n&
P_{q\to q}^{V}\left(x,\alpha_s(\mu^2)\right) -
P_{q\to \overline{q}}^{V}\left(x,\alpha_s(\mu^2)\right) 
\STOP
\end{eqnarray}
In the perturbative expansion of the splitting functions,
\begin{equation}
\label{apexp}
P(x,\alpha_s(\mu^2))=\frac{\alpha_s(\mu^2)}{2\pi}P^{(0)}(x)
+\left(\frac{\alpha_s(\mu^2)}{2\pi}\right)^2P^{(1)}(x)
+{\cal O}\left(\left(\frac{\alpha_s(\mu^2)}{2\pi}\right)^3\right)
\COMMA
\end{equation}
the first two orders have been calculated in \cite{cur1}.
As the formul{\ae} are given in
an implicit form only, in the $x\to 1$ limit, we have 
collected the timelike functions in a ready to use form,
i.e., with the coefficients of the delta functions and the
plus distributions extracted explicitly, 
in \cite{binds}\footnote{The spacelike functions 
    have been treated in a similar
    manner in \cite{ell1}.}.
Here, we will list only the timelike LO splitting functions which enter 
the NLO cross sections via the scale dependences.
\begin{eqnarray}
P_{q\to q}^{V,(0)}(x) &\n=\n& C_F\left[\frac{3}{2}\delta(1-x)
   +2\left(\frac{1}{1-x}\right)_+ -1 -x\right] \COMMA
\label{P0Vqq}
\\
P_{q\to \overline{q}}^{V,(0)}(x) &\n=\n& P_{q\to q}^{S,(0)}(x) = 0
\COMMA
\label{P0Vqqbar}
\\
P_{q\to G}^{(0)}(x) &\n=\n& C_F\left[\frac{1+(1-x)^2}{x}\right]
\COMMA
\label{P0qG}
\\
P_{q\to q}^{(0)}(x) &\n=\n& 2N_F T_R\left[x^2+(1-x)^2\right]
\COMMA
\label{P0Gq}
\\
P_{G\to G}^{(0)}(x) &\n=\n& \left(\frac{11}{6}N_C-\frac{2}{3}N_F T_R
  \right)\delta(1-x) 
\nonumber\\
& & \hspace{1cm}
+2N_C\left[\left(\frac{1}{1-x}\right)_+ 
  +\frac{1}{x}-2+x-x^2\right]
\STOP
\label{P0GG}
\end{eqnarray}
\smallskip

For flexibility in the parametrization at the starting scale, the
evolution of the $D_a$ may be performed in $x$--space,
solving the AP equations in their integral form 
\begin{equation}
D_a^h(x,k)=D_a^h(x,0)+\int_0^kdk'\frac{\alpha_s(k')}{2\pi}
   \sum_b \left[P_{a\to b}(\alpha_s(k')) 
    \otimes D_b^h(k')\right](x)
\label{ap}
\end{equation}
by iteration. 
The variable 
$k=\ln(\mu^2/\mu_0^2)$ replaces $\mu^2$ for ease of notation. 
The first term on the right--hand side of (\ref{ap}) 
is the distribution at the starting scale $\mu_0$. 
\smallskip

Alternatively, the AP--equations can be solved analytically in moment space.
Only the back--transformation has to be performed numerically
in this approach.
Details on the techniques of evolution in $x$-- and $n$--space are
given in appendix~C.

\vspace{2cm}
\subsection{Momentum Sum Rule}
\smallskip

Any parton produced in the hard scattering process fragments with 
probability one into hadrons.
In connection with local parton hadron duality \cite{lphd}, this
is the basis of the concept of jets where all hadrons originating
from the same parton are grouped into a jet that is taken
to bear a one--to--one
correspondence to the primary parton, e.g., carries the same momentum.
If one wants to study details of the jet one has to go beyond this
and must investigate the probability of parton $a$ to fragment into
some specific hadron $h$.
Due to momentum conservation, the momentum fractions 
$\int_0^1 dx x D_a^h(x)$ of the individual hadron species must
add up to one when summing over all hadrons $h$,
\begin{equation}
\label{sumrule}
\sum_h \int_0^1 dx \, x D_a^h(x,Q^2) = 1 .
\end{equation}
When fitting the related PDF's, e.g.\ of the proton, to data,
the related sum rule
\[
\sum_a \int_0^1 dx \, x F_{a/p}(x,Q^2) = 1 
\]
is usually used as a constraint in the fit.
Unfortunately, the sum rules (\ref{sumrule}) can not be employed for
fixing (some of) the parameters of the FF's for charged hadrons,
because the charged hadrons, charged pions or other particles considered
here form an incomplete subset of all hadrons
and equation~(\ref{sumrule}) is valid only for the sum over all hadrons.
For the sum over a subset,
the integral will give
some value $I_a$ for any parton $a$.
These values could be extracted from the ALEPH \cite{ale3}
data --- but to no avail as $I_a \ne I_b$ for two distinct types
of partons $a$ and $b$ will, via mixing in the evolution, give rise to
$Q^2$ dependence of the $I_a$.
As the FF's are parametrized at a lower scale $\mu_0 < m_Z$, 
no information on the relevant $I_a$ values is available.
\smallskip

In \cite{binpk}, we used the fundamental momentum
sum rule as an additional check
of the FF's for charged pions and kaons.
To account for the other hadrons, we had to make a number
of assumptions.
The protons and antiprotons were modeled to data at the $Z$ pole
by a function of the momentum fraction, as given in (\ref{ppbarf}).
Other charged hadrons were neglected.
Of the neutral hadrons, only the pions and kaons were taken
into account.
Based on SU(3) symmetry, we could reasonably assume that they
behave like their charged 
counterparts\footnote{For the pions, it has also been
   found experimentally \cite{ale5}
   that the $\pi^0$ exhibits the same
   spectrum as the average of its charged partners.
   Only for very small $x$, where its momentum is not
   large compared to its mass any more, does their minute mass
   difference show up.}.
The inclusion of other neutral hadrons would have introduced
unnecessary bias. 
When summing over these final states, values close to
one were obtained.
They exhibited a slight decrease with $Q^2$, 
as shown in Table~\ref{tabsumrule}.
This is in agreement with the intuitive picture of the 
opening up of additional channels at higher energies.

\pagebreak
\begin{table}[h]
\begin{center}
\begin{tabular}{|c||c|c|c|c|c|} 
\hline
$a$ & \multicolumn{5}{c|}{$Q$ [GeV]} \\
\cline{2-6} & $\sqrt2$ & 4 & 10 & 91 & 200 \\
\hline
\hline
$u$ & 0.81 (1.06) & 0.86 (1.01) & 0.86 (0.98) & 0.81 (0.89) & 0.79 (0.86) \\
$d$ & 0.93 (1.03) & 0.96 (0.99) & 0.94 (0.96) & 0.88 (0.88) & 0.86 (0.85) \\
$s$ & 1.03 (1.64) & 1.03 (1.50) & 1.01 (1.41) & 0.94 (1.25) & 0.92 (1.20) \\
$c$ & --          & 0.90 (1.10) & 0.89 (1.05) & 0.84 (0.95) & 0.82 (0.92) \\
$b$ & --          & --          & 0.95 (1.12) & 0.90 (1.02) & 0.88 (0.99) \\
\hline
$G$ & 0.92 (1.07) & 0.96 (0.98) & 0.92 (0.91) & 0.81 (0.78) & 0.78 (0.75) \\
\hline
\end{tabular}
\end{center}
\caption{\small
Left-hand side of eq.~(\protect\ref{sumrule}) at NLO (LO) for
$Q=\sqrt 2$, 4, 10, 91, and 200~GeV,
evaluated with the set \protect\cite{binpk}.
For technical reasons, the range of integration is 
restricted to $x \in (0.02,0.98)$.
Hadrons are taken to be either pions, kaons or 
protons/antiprotons as specified above.
\protect\label{tabsumrule}}
\end{table}

\vfill
\clearpage
\section{Particle Production in \ee Annihilation}
\setcounter{equation}{0}
\bigskip

For the extraction of the fragmentation functions from data, the
inclusive particle production (IPP) in \ee annihilation is the
choice process.
It has three big advantages:

(i)
It is particularly clean in the final state, i.e.\ no target
or remnant jets contaminate the measurement.
Hence, the background is fully under control, experimentally.

(ii)
No non--perturbative input is required for the comparison with QCD
calculations (like the parton density functions in
 $p\overline{p}$ or $ep$ scattering),
affording an accuracy in the determination of the FF's that is limited
by the theoretical scale ambiguities (and the experimental errors),
only.

(iii)
The process has been measured over a wide range of energies \cite{eecomp}
and, especially at the $Z$ peak by LEP1, to a very high accuracy of the 
order of a few percent.
\smallskip

For these reasons, this process has traditionally been employed to
fit FF's (see section~4 for a review) and we will follow the
tradition in this work.
One disadvantage of \ee annihilation in this context is, however
that the gluon does not contribute to the cross section in LO, which
leads to a poor correlation of the gluon fragmentation function and the
data on unpolarized IPP.
Some groups have therefore exploited data from other processes,
e.g.\ $p\overline{p} \to hX$, in addition to \ee annihilation
to constrain the gluon fragmentation function $D_G^h$ \cite{ans1}.
\smallskip

The philosophy of the analyses presented here is to salvage as much of
QCD's predictive power as possible by using only the 
 \ee annihilation process in the determination of FF's.
The property of universality can then be tested by applying the
FF's to different processes as will be demonstrated in subsection~5.1.
Following an idea by P.~Nason and B.R.~Webber \cite{nas1}, 
the longitudinal cross section
 $\sigma_L$ is used to constrain the gluon FF instead
in our new fit.
For this
part, the gluon and the quark both contribute with comparable
strength to the IPP.
 $\sigma_L$ vanishes in ${\cal O}(\alpha_s^0)$ and receives
contributions from the quarks as well as the gluons at ${\cal O}(\alpha_s)$.
To reduce the scale dependence, one needs to include the NLO corrections
also in this case, albeit of ${\cal O}(\alpha_s^2)$.
\smallskip

In the following two subsections, the IPP process in \ee annihilation will
be discussed in general and 
the transversely polarized part will be briefly reviewed.
In subsection~3.3, we calculate part of the 
${\cal O}(\alpha_s^2)$ corrections to the longitudinal process,
amending and partly checking the results presented in \cite{rij1,rij2}. 
The section closes with a short discussion of the new results.

\vspace{2cm}
\subsection{The Process}
\smallskip

\begin{figure}[hht!]
\begin{center}
\epsfig{file=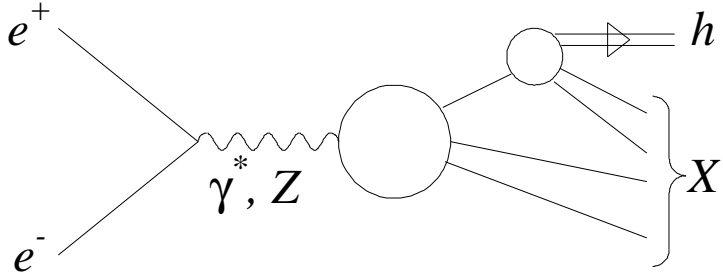 , clip=,width=14cm }
\end{center}
\vspace{-0cm}
\caption[]{\small 
The IPP process in \ee annihilation.
\protect\label{eeprocess}}
\end{figure}

Consider the inclusive production of a hadron $h$ in \ee annihilation,
\begin{equation}
\label{process}
e^+(p_a) \, + e^-(p_b) \to (\gamma^*,Z)(q)\to h(p_h) \, +X 
\STOP
\end{equation}
Here, $h$ denotes the measured outgoing hadron with four--momentum $p_h$,
the intermediate vector boson is either a virtual photon
or a $Z$ with four--momentum $q=p_a+p_b$,
and $X$ stands for the sum of any other debris from the hard scattering 
as depicted in Fig.~\ref{eeprocess}.
 
The virtual $\gamma / Z$ is timelike, i.e., $Q^2 \equiv q^2 > 0$.
For large $Q^2$, the transverse components of the momenta can be neglected.
In the CM system,
the outgoing hadron is characterized completely by its angle with the
beam axis, $\theta$, and the scaling variable
\begin{equation}
\label{xBj}
x=\frac{2 \, p_h \cdot q}{Q^2} \hspace{1.2cm} , 
\hspace{0.4cm} 0<x \le 1 
\STOP
\end{equation}
As the hadrons are considered to be massless, the variable defined above
is the energy fraction of the outgoing hadron and does not coincide
with the momentum fraction that is usually measured in experiments
via the curvature of the particle trajectory in a magnetic field.
For not too small $x$, the difference is negligible but for
 $x$ not much larger than $2m_h/\sqrt{Q^2}$ threshold effects obviously
become important and can not be treated properly within this framework.
\smallskip

\begin{figure}[hht!]
\begin{center}
\epsfig{file=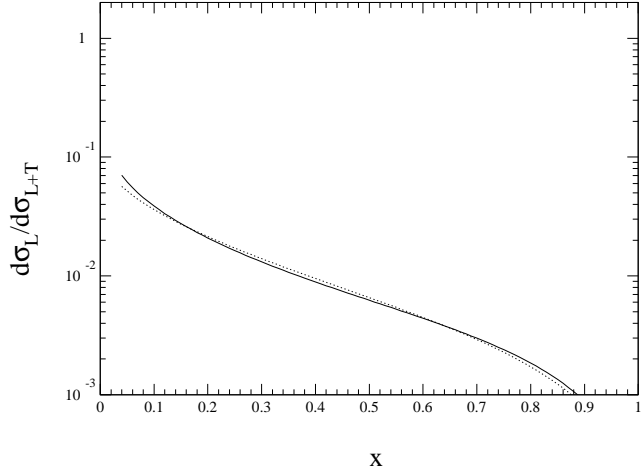,clip=,width=8.7cm}
\end{center}
\vspace{-0.8cm}
\caption[]{\small 
Relative importance of the longitudinal cross section.
The solid (dotted) curves show the NLO (LO)
results for $d\sigma_L/d\sigma_{T+L}$, evaluated at $m_Z$ with our new
set of $h^{\pm}$ FF's.
\protect\label{lrat}}
\end{figure}

The unpolarized differential cross section of process~(\ref{process})
can be decomposed into several contributions with
different tensor structures. 
For inclusive particle production, it is given by \cite{kra1}
\begin{equation}
\label{sigxtheta}
\frac{d^2\sigma^h}{dx\,d{\cos}\theta}=\frac{3}{8}
     (1+{\cos}^2\theta)\frac{d\sigma_T^h}{dx}
     +\frac{3}{4}{\sin}^2\theta\frac{d\sigma_L^h}{dx}
     +\frac{3}{4}{\cos}\theta\frac{d\sigma_A^h}{dx}
\COMMA
\end{equation}
where the subscripts $T$ and $L$ denote the contributions due to
transverse and longitudinal polarizations, respectively, and the asymmetric
term, labeled $A$, is due to the interference of the photon with the $Z$
boson.
Experimentally, the angular structure (\ref{sigxtheta}) has
first been confirmed by the TASSO collaboration \cite{tas5}.
Usually, experiments do not determine $\theta$ distributions for
inclusive particle production measurements but rather
integrate over the angles.
This eliminates the asymmetric term in equation~(\ref{sigxtheta}).
As the angular factors are normalized to their total integral,
the single differential cross section is
\begin{equation}
\label{sigx}
\frac{d\sigma^h}{dx}=\frac{d\sigma_T^h}{dx}+\frac{d\sigma_L^h}{dx}
\STOP
\end{equation}
The asymmetric contribution is therefore disregarded in this work and
the reader is referred to \cite{rij3} for a discussion of $\sigma_A^h$.
The transverse polarization state dominates the cross section,
as demonstrated in Fig.~\ref{lrat}.
The solid (dotted) lines show the ratios of the differential
longitudinal to the unpolarized ($T+L$) cross section in NLO (LO).
We will now turn to the calculation of the differential cross section.
\smallskip

The hadronic cross section is given by the convolution of the
FF's with the partonic cross sections,
\begin{eqnarray}
\label{sigxpol}
\hspace{-.3cm} \frac{1}{N_C\sigma_0} \frac{d\sigma_k^h}{dx} 
&\n=\n&
    N_C \int_x^1 \frac{dz}{z} \left\{ 
      \left[\sum_{i=1}^{N_F}Q_{q_i}(Q^2)\right]
      \left[ \frac{1}{N_F} D_{\Sigma}^h\left(\frac{x}{z},M_f^2\right)
      \frac{1}{N_C\sigma_0} \frac{d\sigma_{k,q}^{\Sigma}}{dz}
\right. \right. \nonumber \\
& & \left. \hspace{.5cm}
      +D_G^h\left(\frac{x}{z},M_f^2\right)
      \frac{1}{N_C\sigma_0} \frac{d\sigma_{k,G}}{dz}
      \right]
\nonumber \\
& & \hspace{.5cm}
      +\sum_{i=1}^{N_F}Q_{q_i}(Q^2) 
      D_{(+),i}^h\left(\frac{x}{z},M_f^2\right)
      \frac{1}{N_C\sigma_0} \frac{d\sigma_{k,q}^{NS}}{dz} 
\nonumber \\
& & \left. \hspace{.5cm}
      +\sum_{i=1}^{N_F}Q_{q_i}^{F}(Q^2) 
      \left[ D_{(+),i}^h\left( \frac{x}{z},M_f^2\right)
      +\frac{1}{N_F}D_{\Sigma}^h\left(\frac{x}{z},M_f^2\right) \right]
      \frac{1}{N_C\sigma_0} \frac{d\sigma_{k,q}^{F}}{dz}
      \right\}
\end{eqnarray}
for polarizations $k=T, L$.
The superscripts $\Sigma$ and $NS$ denote flavor singlet and
flavor non--singlet combinations,
as defined in section~2.2.
Contributions from the class $C_F T_R$ are labeled by $F$,
suggestive of the fact that they vanish for photons due to
Furry's theorem, see below.
The cross section is normalized to the total cross section of
 $\epem \to \mu^+\mu^-$ for massless leptons, 
\begin{equation}
\label{sig0}
\sigma_0=\frac{4\pi\aqed^2}{3Q^2} 
\COMMA
\end{equation}
where $\aqed$ is the electromagnetic coupling constant
and its running is neglected.
 $N_C=3$ counts the color states and takes care of
the additional, unobserved degree of freedom.
In fitting to data, the cross section will commonly be normalized
to the total hadronic cross section,
\begin{equation}
\label{sigtot}
\sigma_{\rm had} = R_{ee} N_C \sum_{i=1}^{N_F}Q_{q_i}(Q^2)\sigma_0 ,
\end{equation}
where the QCD corrections read to order $\alpha_s^2$
\begin{eqnarray}
R_{ee,L} &\n=\n&
   \frac{\alpha_s(\mu_R^2)}{2\pi}\left[2\right]
    +\left(\frac{\alpha_s(\mu_R^2)}{2\pi}\right)^2\left[C_F^2\left(
    -\frac{3}{8}\right)
\right.
\nonumber \\
& & \hspace{3cm} \left.
+N_C C_F \left(-\frac{11}{4}\ln
    \left(\frac{Q^2}{\mu_R^2}\right)-11\zeta(3)+\frac{123}{8}\right)
\right.
\nonumber \\
& & \hspace{3cm} \left.    
+N_F T_R C_F \left(\ln\left(\frac{Q^2}{\mu_R^2}\right)
    +4\zeta(3)-\frac{11}{2}\right)\right]
\COMMA
\label{Reelong}
\\
R_{ee,T} &\n=\n&
   1+{\cal O}(\alpha_s^2)
\COMMA
\label{Reetrans}
\\
R_{ee} &\n\equiv \n&
  R_{ee,L}+R_{ee,T}
\STOP
\label{Reetot}
\end{eqnarray}
Note that only the transverse polarization states contribute
to ${\cal O}(\alpha_s^0)$, whereas the
 ${\cal O}(\alpha_s)$ corrections to the
integrated cross section are solely due to the
longitudinal polarization state. 
 $R_{ee,L}$ has to be taken into account when
studying the renormalization scale dependence of the 
longitudinal cross section.
The $\alpha_s^2$ terms of $R_{ee,T}$ are not needed in our fit.
They can be found in \cite{rij2} where eq.~(\ref{Reetot}) has
been used as an independent check on the NLO (respectively NNLO)
longitudinal and transverse coefficient functions.
\smallskip

The functions $D_{(+),i}^h(x,M_f^2)$ and $D_{\Sigma}^h(x,M_f^2)$
are the combinations defined in subsection~2.2.
In the applications considered in this work,
only pairs of charge--conjugate particles are considered
in the final state, i.e., $h=h^+ + h^-$ (or $h^0 + \overline{h}^0$
for neutral hadrons).
Charge--conjugation invariance of the strong interaction
then ensures $D_{q_i}^h(x,M_f^2) = D_{\overline{q}_i}^h(x,M_f^2)$ and
$D_{(-),i}^h(x,M_f^2)$ vanishes. 
The summation index $N_F$ denotes the number of (active) flavors in
the VFNS.
At NLO, $M_f$ defines the scale where the divergence associated with
the collinear radiation off parton $a$ is to be subtracted, 
cf.\ subsection~2.1.
\smallskip

\begin{figure}[hht!]
\begin{center}
\epsfig{file=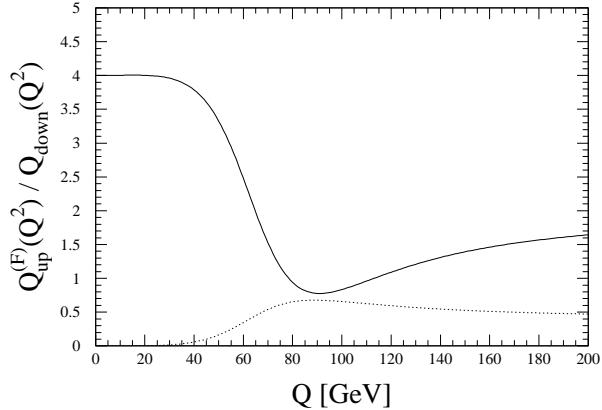,clip=,width=8cm}
\end{center}
\caption[]{\small 
Dependence of the electroweak charges on the virtuality $Q^2$ of
the vector boson.
The solid (dotted) lines show the ratios of $Q_{\rm up}(Q^2)$
 ($Q_{\rm up}^F(Q^2)$) over $Q_{\rm down}(Q^2)$, respectively.
At the $Z$ pole,
the relative weight of $Q_{\rm up}^F(Q^2)$ becomes maximal, whereas
 $Q_{\rm up}(Q^2)$ has its minimum there.
\protect\label{charge}}
\end{figure}

The additional factors
\begin{eqnarray}
\label{cplgs1}
Q_{q_i}(Q^2) &\n \equiv \n& 
     \left( v_e^{\gamma} \right)^2 \left( v_f^{\gamma} \right)^2 
     -2 v_e^{\gamma} v_e^Z v_f^{\gamma} v_f^Z \rho_1(Q^2)  
\nonumber
\\
& & \hspace{1.5cm}
     +\left[ \left( v_e^Z \right)^2 + \left( a_e^Z \right)^2 \right] 
        \left[ \left( v_f^Z \right)^2 + \left( a_f^Z \right)^2 \right]
        \rho_2(Q^2)
\COMMA
\\ 
Q_{q_i}^{F}(Q^2) &\n \equiv \n& 
     \left[ \left( v_e^Z \right)^2 + \left( a_e^Z \right)^2 \right] 
        \left[ \sum_{l=1}^{N_F} a_{q_l}^Z \right]
        a_{q_i}^Z \rho_2(Q^2)
\end{eqnarray}
are due to the different couplings of the quarks to the vector bosons.
The functional dependence on the CM
energy is plotted in Fig.~\ref{charge}
for the ratios $Q_{\rm up}(Q^2)/Q_{\rm down}(Q^2)$ (solid) 
and $|Q_{\rm up}^F(Q^2)|/Q_{\rm down}(Q^2)$ (dotted).
For energies $Q^2 \ll m_Z^2$ the propagator terms $\rho_1(Q^2)$
(for the interference of $\gamma^*$ and $Z$) and $\rho_2(Q^2)$
(two $Z$ propagators) given by\footnote{
   The expressions have been derived in the narrow width
   approximation. Deviations from improved formul{\ae} that take into
   account the energy dependence of the width \cite{con1} are
   small compared to the strong interaction effects under consideration.}
\begin{eqnarray}
\label{rho}
\rho_1(Q^2) &\n=\n& \frac{1}{\left(4 \sin^2\theta_W \cos^2\theta_W
                           \right)}
                   \frac{Q^2 \left( m_Z^2 - Q^2 \right)}
                       {\left(m_Z^2 - Q^2\right)^2 + m_Z^2\Gamma_Z^2}
\SPACE {\rm and}
\\
\rho_2(Q^2) &\n=\n& \frac{1}{\left( 4 \sin^2\theta_W \cos^2\theta_W
                           \right)^2}
                   \frac{\left( Q^2 \right)^2}
                       {\left(m_Z^2 - Q^2\right)^2 + m_Z^2\Gamma_Z^2}
\end{eqnarray}
are small and can be neglected.
The usual couplings are recovered in that limit,
 $Q_{q_i}(Q^2) \to e_{q_i}^2$,
and the diagrams of class $C_F T_R$ that give rise 
to $d\sigma_{k,q}^F$
do not contribute to the cross
section in accord with Furry's theorem \cite{furry}.
The complete set of vector and axial couplings of the photon and
the $Z$ boson, respectively, is given by

\begin{table}[!hhh]
\begin{center}
\begin{tabular}{ll}
 $v_e^{\gamma}=-1$                     & $a_i^{\gamma}=0$    \\
 $v_{\rm{up}}^{\gamma}=+\frac{2}{3}$   & for any fermion $i$ \\
 $v_{\rm{down}}^{\gamma}=-\frac{1}{3}$ &                     \\
\vspace{0.5cm} \\
 $v_e^Z=T_{3,e}-2v_e^{\gamma} \sin^2 \theta_W$ 
        \hspace{1.5cm} & $a_e^Z=T_{3,e}$ \\
 $v_q^Z=T_{3,q}-2v_q^{\gamma} \sin^2 \theta_W$ 
        \hspace{1.5cm} & $a_q^Z=T_{3,q}$ \\
\end{tabular}
\end{center}
\label{t_cplgs1}
\end{table} 
\noindent
where $q$ stands for either an up or a down type quark
and where $\theta_W$ is the Weinberg angle.
The third components of the strong isospin are
\begin{eqnarray}
T_{3,e}        &\n=\n& -\frac{1}{2} \COMMA  \nonumber \\
T_{3,\rm{up}}  &\n=\n& +\frac{1}{2} \COMMA  \nonumber \\
T_{3,\rm{down}}&\n=\n& -\frac{1}{2} \STOP  \nonumber
\end{eqnarray}
\smallskip

The bare, unrenormalized partonic 
cross sections $d\overline{\sigma}_k/dx$ are calculated via
\begin{eqnarray}
\label{sigpart}
d\overline{\sigma}_{k,a}^{(l)} &\n=\n&
\frac{1}{{\rm flux}} |\overline{M}(\epem \to (\gamma^*,Z) 
\to a_1 ... a_l)|^2 \PS^{(l)}
\nonumber
\\
&\n=\n& \frac{1}{2Q^2}\frac{1}{4Q^4}\widetilde{L}_{\mu\nu}H^{\mu\nu}
\PS^{(l)}
\COMMA
\end{eqnarray}
where $\PS^{(l)}$ denotes the phase space of an $l$--particle
final state $a_1...a_l$.
The flux factor is $1/(2Q^2)$, and $1/(4Q^4)$ accounts for the
average over initial spins and for the two photon (or Z) propagators.
The $\theta$--dependence of the lepton tensor is factored out
according to (\ref{sigxtheta}) so that instead of the
full lepton tensor, which gives the sum of
$\sigma_T + \sigma_L$, just
\begin{equation}
\label{lmunu}
\widetilde{L}_{\mu\nu}=4e^2\frac{Q^2}{3}
\sum_{\lambda}
\epsilon_{\mu}^{(\lambda)}\epsilon_{\nu}^{(\lambda)}
\end{equation}
must be contracted with the hadron tensor in eq.~(\ref{sigpart}).
The partonic cross section can then be written in the form
\begin{eqnarray}
\label{sigpart2}
d\overline{\sigma}_{k,a}^{(l)} &\n=\n&
\frac{e^2}{6Q^4} 
\sum_{\lambda}\epsilon_{\mu}^{(\lambda)}\epsilon_{\nu}^{(\lambda)}
H^{\mu\nu}
\PS^{(l)}
\nonumber
\\
&\n=\n& \sigma^{(2)} \left[
\sum_{\lambda}\epsilon_{\mu}^{(\lambda)}\epsilon_{\nu}^{(\lambda)}
\widetilde{H}^{\mu\nu}
\right]
\widetilde{\PS}^{(l)}
(4\pi\overline{\alpha}_s)^{l-2}
\STOP
\end{eqnarray}
The dimensional regularization scheme \cite{tho1,dimreg} is used
to keep track of the singularities in intermediate steps of
the calculation in $n=4-2\ve$ dimensions.
The modified entities in (\ref{sigpart2}) are given by\footnote{
   As usual, the purely leptonic vertex is calculated in 4 dimensions
   from the start so that the electromagnetic couplings give a factor
   of $g^4 = (4\pi\aqed)(4\pi\aqed\mu^{2\ve})$.},
\begin{eqnarray}
\label{sig2n}
\sigma^{(2)} &\n=\n& \sigma_0 \left( \frac{4\pi\mu^2}{Q^2} \right)^{\ve}
                 \frac{\Gamma(2-\ve)}{\Gamma(2-2\ve)}
\COMMA
\\
\widetilde{H}^{\mu\nu} &\n=\n& 
e^{-2}(\overline{g}_s)^{-2(l-2)}
\frac{1}{N_C(1-\ve)}H^{\mu\nu}
\hspace{.7cm} {\rm with} \hspace{.7cm}
\overline{g}_s^2 = 4\pi\overline{\alpha}_s\mu^{2\ve}
\label{hmunun} \COMMA\\
\widetilde{\PS}^{(l)} &\n=\n& \frac{1}{A_0} \mu^{2\ve(l-2)} \PS^{(l)}
\label{psln}
\STOP
\end{eqnarray} 
The mass parameter $\mu$ serves to keep the 
unrenormalized $\overline{\alpha}_s$ dimensionless
also in $n \neq 4$ dimensions.
The factor 
\begin{equation}
\label{A0def}
A_0 = \frac{Q^2}{2\pi} \left[ \frac{\Gamma(1-\ve)}{\Gamma(2-2\ve)}
    \left(\frac{4\pi}{Q^2}\right)^{\ve}\right] 
\end{equation}
will appear in the expressions for the $l$--particle
phase spaces and will thus cancel in $\widetilde{\PS}^{(l)}$.
For $k=T$ the sum in (\ref{sigpart2}) runs over the
two transverse polarizations $\lambda=1,2$ and for $k=L$
just $\lambda=0$ contributes.
In the following, we define 
\begin{equation}
\label{me}
\left[
 \sum_{\lambda}\epsilon_{\mu}^{(\lambda)}\epsilon_{\nu}^{(\lambda)}
 \widetilde{H}^{\mu\nu}
\right]
\equiv |\overline{\widetilde{M}}|^2
\COMMA
\end{equation}
where the couplings have been factored out of the
hadronic matrix elements in (\ref{hmunun}).
The factor $1/N_C$ will be absorbed in the color factors, and 
the additional factor $1/(1-\ve)$
arises from the normalization to the Born--level
quark--antiquark cross section in $n$ dimensions, $\sigma^{(2)}$,
following the notation of \cite{fab1}.
The factor $\sigma^{(2)}$ is universal, i.e., independent of
the number of particles in the final state.
Since the final result will be obtained by $\ve \to 0$,
$\ve$ might be as well set to zero
right away in this common factor
and $\sigma^{(2)}$ then cancels with the 
normalization factor in (\ref{sigxpol}).
The same holds for the factor $1/(1-\ve)$, leading to
\begin{equation}
\label{sigpart3}
\frac{1}{
  \sigma_0}\frac{d\overline{\sigma}_{k,a}^{(l)}}{dx}
=(4\pi\overline{\alpha}_s)^{l-2} |\overline{\widetilde{M}}|^2 
\frac{d}{dx} \widetilde{\PS}^{(l)}
\STOP
\end{equation}
The renormalization of $\overline{\alpha}_s$, and the
factorization of the poles in (\ref{sigpart3}) into the
(singular) bare FF's will be performed in
subsection~3.3.3.
This will introduce the renormalization scale $\mu_R$ and the
factorization scale $M_f$.

\vspace{2cm}
\subsection{Transverse Cross Section to ${\cal O}(\alpha_s)$}
\smallskip

The polarization tensor for the transversely polarized photon reads 
\begin{equation}
\label{Tpol}
\sum_{\lambda=1,2}
\epsilon_{\mu}^{(\lambda)}\epsilon_{\nu}^{(\lambda)}
=
-g_{\mu\nu}-\epsilon_{\mu}^{(0)}\epsilon_{\nu}^{(0)}
\COMMA
\end{equation}
in the Feynman gauge.
The longitudinal polarization tensor that is
subtracted in (\ref{Tpol}),is given in the next subsection.
\smallskip

The right hand side of (\ref{sigpart3}) is expressed in the form
of the coefficient functions $\COFm_a^{(m)}$ of parton $a$
in $m$-th order. 
The leading order does not receive virtual corrections and
no singularities need to be factored off
so that the coefficient functions are
in that case just the real matrix elements.
For higher orders, things are slightly more involved
and will be discussed for the longitudinal case.
Here, just the final result in the {\msbar}--scheme 
\cite{msbar} is given. 
\begin{eqnarray}
\frac{1}{N_C\sigma_0}
   \frac{d\sigma_{T,q}^{NS}}{dz}
&\n=\n&  \COFm_{T,q}^{NS,(0)}(z)
+\frac{\alpha_s(\mu_R^2)}{2\pi}
   \left[P_{q\to q}^{(0)}(z)\ln{\left(\frac{Q^2}{M_f^2}\right)}
     +\COFm_{T,q}^{NS,(1)}(z)\right]
\label{sigTNSren}
\COMMA
\\
\frac{1}{N_C\sigma_0}
   \frac{d\sigma_{T,q}^{PS}}{dz}
&\n=\n& 0
\label{sigTSren}
\COMMA
\\
\frac{1}{N_C\sigma_0}
   \frac{d\sigma_{T,q}^{F}}{dz}
&\n=\n& 0
\label{sigTFren}
\COMMA
\\
\frac{1}{N_C\sigma_0}
   \frac{d\sigma_{T,G}}{dz}
&\n=\n& \frac{\alpha_s(\mu_R^2)}{2\pi}
   \left[2P_{q\to G}^{(0)}(z)\ln{\left(\frac{Q^2}{M_f^2}\right)}+
\COFm_{T,G}^{(1)}(z)\right]
\label{sigTGren}
\STOP
\end{eqnarray}
The strong coupling is evaluated at the renormalization scale $\mu_R$.
For the choice $\mu_R^2=M_f^2=Q^2$, the logarithms in (\ref{sigTNSren}),
(\ref{sigTGren})
vanish and the coefficient functions alone determine
the cross sections.
The above expressions are given to NLO which 
is ${\cal O}(\alpha_s)$ in the transverse case.
The dependence on the factorization scale is reduced at NLO.
Furthermore, the $K$~factor for the NNLO,
\begin{equation}
\label{kfactor}
K^{NNLO} \equiv \frac{{\rm Observable}^{NNLO}}
     {{\rm Observable}^{NLO}} 
\COMMA
\end{equation}
is not expected to differ significantly from 1.
For example,
the transverse part of the total cross section to NLO,
evaluated with $\alpha_s(m_Z)=0.12$
is from equations (\ref{Reelong}) through (\ref{Reetot})
\begin{equation}
\label{Reet}
R_{ee,T}^{NLO} \equiv \frac{\sigma_T^{NLO}}{\sigma_{\rm had}} = 
 \frac{1}{1+\frac{\alpha_s}{\pi}} = 0.963
\STOP
\end{equation}
This compares favorably to \cite{opa3}
\begin{equation}
\label{Reetexp}
R_{ee,T}^{\rm exp.} = 0.943 \pm 0.005 \STOP
\end{equation}
\smallskip

The coefficient functions have been calculated 
to  ${\cal O}(\alpha_s^2)$ \cite{rij2} but the NLO expressions
are sufficient for our purposes 
and have the advantage that the splitting functions 
can be implemented consistently at this order.

\vspace{2cm}
\subsection{Longitudinal Cross Section to ${\cal O}(\alpha_s^2)$}
\smallskip

As can be seen from (\ref{sigxtheta}),
the longitudinal polarization state of the intermediate vector boson
is selected by the angular distribution of the hadron produced in the
inclusive process.
In the framework considered in this thesis, the three--momentum of the 
hadron is aligned with that of its parent parton.
Hence, the momentum of the parton produced in the hard scattering
is linked to the gauge boson polarization.
The fragmenting parton is labeled 3.
The polarization vector reads
\begin{equation}
\label{pol2}
\epsilon_{\mu}^{(0)} = \frac{p_{3 \mu}}{p_{3 0}}
         - \frac{q_{\mu}}{q_0}
\end{equation}
and due to current conservation the polarization tensor is then
effectively given by
\begin{equation}
\label{pol3}
\epsilon_{\mu}^{(0)}\epsilon_{\nu}^{(0)} = 
Q^2 \frac{p_{3 \mu} p_{3 \nu}}{(p_3 \cdot q)^2}
=\frac{4}{Q^2x_3^2}  p_{3 \mu} p_{3 \nu}
\STOP
\end{equation}
\smallskip

At ${\cal O}(\alpha_s^0)$, only the transversely 
polarized photon and $Z$ boson contribute to the cross section.
To ${\cal O}(\alpha_s)$, the longitudinal cross 
sections of the subprocesses are given by
\begin{eqnarray}
\label{sigl}
\frac{1}{N_C\sigma_0}
\frac{d\sigma_{L,q}^{NS}}{dz}
&\n=\n& \frac{\alpha_s(\mu_R^2)}{2\pi} \COFm_{L,q}^{NS,(1)}(z)
\COMMA
\\
\frac{1}{N_C\sigma_0}
\frac{d\sigma_{L,q}^{PS}}{dz}
&\n=\n& 0
\COMMA
\\
\frac{1}{N_C\sigma_0}
\frac{d\sigma_{L,q}^{F}}{dz}
&\n=\n& 0
\COMMA
\\
\frac{1}{N_C\sigma_0}
  \frac{d\sigma_{L,G}}{dz}
&\n=\n& \frac{\alpha_s(\mu_R^2)}{2\pi} \COFm_{L,G}^{(1)}(z)
\STOP
\end{eqnarray}
Inserting this together with the coefficient functions
from appendix~D in (\ref{sigxpol}) leads to 
\begin{eqnarray}
\label{sigmal}
& &
 \frac{1}{N_C\sigma_0}\frac{d\sigma_{L}^{h}}{dx}=
 N_C C_F\frac{\alpha_s(\mu_R^2)}{2\pi}\int_x^1\frac{dz}{z}
 \left\{
 \sum_{i=1}^{N_F}Q_{q_i}(Q^2) D_{(+),i}^h\left(\frac{x}{z},M_f^2\right)
\right.
\nonumber\\
& &
\hspace{3cm} \left.
   + \left[ \sum_{i=1}^{N_F}Q_{q_i}(Q^2) \right]
   4\, \frac{1-z}{z}D_{G}^h\left(\frac{x}{z},M_f^2\right)
 \right\}
 +{\cal O}(\alpha_s^2)
\STOP
\end{eqnarray}
Since eq.~(\ref{sigmal}) is only given to LO in $\alpha_s$, 
the $\alpha_s$
value here is not expected to be the same as the one in the NLO
relations, i.e., when the NLO expressions are inserted
in (\ref{sigxpol}).
ALEPH \cite{ale3} and OPAL \cite{opa3} have recently
presented data on $d\sigma_L/dx$.
In Fig.~\ref{sl0a}, eq.~(\ref{sigmal}) is evaluated 
with the NLO set of \cite{binpk} and two-loop
 $\alpha_s$ and compared to the ALEPH data \cite{ale3}.
The result obtained (full curve) falls short of 
the data by a factor of two.
At this point, one should keep in mind that 
eq.~(\ref{sigmal}) is a LO
prediction, and one should be prepared to 
allow for a $K$ factor, which
is typically larger than one.
This $K$ factor may be simulated by changing the 
renormalization and factorization scales.
When choosing low scales of $\mu_R=M_f=20$~GeV, satisfactory agreement 
with the data is found.
\smallskip

In fact, already the total QCD corrections, $R_{ee,L}=1-R_{ee,T}$, fall
well short of the experimental data in LO,
\begin{equation}
\label{Reel}
R_{ee,L}^{LO} = 0.037 \hspace{.6cm} {\rm , whereas} \hspace{.6cm}
R_{ee,L}^{\rm exp.} = 0.057 \pm 0.005 \hspace{.6cm} {\cite{opa3}} 
\STOP
\end{equation}
\smallskip

\setlength{\unitlength}{1cm}
\begin{figure}[hht!]
\begin{center}
\begin{picture}(12.5,7.5)
\put(2.4,3.4){\rotateleft{$\rm \frac{1}{\sigma_{\rm had}}
      \frac{d\sigma_L^{h^{\pm}}}{dx}$}}
\put(7.0,0.2){$\rm x$}
\put(3.0,0.7){\makebox{
\epsfig{file=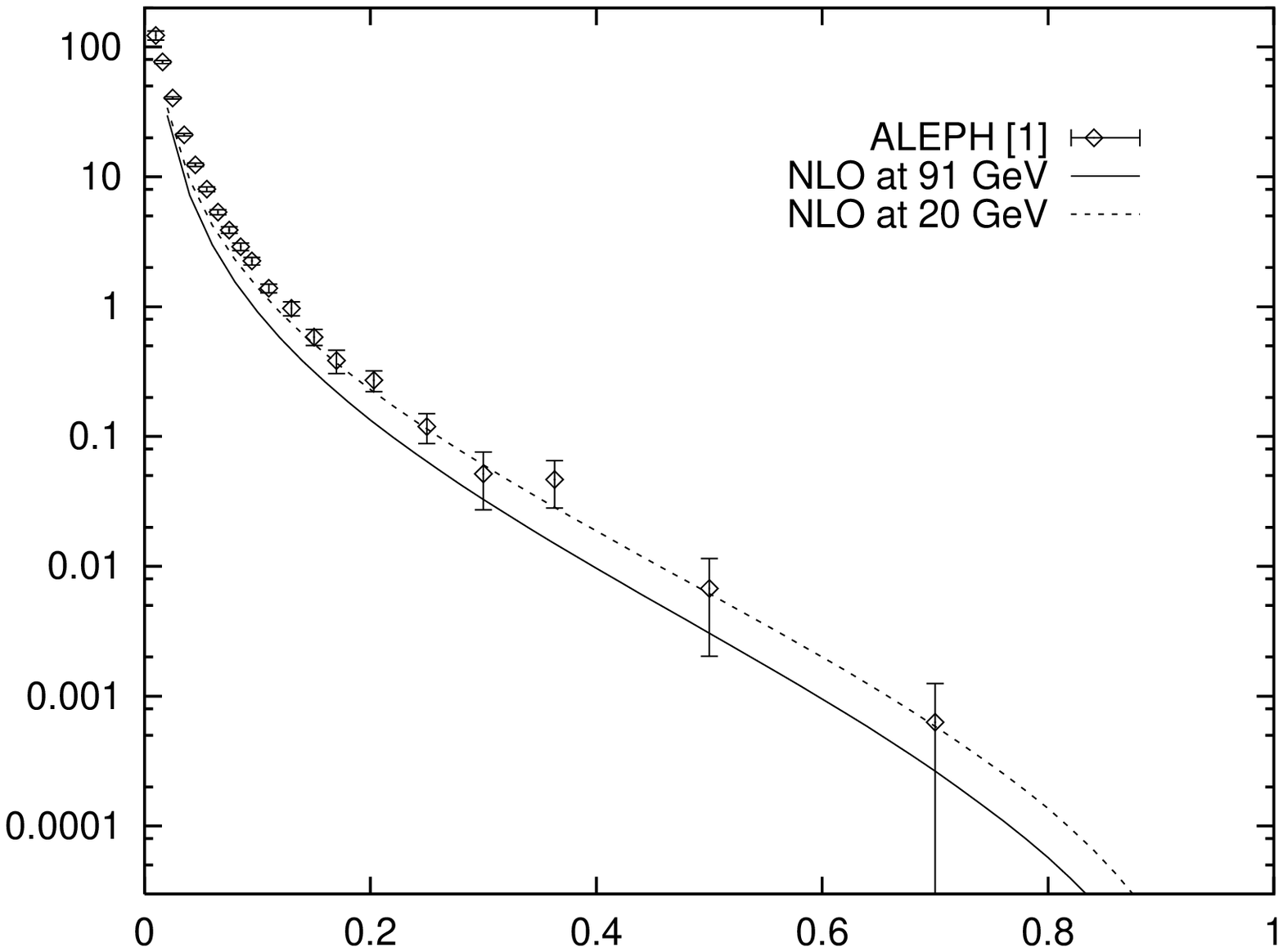,clip=,width=7.3cm,height=6.8cm}}}
\end{picture}
\end{center}
\vspace{-0.5cm}
\caption[]{\small 
Scale dependence of the longitudinal cross section in \ee
annihilation to LO.
While the LO result with $M_f=m_Z$ (solid) falls short of the
ALEPH data \cite{ale3}
by a significant amount, a very low scale of $M_f=20$~GeV
(dotted) leads to satisfactory agreement.
The plot is taken from \protect\cite{binpk}, where the 
calculation with the leading order coefficient functions was
misleadingly labeled NLO.
\protect\label{sl0a}}
\end{figure}

The inclusion of the subleading 
corrections to the longitudinal cross section reduces
the scale dependence, as has already been argued.
Moreover, only at NLO do the coefficient functions acquire a
scheme dependence.
Consistently using the NLO Altarelli--Parisi kernels with the
NLO longitudinal cross section leads to a partial cancellation
of this scheme dependence and thus to a better
perturbative stability.
Our policy therefore is to calculate the longitudinal
cross section to NLO when comparing with longitudinally polarized
particle production data and to calculate the unpolarized cross section,
which is the sum of the longitudinal and the transverse parts, also
to NLO, when comparing to unpolarized data.
In the latter case, the NLO contribution to the transverse part is
of ${\cal O}(\alpha_s)$ so that the  longitudinal part is
consistently added at the same order of $\alpha_s$, although
this is the leading contribution.
\smallskip

The $\COFm_L^{(2)}$ functions and the formul{\ae} for the
cross sections to ${\cal O}(\alpha_s^2)$ 
have been presented in \cite{rij1} for the process
 $\epem \to \gamma^* \to h X$.
In the next section,
we will complete the calculation with the class $C_F T_R$
that contributes to $\epem \to Z \to h X$ in the case of an
odd number of active flavors.
For a cross check we will also repeat the calculation of 
class $N_F T_R C_F$.
\smallskip

The two classes treated here belong 
to the process $\gamma^* / Z \to q\overline{q}q\overline{q}$
that does not receive virtual corrections 
to order ${\cal O}(\alpha_s^2)$
-- this order is the LO for this final state\footnote{
   In addition, the two--loop virtual corrections do
   not contribute to $\sigma_L$ at all, as 
   one of the vector bosons couples directly to the
   observed quark/antiquark in the relevant cut--diagrams,
   leading to vanishing matrix elements by current conservation.}.
Hence, only the real corrections must be calculated in the next
subsections.

\vspace{1cm}
\subsubsection{Matrix Elements for the Real Corrections}

In this section all matrix elements are given for the 
process $\gamma^*/Z \to q\overline{q}q\overline{q}$
to ${\cal O}(\alpha_s^2)$ in the dimensional 
regularization scheme\footnote{
   The calculation of the gamma traces has been
   performed with the help of {\tt REDUCE} \cite{red}.
   For color class $C_F T_R$, the traces involve $\gamma_5$--matrices
   that have been implemented in \cite{red} for four 
   dimensions only.
   This is sufficient for the present analysis, 
   as no poles appear in class $C_F T_R$.
   One may obtain the $n$--dimensional results with the use of
   the Breitenlohner--Maison scheme \cite{tho1,bm} for $\gamma_5$
   that has been implemented in \cite{tracer,mat1}.}.
Those for the process $\gamma^*/Z \to q\overline{q}GG$
can be found in \cite{ali1}.
\begin{figure}[hht!]
\begin{center}
\epsfig{file=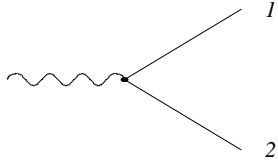 , clip=,width=14cm }
\end{center}
\vspace{-0.4cm}
\caption[]{\small 
Born--amplitude for the inclusive production of a parton $l$ 
in $\gamma^*/Z \to l X$.
\protect\label{bornamp}}
\end{figure}
 
\begin{figure}[hht!]
\begin{center}
\epsfig{file=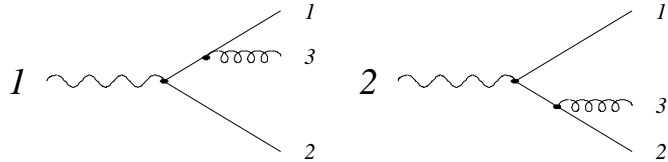 , clip=,width=14cm }
\end{center}
\vspace{-0.4cm}
\caption[]{\small 
Amplitudes contributing to ${\cal O}(\alpha_s)$.
\protect\label{loamp}}
\end{figure}

The Born approximation to the \ee annihilation process is given
in Fig.~\ref{bornamp}.
As has been pointed out, it is of order $\alpha_s^0$.
Moreover, it does not contribute to the longitudinal cross section.
To ${\cal O}(\alpha_s)$, the amplitudes in Fig.~\ref{loamp}
contribute to the cross section, with color factor $C_F$.
As this is the leading order, there are no virtual corrections.

For the factorization later on one also needs the ${\cal O}(\ve)$
corrections to the matrix elements of the three
particle final states that are given in the literature to
 ${\cal O}(\ve^0)$, only.
In $n$ dimensions, the matrix elements are 
-- apart from the color factor $C_F$ -- given by
\small
\begin{eqnarray}
Q22 &\n=\n&  0 \\
Q21 &\n=\n&  0 \\
Q11 &\n=\n&  16(1-\ve )\frac{y_{12}}{x_1^2} \label{mes0Q} 
\end{eqnarray}
for the fragmentation of parton 1.
As this is a quark (or antiquark), the elements
are named $Qkl$ for the interference of amplitudes $k$ and $l$
in Fig.~\ref{loamp}.
The momentum of the fragmenting parton defines the azimuthal
angle $\theta$ so that the results for the matrix elements
differ if some other parton is selected.
When parton 3, a gluon, fragments, the matrix elements are 
named $G22$ and so on and read
\begin{eqnarray}
G22 &\n=\n&  0 \\
G21 &\n=\n&  32 \frac{y_{12}}{x_3^2} \label{mes0G} \\
G11 &\n=\n&  0 
\end{eqnarray}
\normalsize
$x_i\equiv 2 p_i \! \cdot \! q/Q^2$ is just the 
fraction of the total energy that is carried by parton $i$.
The $y_{ij}$ are normalized dot products of the momenta
of particles $i$ and $j$, c.f.\ (\ref{ydefs}) below.
\smallskip

\begin{figure}[hht!]
\begin{center}
\epsfig{file=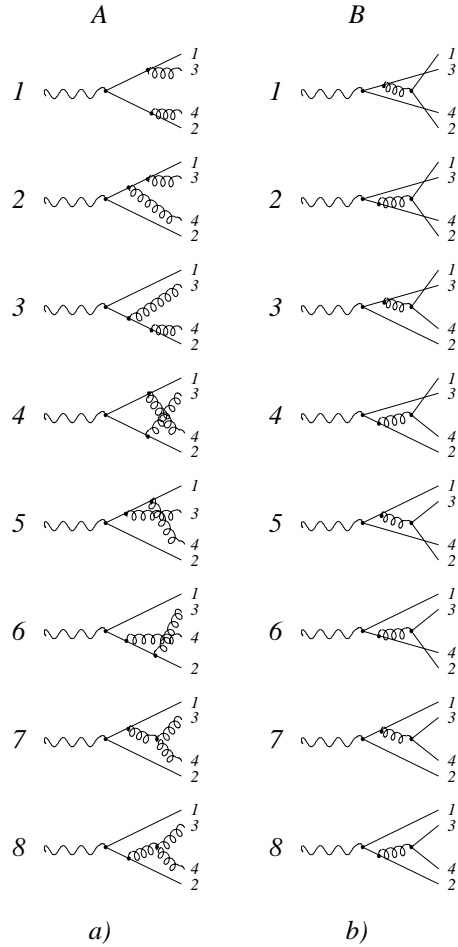 , clip=,width=8.0cm }
\end{center}
\vspace{-0.4cm}
\caption[]{\small 
Amplitudes contributing to  ${\cal O}(\alpha_s^2)$,
a)
for the $q\overline{q}GG$ and 
b)
for the $q\overline{q}q\overline{q}$ final state.
Parton 1 (and 3 for b)) is a quark and 2 (and 4) an antiquark.
\protect\label{nloamp}}
\end{figure}

To order $\alpha_s^2$, many amplitudes contribute to the
real corrections to the longitudinal cross section.
Following the conventions of \cite{ali1},
they are shown in Fig.~\ref{nloamp}a for the $q\overline{q}GG$
and in Fig.~\ref{nloamp}b for the $q\overline{q}q\overline{q}$
final--states, respectively.
\smallskip

When the amplitudes of the process \ref{nloamp}a are added and squared,
the interference (and diagonal) terms come with three
distinct color factors.
Those of process \ref{nloamp}b also come with three different
factors.
Following \cite{ell2}, they will be grouped together
according to their color factors, giving 6 contributions to
the longitudinal cross section, labeled A through F, as
listed in Tables~\ref{tabproca} and \ref{tabprocb}.
\smallskip

\begin{table}[hht!]
\begin{center}
{
\small
\begin{tabular}{|c|llll|llll|lllll|}
\hline
perm. of& \multicolumn{4}{|c|}{class A} &\multicolumn{4}{|c|}{class B} &
\multicolumn{5}{|c|}{class C} \\
first row: & \multicolumn{4}{|c|}{$C_F^2$} 
&\multicolumn{4}{|c|}{$C_F (C_F-{\scriptstyle{\frac{1}{2}}} N_C)$} &
\multicolumn{5}{|c|}{$C_F N_C$} \\
\hline
 & A11 & A32 & A21 & A22 & A42 & A52 & A53 & A41 & 
A71 & A72 & A82 & A77 & A87 \\
(34) & A44 & A65 & A54 & A55 & A51 & & A62 & & 
A74 & A75 & A85 & & \\
(12) & & & A64 & A66 & A61 & & & & A84 & A86 & A76 & A88 & \\
(12)(34) & & & A31 & A33 & A43 & A63 & & & A81 & A83 & A73 & & \\
\hline
\end{tabular}
\normalsize
}
\end{center}
\caption[]{\small
The color classes that contribute to the process
$\epem \to q\overline{q} GG$.
\protect\label{tabproca}}
\end{table}

\begin{table}[hht!]
\begin{center}
{
\small
\begin{tabular}{|c|lll|llllll|lll|}
\hline
perm. of& \multicolumn{3}{|c|}{class D} &\multicolumn{6}{|c|}{class E} &
\multicolumn{3}{|c|}{class F} \\
first row: & \multicolumn{3}{|c|}{$N_F T_R C_F$} 
&\multicolumn{6}{|c|}{$C_F (C_F-{\scriptstyle{\frac{1}{2}}} N_C)$} &
\multicolumn{3}{|c|}{$C_F T_R$} \\
\hline
         & B77 & B88 & B87 & B83 & B76 & B73 & B86 & B84 & B75 
& B81 & B82 & B53 \\
(24)     & B55 & B66 & B65 & B61 & B85 & B51 & & B62 & 
& B63 & B64 & B71 \\
(13)     & B33 & B44 & B43 & B74 & B32 & & B42 & & B31 
& B54 & & \\ 
(24)(13) & B11 & B22 & B21 & B52 & B41 & & & & 
& B72 & & \\
\hline
\end{tabular}
\normalsize
}
\end{center}
\caption{\small
The color classes that contribute to the process
$\epem \to q\overline{q} q\overline{q}$.
\protect\label{tabprocb}}
\end{table}

The Tables are organized in columns of matrix elements that can be 
transformed into each other by interchanging a pair of partons
in the final state, $(12)$ for example 
represents the exchange of partons 1 and 2.
Only part of the permutations in the tables are actually
allowed for the longitudinal matrix elements, 
depending on which parton is singled out in the final state.
For the fragmentation of parton 3, only the permutations
(12), respectively (24) are permitted, whereas for the
fragmentation of parton 1 the permutations
(34) and (13), respectively, are applicable.
\smallskip

The color factors are obtained from the color traces \cite{sch1}
\begin{eqnarray}
\label{colortraces}
\sum_a tr \left\{ T_a T_a \right\} &\n=\n& N_C C_F 
\\
tr \left\{ T_a T_b \right\} &\n=\n& T_R \delta_{ab} 
\\
-i f_{abc} tr \left\{ T_a T_b T_c \right\} &\n=\n& \frac{1}{2} N_C^2 C_F 
\\
tr \left\{ T_a T_b T_a T_b \right\} &\n=\n& N_C C_F \left( C_F 
     - \frac{1}{2} N_C \right)
\\
tr \left\{ T_a T_b T_b T_a \right\} &\n=\n& N_C C_F^2 
\STOP
\label{colr}
\end{eqnarray}
In the case of the color gauge group SU($N_C$), the Casimir operator 
of the fundamental representation has the eigenvalue
$C_F=T_R\left(N_C^2-1\right)/(N_C)$ and 
the trace normalization is $T_R=1/2$ \cite{mar7}.
\smallskip

For the process $\epem \to q\overline{q}GG$ the complete
set of NLO coefficient functions is given in \cite{rij1}.
We will hence only be concerned with the process
$\epem \to q\overline{q} q\overline{q}$ in the following.
This process does not receive virtual corrections.
We will recalculate the color class D as
an independent check of the results presented in \cite{rij1}
and also consecutively calculate class F that is needed 
for the comparison to LEP1 data.
\smallskip

\begin{figure}[hht!]
\begin{center}
\epsfig{file=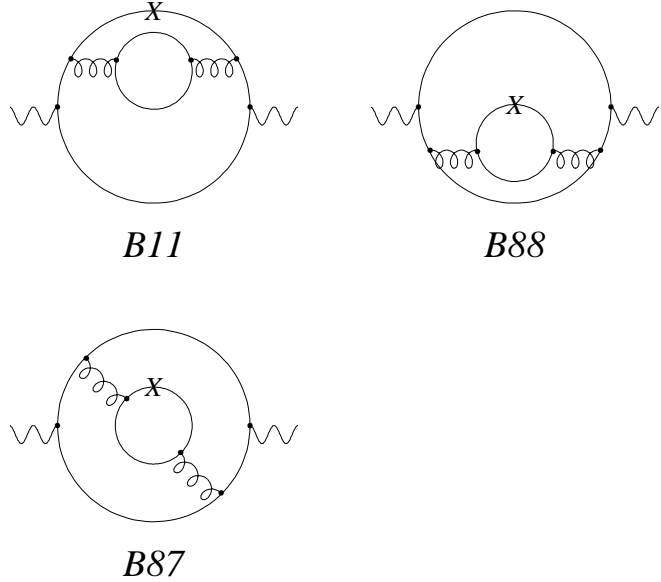 , clip=,width=12cm }
\end{center}
\vspace{-0.3cm}
\caption[]{\small 
Cut--diagrams for \ee annihilation at ${\cal O}(\alpha_s^2)$,
color class $N_F T_R C_F$.
The cuts are not indicated -- they run vertically through the
center of each diagram so that the final state consists of four
quarks.
The fragmenting quark (3) is indicated by an $X$.
\protect\label{cutdiagD}}
\end{figure}

Fig.~\ref{cutdiagD} shows the cut--diagrams of class~D.
Any other matrix element can be obtained from those
topologies via crossing, as specified in eq.~(\ref{mesd})ff.
The fragmenting quark, marked by an $X$,
couples only indirectly to the vector
boson in the cut--diagrams $B88$ and $B87$.
The intermediate gluon is a flavor singlet so that
those matrix elements give rise to pure singlet contributions.
These will be denoted by $PS$.
Together with the non--singlet $NS$
they make up the singlet, $\Sigma$. 
The matrix elements for class $N_F T_R C_F$ are -- apart from the common
color factor --
\small
\begin{eqnarray}
\label{mesd}
B88 &\n=\n&  \frac{32}{Q^2} 
        \frac{(2y_{23}^2+2y_{23}y_{34}+y_{34}^2-\ve y_{34}^2
      )y_{13}y_{24}}{x_3^2y_{234}^2y_{34}^2} \\
B87 &\n=\n&  \frac{32}{Q^2} 
       \frac{(\ve y_{34}^2+2y_{13}y_{23}+y_{13}y_{34}+y_{23}y_{34}
      )(y_{12}y_{34}-y_{13}y_{24}-y_{14}y_{23})}
       {x_3^2y_{134}y_{234}y_{34}^2} \\
B11 &\n=\n&  \frac{32}{Q^2} 
       \frac{(y_{13}^2+y_{23}^2-\ve y_{13}^2-2\ve y_{13}y_{23}-\ve 
      y_{23}^2)y_{34}}{x_3^2y_{123}^2y_{12}} \\
B33 &\n=\n& cr24(B11) \\
B77 &\n=\n& cr12(B88) \\
B66 &\n=\n& cr24(B88) \\
B65 &\n=\n& cr24(B87) \\
B55 &\n=\n& \Pi_{(142)}(B88) \label{permute} \\
B44 &\n=\n&  0  \\
B43 &\n=\n&  0  \\
B22 &\n=\n&  0  \\
B21 &\n=\n&  0  
\STOP
\end{eqnarray}
\normalsize
The operations $cr12(M)$ etc.\ denote the interchange of two
partons in the matrix element $M$.
The $\Pi_{(142)}$ in eqn.~(\ref{permute}) represents the 
cyclic permutation of the indices $1,4,2$
in the manner $1 \to 4$, $4 \to 2$, and $2 \to 1$.
The matrix elements for the class $C_F(C_F- 1/2 N_C)$ are given by
\small
\begin{eqnarray}
\label{mese}
B86 &\n=\n&  \frac{64}{Q^2}
    \frac{(2y_{23}^2+2y_{23}y_{34}+y_{34}^2-\ve y_{34}^2
 )y_{13}y_{24}}{x_3^2y_{23}y_{234}^2y_{34}} \\
B84 &\n=\n&  0  \\
B83 &\n=\n&  \frac{32}{Q^2}
   \left(\ve y_{12}y_{13}y_{34}^2+\ve y_{12}y_{34}^3
 +\ve y_{13}^2y_{24}y_{34}-\ve y_{13}y_{14}y_{23}y_{34}
+\ve y_{13}y_{24}y_{34}^2
\right.
\nonumber\\
& & \left. \hspace{.6cm}
-\ve y_{14}y_{23}y_{34}^2-y_{12}
 y_{13}y_{23}y_{34}-y_{12}y_{13}y_{34}^2+y_{12}y_{23}y_{34}^2-y_{13}^2
 y_{23}y_{24}
\right.
\nonumber\\
& & \left. \hspace{.6cm}
-y_{13}^2y_{24}y_{34}+y_{13}y_{14}y_{23}^2+y_{13}y_{14}y_{23}
 y_{34}-3y_{13}y_{23}y_{24}y_{34}
\right.
\nonumber\\
& & \left. \hspace{.6cm}
-2y_{13}y_{24}y_{34}^2-y_{14}y_{23}^2
 y_{34}
\right)
\frac{1}{x_3^2y_{134}y_{14}y_{234}y_{34}} \\
B76 &\n=\n&  \frac{32}{Q^2}
\frac{(\ve y_{34}^2+2y_{13}y_{23}+y_{13}y_{34}+y_{23}y_{34}
 )(y_{12}y_{34}-y_{13}y_{24}-y_{14}y_{23})}{x_3^2y_{134}y_{23}y_{234}
 y_{34}} \\
B75 &\n=\n&  -\frac{32}{Q^2}
  \left(\ve y_{12}y_{34}^2-\ve y_{13}y_{24}y_{34}+
 \ve y_{14}y_{23}y_{34}+y_{12}y_{13}y_{34}-4y_{13}^2y_{14}
\right.
\nonumber\\
& & \left. \hspace{.6cm}
-y_{13}^
 2y_{24}-3y_{13}y_{14}y_{23}-2y_{13}y_{14}y_{34}-2y_{14}y_{23}y_{34}
\right)
\frac{1}{x_3^2y_{123}y_{134}y_{34}} \\
B74 &\n=\n&  0  \\
B73 &\n=\n&  -\frac{32}{Q^2}
    \left(\ve y_{12}y_{13}y_{34}^2+\ve y_{12}y_{34}^3-
 \ve y_{13}^2y_{24}y_{34}+\ve y_{13}y_{14}y_{23}y_{34}
-\ve y_{13}y_{24}y_{34}^2
\right.
\nonumber\\
& & \left. \hspace{.6cm}
+\ve y_{14}y_{23}y_{34}^2+y_{12}
 y_{13}^2y_{34}+y_{12}y_{13}y_{34}^2-y_{13}^3y_{24}+y_{13}^2y_{14}y_{23}
\right.
\nonumber\\
& & \left. \hspace{.6cm}
 -y_{13}^2y_{24}y_{34}-3y_{13}y_{14}y_{23}y_{34}-2y_{14}y_{23}y_{34}^2
\right)
\frac{1}{x_3^2y_{134}^2y_{14}y_{34}} \\
B31 &\n=\n&  -\frac{32}{Q^2}\left(2\ve y_{13}^2y_{14}+2\ve y_{13}y_{14}
 y_{23}+2\ve y_{13}y_{14}y_{34}+2\ve y_{14}y_{23}y_{34}
\right.
\nonumber\\
& & \left. \hspace{.6cm}
 +y_{12}y_{13}y_{34}-y_{12}y_{34}^2-2y_{13}^2y_{14}
-y_{13}^2y_{24}
\right.
\nonumber\\
& & \left. \hspace{.6cm}
 -y_{13}y_{14}y_{23}+y_{13}y_{24}y_{34}-y_{14}y_{23}y_{34}
\right)
\frac{y_{23}}{x_3^2y_{12}
 y_{123}y_{134}y_{14}} \\
B62 &\n=\n&  0  \\
B61 &\n=\n&  cr12(B83) \\
B52 &\n=\n&  0  \\
B51 &\n=\n&  cr12(B73) \\
B42 &\n=\n&  0  \\
B41 &\n=\n&  0  \\
B32 &\n=\n&  0  \\
B85 &\n=\n&  cr12(B76)
\end{eqnarray}
\normalsize
and those of class $C_F T_R$ are
\small
\begin{eqnarray}
\label{mesf}
B81 &\n=\n&  \frac{32}{Q^2}
     \frac{(y_{14}y_{23}-y_{12}y_{34}-y_{13}y_{24})
     (y_{13}y_{23}+y_{13}y_{34}+
      y_{23}^2)}{x_3^2y_{12}y_{123}y_{234}y_{34}} + {\cal O}(\ve) 
\\
B82 &\n=\n&  0 
\\
B53 &\n=\n&  \Pi_{(142)}(B81)
\\
B63 &\n=\n&  cr24(B81)
\\
B64 &\n=\n&  0 
\\
B71 &\n=\n&  cr12(B81)
\\
B54 &\n=\n&  0  
\\
B72 &\n=\n&  0 
\COMMA
\end{eqnarray}
\normalsize
where $Bij$ stands for the interference term of amplitude $i$
with amplitude $j$ of Fig.~\ref{nloamp}b.
To account for the $Bij$, $i<j$ that are not listed, the
off--diagonal matrix elements are weighted by a factor of two.
Note that the matrix elements listed above are for the fragmentation of 
parton~3.
As there are two quarks/antiquarks in the final state, the cross section
receives an additional factor of 2. 
\smallskip

\begin{figure}[hht!]
\begin{center}
\epsfig{file=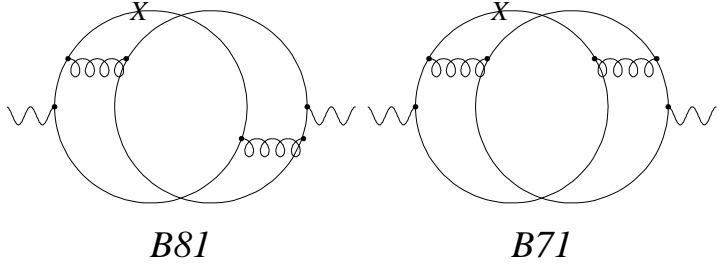 , clip=,width=12cm }
\end{center}
\vspace{-0.3cm}
\caption[]{\small 
Cut--diagrams for \ee annihilation at ${\cal O}(\alpha_s^2)$,
color class $C_F T_R$.
\protect\label{cutdiagF}}
\end{figure}

In Fig.~\ref{cutdiagF} the cut--diagrams of the color class~F are
displayed.
They differ from the cut--diagrams of any of the other classes
(only partially given here, see \cite{ell2} for a complete set)
in the number of gauge boson couplings per loop.
While any fermion loop in the other classes couples to
an even number of bosons, the loops 1 and 2 in class~F each couple to
three bosons, namely two gluons and one electroweak gauge boson
 $V_1$ and $V_2$, respectively.
For the case of $V_1=V_2= \gamma$, the sum over the matrix elements
vanishes due to Furry's theorem \cite{furry}.
This cancellation comes about because there is essentially only one
matrix element, $B81$, and the others can be obtained via
crossing.
However, permuting the momenta 1 and 2 
in $B71$ gives a relative minus sign
to $B81$ so that the two cancel, and analogously for the pair
 $B63$ and $B53$.
When considering just one of the two vector bosons to be a
 $Z$--boson the same cancellation is at work and only the diagrams with
 $V_1=V_2=Z$ survive.
Due to the axial part of the $Zqq$ coupling (the vector part vanishes
in the same manner as outlined above), the matrix elements are
different from those with $V_i=\gamma$.
The main change is that they acquire an additional minus sign
when crossed which spoils the cancellation.
Apart from the minus sign, though, the symmetry is still intact
and thus only $B81$ needs to be kept with an extra factor of four.
As usual, parton 3 is taken to fragment.
The electroweak coupling is then given by
\[
\left[\left(v_e^Z\right)^2+\left(a_e^Z\right)^2\right]
    a_{q_3}^Za_{q_1}^Z
\COMMA
\]
where the factor in square brackets accounts for the coupling on the
leptonic side.
As $q_1$ is unobserved one can right away sum over all
 $q_i, \, \, i=1...N_F$.
Because of $a_{q_{up}}=-a_{q_{down}}$, the up-- and down--type
unobserved loops cancel.
This would lead to a total cancellation 
if the quarks were truly massless.
Albeit that is not the case, and at LEP1 the top quark
is not active in the VFNS framework, but the bottom must
be taken into account.
So the one graph that survives these numerous cancellations is
 $4 B81$ with two $Z$'s and one $b$ loop.

The various interference terms in each
color class are finally grouped together.
The matrix element for any specific color class then reads
\begin{equation}
\label{meamp}
|\overline{\widetilde{M}}|^2 = 
N_C F_{\rm color} \,  \sum_{i \ge j} \, Bij 
\COMMA
\end{equation}
e.g., $|\overline{\widetilde{M}}|_F^2 = 
    C_F T_R  (4  B81)$ for the class $F$.
\smallskip

The $x_i$, $y_{ij}$ and $y_{ijk}$ that appear in the matrix elements
are simple functions of dot--products
of the four--momenta of the intermediate vector boson and the
four final--state particles, $q$ and $p_i$, respectively
($i=1,..,4$):
\begin{eqnarray}
\label{ydefs}
x_i &\n \equiv \n& \frac{2 \, p_i \cdot q}{Q^2} \hspace{.5cm}
, \hspace{.3cm} x_i \in (0,1]
\hspace{.5cm} {\rm and}   \hspace{.5cm} \sum_i x_i =2
\\
y_{ij} &\n \equiv \n& \frac{(p_i+p_j)^2}{Q^2} = 
     \frac{2p_i \cdot p_j}{Q^2}
\hspace{.5cm} , \hspace{.3cm} y_{ij} \in [0,1]
\hspace{.5cm} {\rm and} \hspace{.5cm}
\sum_{i<j} y_{ij} = 1 
\\
y_{ijk} &\n \equiv \n& \frac{(p_i+p_j+p_k)^2}{Q^2} = 
y_{ij}+y_{ik}+y_{jk}
\hspace{.2cm} , \hspace{.5cm} y_{ijk} \in [0,1]
\STOP
\end{eqnarray}
These definitions apply to any final state of $l$ massless partons.
In addition, for any specific $l$
there are a number of useful relations between these variables
that read for $l=4$:
\begin{eqnarray}
\label{yrel3}
y_{ij} &\n=\n& 1-x_k-x_m+y_{km}
\\
y_{ij}+y_{ik}+y_{im} &\n=\n& 1-y_{jkm} 
\\
y_{ijk} &\n=\n& 1-x_m \label{kinx3}
\\
y_{ijk}+y_{ijm}+y_{jkm} &\n=\n& 2 \label{kiny3}
\COMMA
\end{eqnarray}
where $i,j,k,m$ represent any permutation of $1,2,3,4$.
\smallskip

For a cross check, also the unpolarized results (i.e.\ where the
hadron tensor is contracted with $-g_{\mu\nu}$) have been calculated 
to all orders in $\ve$ and
perfect agreement has been found with independent calculations 
in $n$ dimensions \cite{poe1}.
The $n$ dimensional results for the classes A through E have
already been presented in \cite{kra3}.
\smallskip

In the next two subsections, we will perform the phase space integration 
over the matrix elements of class~D as a cross check
and we will complete the set of coefficient functions 
by the calculation of the $\COFm_{L,q}^{F,(2)}$.

\vspace{1cm}
\subsubsection{Phase Space Integration}

In this subsection, the integration over the degrees of freedom of the
unobserved final--state particles will be performed.
To this end, the phase space factors for $k$--particle final states
are given for $k=2...4$ in the center--of--mass system
of partons 1 and 2:
\begin{equation}
\label{ps2}
\PS^{(2)}= A_0 \, \int dy_{12} \delta(1-y_{12})\frac{1}{4 Q^2}
\end{equation}
\begin{eqnarray}
\label{ps3}
\PS^{(3)} &\n=\n& A_0 \, \frac{1}{64 \pi^2} \left[ \frac{1}{\Gamma(1-\ve)}
\left(\frac{4\pi}{Q^2}\right)^{\ve}\right]
\nonumber\\
& &  \int d y_{12} d y_{23} y_{12}^{-\ve} y_{23}^{-\ve}
(1-y_{12}-y_{23})^{-\ve} \Theta(1-y_{12}-y_{23})
\end{eqnarray}
\begin{eqnarray}
\label{ps4}
\PS^{(4)} &\n=\n& A_0 \, \frac{Q^2}{16}  \frac{1}{64 \pi^4} \, N_{\rm stat} 
\left[ \frac{1}{\Gamma(1-2\ve)}\left( \frac{4\pi}{Q^2}\right)^{2\ve}
\frac{1}{(1-2\ve)} \right]
\nonumber\\
& &  \int dy_{124}dy_{123}dy_{12}
(y_{123}y_{124}-y_{12})^{-\ve}(y_{12}+1-y_{124}-y_{123})^{-\ve}y_{12}^{-\ve}
\nonumber\\
& & \hspace{1cm} \Theta(y_{12}) \Theta(y_{123}y_{124}-y_{12})
\Theta(y_{12}+1-y_{124}-y_{123})
\nonumber\\
& & \hspace{1cm}
\int_0^1 \frac{dv}{N_v} v^{-\ve}(1-v)^{-\ve}
\int_0^{\pi}\frac{d\phi}{N_{\phi}}\sin^{-2\ve}\phi
\COMMA
\end{eqnarray}
where $A_0$ is defined in (\ref{A0def}).
The expressions for $\PS^{(2)}$ and $\PS^{(3)}$ have also been derived
independently and perfect agreement with \cite{ell2} has been
achieved in $n$ dimensions.
The four--particle final state is determined by three invariants
and the two angles $\theta$ and $\phi$.
The angular integration over $\theta$ is written in terms
of $v=1/2(1-\cos{\theta})$. 
The normalization factors for the angular integrations in $\PS^{(4)}$
are given in appendix~B and the statistical factor $N_{\rm stat}$
is due to the quantum statistics of the final--state particles.
For the $q\overline{q}GG$ final state, $N_{\rm stat}=1/2$
and for the $q\overline{q}q\overline{q}$ 
final state, $N_{\rm stat}=1/4$.
The two-- and three--particle phase spaces are relatively simple
and only the four--particle phase space that is the central
ingredient in the calculation of the NLO coefficient functions
requires some further attention.
\smallskip

The coordinate system can be chosen so that,
of the irreducible set of five invariants $x_3$, $y_{123}$, $y_{12}$, 
 $y_{234}$, and $y_{23}$, only $y_{23}$ depends 
on $\phi$ and only the last two depend on $v$.
The angular integrations over the matrix elements $B88$ and $B11$ can then
easily be performed, the required integrals are given in appendix~B.
From (\ref{kiny3}) and (\ref{kinx3}), $y_{134}=1+x_3-y_{123}-y_{234}$ and
the denominator of $B87$ decomposes into inverse monomials of $y_{234}$,
\begin{equation}
\label{decomp2}
\frac{1}{y_{134}y_{234}} = \frac{1}{1+x_3-y_{123}} \left(
    \frac{1}{y_{134}} + \frac{1}{y_{234}} \right)
\STOP
\end{equation}
In the same manner, also the denominator of $B81$ gets decomposed.
Moreover, in the case of $B87$, the matrix element is symmetric under 
 $(1 \leftrightarrow 2)$, and so, by construction, is $\PS^{(4)}$
so that the two terms arising from (\ref{decomp2}) can be treated
collectively as
\begin{equation}
\label{b87new}
B87 =  \frac{64}{Q^2} 
      \frac{(\ve y_{34}^2+2y_{13}y_{23}+y_{13}y_{34}+y_{23}y_{34}
      )(-y_{12}y_{34}+y_{13}y_{24}+y_{14}y_{23})}
       {x_3^2(1+x_3-y_{123})y_{234}y_{34}^2} \\
\STOP
\end{equation}
\smallskip

\begin{figure}[hht!]
\begin{center}
\epsfig{file=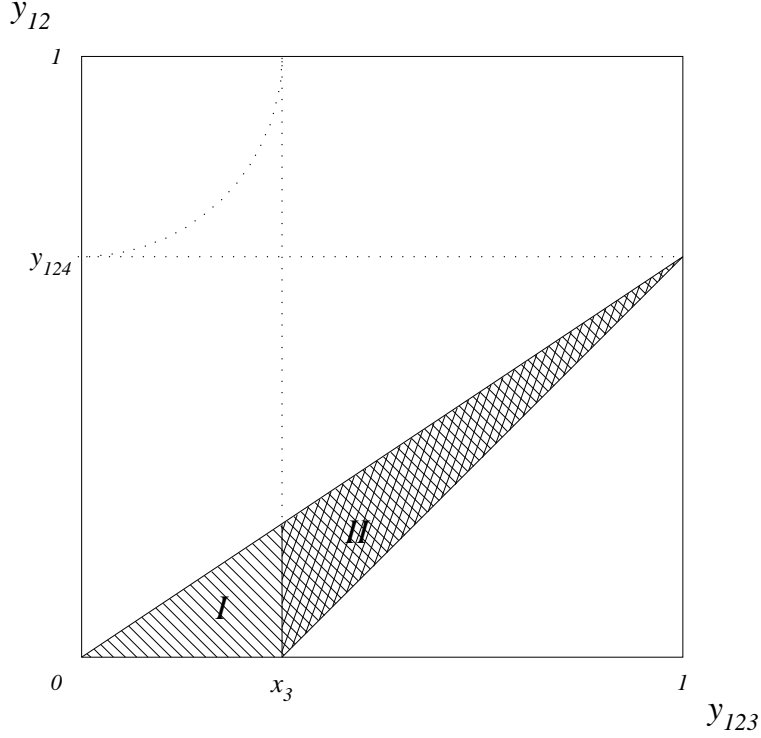 , clip=,width=12cm }
\end{center}
\vspace{-0.2cm}
\caption[]{\small 
Schematic drawing of the four particle phase space.
The shaded areas are the kinematically allowed regions.
In the hatched region ($I$) $y_{12}$ is the singular
variable, whereas for the cross--hatched region ($II$) the
combination $y_{123}-x_3-y_{12}$ gives rise to poles in $\ve$.
\protect\label{ps}}
\end{figure}

The two remaining energy integrations over $y_{123}$ and $y_{12}$ are
restricted by the kinematics (see (\ref{yrel3}) ff.), the boundaries are
specified by the Heaviside functions in (\ref{ps4}).
The resulting triangle is decomposed into two distinct regions,
 $I$ and $II$, as sketched in Fig.~\ref{ps}.
These regions are treated separately and the results are then added.
The respective integration measures can be written as
\begin{eqnarray}
{\rm \bf I}_I \left[f(x_3,r,z)\right] &\n\equiv \n& 
  x_3^{2-3\ve} (1-x_3)^{1-2\ve} \int_0^1 dr r^{1-2\ve}(1-r)^{-\ve}
\nonumber \\
& & \hspace{0.8cm}
  \int_0^1 dz z^{-\ve} (1-z)^{-\ve} 
  \left( 1+\frac{(1+x_3)r}{1-r}z\right)^{-\ve} \, f(x_3,r,z)
\label{meas1}
\COMMA
\end{eqnarray}
where the remaining invariants in terms of the new variables are
\begin{eqnarray}
\label{m1v1}
y_{12} &\n=\n& (1-x_3)x_3rz
\SPACE {\rm and}
\\
\label{m1v2}
y_{123} &\n=\n& x_3r
\end{eqnarray}
for region I.
In an analogous way
\begin{eqnarray}
{\rm \bf I}_{II} \left[ f(x_3,r,z) \right] &\n=\n& 
 (1-x_3)^{2-3\ve} x_3^{1-2\ve} \int_0^1 dr r^{-\ve}(1-r)^{1-2\ve}
\nonumber \\
& & \hspace{0.8cm}
  \int_0^1 dz z^{-\ve} (1-z)^{-\ve} 
  \left( 1+\frac{(1-r)x_3}{r}z\right)^{-\ve} \, f(x_3,r,z)
\label{meas2}
\COMMA
\end{eqnarray}
with
\begin{eqnarray}
\label{m2v1}
y_{12} &\n=\n& (1-x_3)r\left(1+\frac{x_3(1-r)}{r}z\right)
\SPACE {\rm and}
\\
\label{m2v2}
y_{123} &\n=\n& 1-(1-x_3)(1-r)
\STOP
\end{eqnarray}
\smallskip

From Fig.~\ref{ps} it is obvious that in region~$I$ only $y_{12}$ and
in region~$II$ only $(y_{123}-x_3-y_{12})$ give rise to 
singularities in (\ref{ps4}).
From the definition of the new variables in (\ref{m1v1}), (\ref{m1v2})
and (\ref{m2v1}), (\ref{m2v2}), respectively, one sees that in
both regions the possible singularities then turn up in the $z$ integration
as
\begin{equation}
\label{energysing}
\int_0^1dz \, z^{-1-\ve} \left(1+\ve g(z,x_3,r)\right)
\COMMA
\end{equation}
where $g(z,x_3,r) \to 0$ for $z \to 0$.
The solution of the energy integrals both on and off singularities
is briefly sketched in appendix~B.
The last integration, of course, corresponds to that over the momentum
fraction $x$ of the outgoing parton and is not performed.
Information on the chosen coordinate system is obliterated by the
phase space integration and the subscript of $x$ will be dropped in the 
following.
\smallskip

The integration over the matrix elements of class~D yields a nonsinglet
contribution (from $B11$) and a pure singlet contribution.
Of course, this is not the only class to contribute to the
nonsinglet coefficient functions.
However, the factorization works separately for each color factor
so that it is possible to project out the terms proportional
to $N_F T_R C_F$ and check the calculation for this part,
independently.
Together with the integration over the ${\cal O}(\alpha_s)$
matrix elements to order $\ve$, we obtain
\begin{eqnarray}
\frac{1}{N_C\sigma_0}\frac{d\overline{\sigma}_{L,q}^{NS}}{dx} &\n=\n& 
     \frac{\overline{\alpha}_s}{2\pi} 
     S_{\ve} \left(\frac{\mu^2}{Q^2}\right)^{\ve}
     \left\{ \COFm_{L,q}^{NS,(1)}(x) + \ve a_{L,q}^{(1)}(x) \right\}
\nonumber \\
& & \hspace{1cm}
     +\left(\frac{\overline{\alpha}_s}{2\pi}\right)^2 
     S_{\ve}^2 \left(\frac{\mu^2}{Q^2}\right)^{2\ve}
     \left\{ \frac{1}{\ve}\frac{\beta_0}{2}\COFm_{L,q}^{NS,(1)}(x)
     +\COFm_{L,q}^{NS,(2)}(x)
\nonumber \right. \\
& & \left. \hspace{1.8cm}   
     + \frac{\beta_0}{2}a_{L,q}^{(1)}(x) 
     + {\rm U}_{N_F=0}(x) \right\} +{\cal O}(\overline{\alpha}_s^3)
\label{baresigNS}
\\
\frac{1}{N_C\sigma_0}\frac{d\overline{\sigma}_{L,q}^{PS}}{dx} &\n=\n&
     \left(\frac{\overline{\alpha}_s}{2\pi}\right)^2 
     S_{\ve}^2 \left(\frac{\mu^2}{Q^2}\right)^{2\ve}
     \left\{ -\frac{1}{2\ve}\left[ P_{G\to q}^{(0)} 
     \otimes \COFm_{L,G}^{(1)}\right](x)
     +\COFm_{L,q}^{PS,(2)}(x) 
\nonumber \right. \\
& & \left. \hspace{1.8cm}   
     -\frac{1}{2}\left[ P_{G\to q}^{(0)} \otimes a_{L,G}^{(1)}\right](x) 
     + {\rm U}_{N_F=0}'(x) \right\} +{\cal O}(\overline{\alpha}_s^3)
\label{baresigPS}
\\
\label{baresigG}
\frac{1}{N_C\sigma_0}\frac{d\overline{\sigma}_{L,G}}{dx} &\n=\n& 
     \frac{\overline{\alpha}_s}{2\pi} 
     S_{\ve} \left(\frac{\mu^2}{Q^2}\right)^{\ve}
     \left\{ \COFm_{L,G}^{(1)}(x) + \ve a_{L,G}^{(1)}(x) \right\}
     +{\cal O}(\overline{\alpha}_s^2)
\COMMA
\end{eqnarray}
where $S_{\ve}$ is defined in (\ref{Sepsdef})
and the coefficient functions $\COFm^{(1)}$ and the $a^{(1)}$ 
are given in appendix~D.
We have expressed the order $\alpha_s^2$ corrections as sums
of the NLO coefficient functions plus terms 
involving the $a_{L,q}^{(1)}$, respectively $a_{L,G}^{(1)}$
that will cancel with the ${\cal O}(\ve)$ corrections to the
lower order terms.
The details of these cancellations will be discussed in the
next subsection in connection with the renormalization and
factorization procedures.
Only the class~D of the $\COFm_{L,q}^{(2)}$ has been calculated so that 
above equations are determined only for the contributions with
color factor $N_F T_R C_F$ as indicated by the addition
of an unknown contribution U that does not depend on $N_F$.
For this color class, as well as for the LO functions, we find
perfect agreement with \cite{rij1,rij2}.
The complete set of functions $\COFm_L^{(2)}$ 
%__nocit__%from \cite{rij1}
is given in appendix~D. 
\smallskip

The color class~F has been calculated here for the first time.
It receives only ${\cal O}(\alpha_s^2)$ contributions so that the
result is regular,
\begin{equation}
\label{baresigF}
\frac{1}{N_C\sigma_0}\frac{d\overline{\sigma}_{L,q}^{F,(2)}}{dx} = 
     \left(\frac{\overline{\alpha}_s}{2\pi}\right)^2 
     S_{\ve}^2 \left(\frac{\mu^2}{Q^2}\right)^{2\ve}
     \COFm_{L,q}^{F,(2)}(x)
\STOP
\end{equation}
\smallskip

In the following subsection, the bare cross sections will be factorized in 
the {\msbar}--scheme and the coupling constant renormalization will
be performed.

\vspace{1cm}
\subsubsection{{\msbar} Factorization and Renormalization}

As the cross section (\ref{baresigF}) does not
contain singularities, it does not get factorized.
The coupling constant renormalization is also trivial as this
is the leading contribution to $d\sigma^{F}$ and therefore
the substitution (see, e.g.\ \cite{gon1})
\begin{equation}
\label{ccren}
\frac{\overline{\alpha}_s}{2\pi} \longrightarrow
  \frac{\alpha_s(\mu_R^2)}{2\pi} \left[1-\frac{\alpha_s(\mu_R^2)}{2\pi}
  \frac{\beta_0}{2}\frac{1}{\ve}S_{\ve}\left(\frac{\mu^2}{\mu_R^2}\right)^{\ve}
  \right]
\end{equation}
just introduces the scale $\mu_R$,
\begin{equation}
\label{sigFren}
\frac{1}{N_C\sigma_0}\frac{d\sigma_{L,q}^{F,(2)}}{dz} = 
   \left(\frac{\alpha_s(\mu_R^2)}{2\pi}\right)^2 \COFm_{L,q}^{F,(2)}(z)
\COMMA
\end{equation}
where now the variable $z$ is used, 
just to be compatible with the
notation in (\ref{sigxpol}).
For the nonsinglet part of class~D, the 
coupling constant renormalization actually serves to remove the
$1/\ve$ singularity and the term proportional to $a_{L,q}^{(1)}$.
The LO functions $\COFm_{L,q}^{NS,(1)}$ and $\ve a_{L,q}^{(1)}$ appear
in the subleading corrections of (\ref{baresigNS}) with a common
factor, which is (apart from a $\scriptstyle{(-)}$ sign) 
the same as in (\ref{ccren}).
In the case of the pure singlet the renormalization only introduces
a scale dependence in the strong coupling.
The additional terms in (\ref{baresigPS}) are again proportional to
the LO, but they are now removed by the factorization with
the transition functions (\ref{defGNS}) and (\ref{defGGq})
according to (\ref{factSig}).
In the {\msbar}--scheme, the finite 
terms $\ln{(4\pi)} -\gamma_{\tt E}$ arising from the
expansion of the factor $S_{\ve}$ are absorbed together with the
singularity.
Again just caring about terms proportional to $N_F$ in 
the ${\cal O}(\alpha_s^2)$,
\begin{eqnarray}
\hspace{-.4cm}
\frac{1}{N_C\sigma_0}\frac{d\sigma_{L,q}^{NS}}{dz} &\n=\n&
\frac{\alpha_s(\mu_R^2)}{2\pi}\COFm_{L,q}^{NS,(1)}(z)
\nonumber\\ && \hspace{.3cm}
   +\left(\frac{\alpha_s(\mu_R^2)}{2\pi}\right)^2 \left\{-\frac{1}{2}
   \beta_0 \COFm_{L,q}^{NS,(1)}(z)
   \ln{\left(\frac{Q^2}{\mu_R^2}\right)}+\COFm_{L,q}^{NS,(2)}(z)\right\}
\label{sigNSrenD}
\\
\nonumber\\
\hspace{-.4cm}
\frac{1}{N_C\sigma_0}\frac{d\sigma_{L,q}^{PS}}{dz} &\n=\n&
\left(\frac{\alpha_s(\mu_R^2)}{2\pi}\right)^2 \left\{\frac{1}{2}
    \left[P_{G\to q}^{(0)} \otimes \COFm_{L,G}^{(1)}\right](z)
    \ln{\left(\frac{Q^2}{M_f^2}\right)}+\COFm_{L,q}^{PS,(2)}(z)\right\},
\label{sigPSrenD}
\end{eqnarray}
where the logarithmic terms come from the expansion of the 
factors $(\mu^2/\mu_R^2)^{\ve}$,
 $(\mu^2/Q^2)^{\ve}$, and $(\mu^2/M_f^2)^{\ve}$, respectively, 
in powers of $\ve$ according to
\begin{equation}
\label{logexp}
z^{\ve}=1+\ve\ln(z)+{\cal O}\left(\ve^2\right)
\STOP
\end{equation}
The complete set of 
longitudinal partonic cross sections is given in \cite{rij1}.
There, the factorization and renormalization scales are
identified.
They will be disentangled here by the replacement
\begin{equation}
\label{alpharep}
\alpha_s(M_f^2) \to \alpha_s(\mu_R^2)\left[1+\frac{\alpha_s(\mu_R^2)}{2\pi}
   \frac{\beta_0}{2}\ln{\left(\frac{\mu_R^2}{M_f^2}\right)}\right] 
\COMMA
\end{equation}
The perturbative expansion is performed in powers of $\alpha_s/4\pi$ in
\cite{rij1}, whereas in this work 
the expansion parameter is $\alpha_s/2\pi$.
In our notation, the ${\cal O}(\alpha_s^2)$ coefficient functions thus
get an extra factor of $1/4$.
The expressions are given explicitly in the appendix. 
The cross sections are defined by
\begin{eqnarray}
\frac{1}{N_C\sigma_0}\frac{d\sigma_{L,q}^{NS}}{dz} 
&\n=\n& \frac{\alpha_s(\mu_R^2)}{2\pi}
   \COFm_{L,q}^{NS,(1)}(z)
   +\left(\frac{\alpha_s(\mu_R^2)}{2\pi}\right)^2
   \left\{\COFm_{L,q}^{NS,(2)}(z) 
\right. 
\nonumber \\ && \hspace{.8cm}
\left. 
   +C_F^2\left(2\ln(1-z)-\ln(z)+\frac{1}{2}+z\right)
   \ln\left(\frac{Q^2}{M_f^2}\right)
\right. 
\nonumber \\ && \hspace{.8cm}
\left. 
      +\left[N_CC_F\left(-\frac{11}{6}\right)+N_F T_R C_F\left(
      \frac{2}{3}\right)\right]\ln\left(\frac{Q^2}{\mu_R^2}\right)
\right\}
\label{sigNSren}
\\
\frac{1}{N_C\sigma_0}\frac{d\sigma_{L,q}^{PS}}{dz} 
&\n=\n& \left(\frac{\alpha_s(\mu_R^2)}{2\pi}\right)^2
   \left\{\COFm_{L,q}^{PS,(2)}(z) 
\right.
\nonumber \\ && \hspace{.8cm}
\left.
      +N_F T_R C_F\left[4\ln(z)+\frac{8}{3}\frac{1}{z}-4z
      +\frac{4}{3}z^2\right]\ln\left(\frac{Q^2}{M_f^2}\right)
\right\}
\label{sigPSren}
\\
\frac{1}{N_C\sigma_0}\frac{d\sigma_{L,q}^{\Sigma}}{dz}
&\n=\n& 
\frac{1}{N_C\sigma_0}\frac{d\sigma_{L,q}^{NS}}{dz} 
+\frac{1}{N_C\sigma_0}\frac{d\sigma_{L,q}^{PS}}{dz} 
\label{sigSren}
\\
\frac{1}{N_C\sigma_0}\frac{d\sigma_{L,G}}{dz} 
&\n=\n& \frac{\alpha_s(\mu_R^2)}{2\pi}
   \COFm_{L,G}^{(1)}(z)
   +\left(\frac{\alpha_s(\mu_R^2)}{2\pi}\right)^2
   \left\{\COFm_{L,G}^{(2)}(z) 
\right. 
\nonumber \\ && \hspace{.8cm} 
\left. 
      +\left[C_F^2\left(4\ln(z)-2+4\frac{1}{z}-2z\right)
          +N_F T_R C_F\left(-\frac{8}{3}\frac{1}{z}+\frac{8}{3}\right)
\right. \right.
\nonumber \\ && \hspace{1.4cm} 
\left. \left. 
      +N_CC_F\left(\left(-8+8\frac{1}{z}\right)\ln(1-z)
      -\left(8+8\frac{1}{z}\right)\ln(z)
\right. \right. \right.
\nonumber \\ && \hspace{1.4cm} 
\left. \left. \left.
      +\frac{38}{3}-\frac{46}{3}\frac{1}{z}
      +4z-\frac{4}{3}z^2\right)
      \right]\ln\left(\frac{Q^2}{M_f^2}\right) 
\right.
\nonumber \\ && \hspace{.8cm} 
\left.
      +\left[N_CC_F\left(-\frac{22}{3}\frac{1}{z}+\frac{22}{3}\right)      
\right. \right.
\nonumber \\ && \hspace{1.4cm} 
\left. \left.
      +N_F T_R C_F\left(\frac{8}{3}\frac{1}{z}-\frac{8}{3}\right)
      \right]\ln\left(\frac{Q^2}{\mu_R^2}\right)
\right\}
\label{sigGren}
\end{eqnarray}
in addition to eq.~(\ref{sigFren}).
The convolutions with the $\COFm^{(1)}$ (cf.\ (\ref{sigPSrenD}))
and the terms arising from the renormalization (cf.\ (\ref{sigNSrenD}))
are evaluated and given as functions of $z$ in above equations.
When the scales are identified with $Q^2$, as in the fits
discussed below, the equations (\ref{sigFren})
and (\ref{sigNSren}) -- (\ref{sigGren}) reduce to the generic form
\begin{equation}
\label{sigrengen}
\frac{1}{N_C\sigma_0}\frac{d\sigma_{L,a}^{\lb}}{dz}=
   \frac{\alpha_s(Q^2)}{2\pi}\COFm_{L,a}^{\lb ,(1)}(z)
   +\left(\frac{\alpha_s(Q^2)}{2\pi}\right)^2\COFm_{L,a}^{\lb ,(2)}(z)
   +{\cal O}\left((\alpha_s(Q^2))^3\right)
.
\end{equation}
The correct implementation of the coefficient functions in our
computer program has been checked numerically by
comparison of eq.~(\ref{Reelong}) with the integral
\begin{equation}
\label{Reedef}
R_{ee,L} \equiv \int_0^1 dz z \left[ \COFm_{L,q}^{\Sigma}\left(
   z,\frac{Q^2}{\mu_R^2}\right)
  +\frac{1}{2}\COFm_{L,G}\left(z,\frac{Q^2}{\mu_R^2}\right) \right]
\STOP
\end{equation}

\vspace{2cm}
\subsection{Discussion}
\smallskip

After thus having reassured ourselves of the correctness of our
numerical results, we will conclude with a short discussion of
the NLO corrections to the longitudinal cross section.
\smallskip

\begin{figure}[hht!]
\begin{center} 
\begin{picture}(12.5,5.5)
\put(-1.2,-.3){\makebox{
\epsfig{file=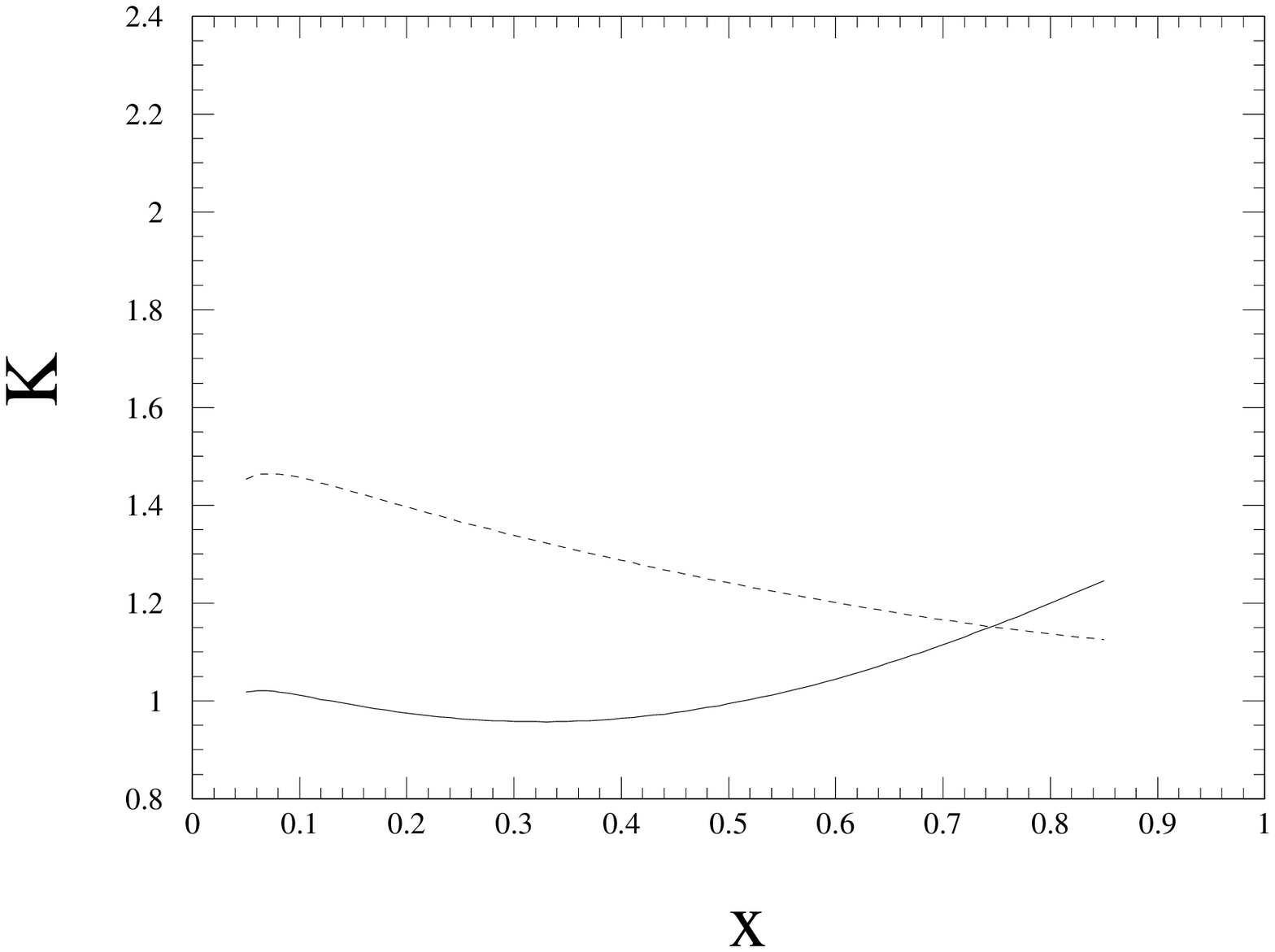,clip=,width=6.9cm}}}
\put(6.4,-.3){\makebox{
\epsfig{file=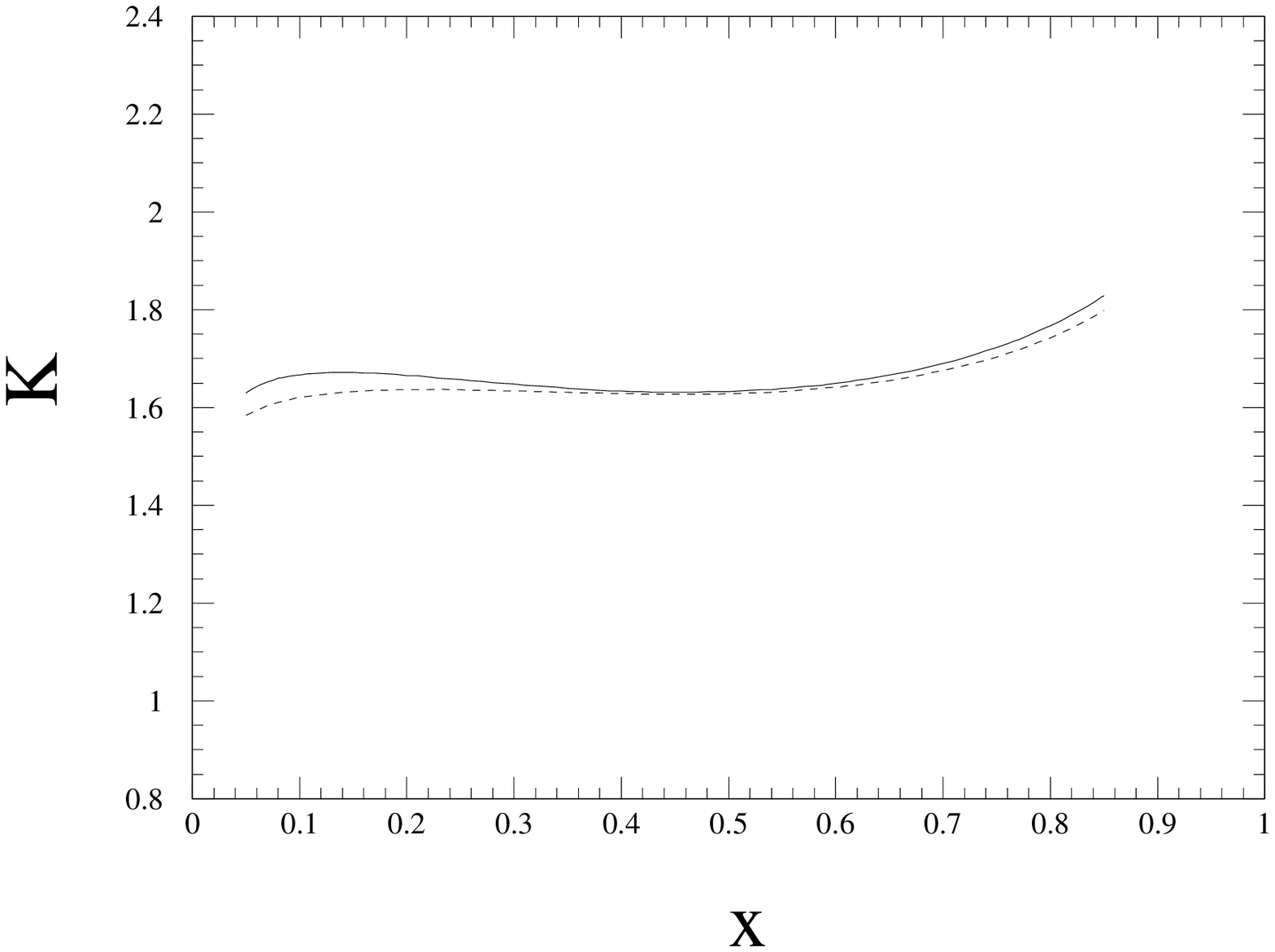,clip=,width=6.9cm}}}
\put(-0.7,0.2){a)}
\put(6.8,0.2){b)}
\end{picture}
\end{center} 
\vspace{-.6cm} 
\caption[]{\small  
The K factors for the longitudinal cross section,
evaluated with our new charged hadron set (solid)
and with the charged hadron set, as constructed in section~4.1
(dashed).
a)
shows the ratio of the cross sections
evaluated consistently in NLO and LO, respectively.
In plot
b), 
the NLO FF's are used in both the numerator and the denominator and 
only the coefficient functions are changed.
\protect\label{kfac}} 
\end{figure}

First, we are now in a position to assess the size of the
K factor for the NLO.
In the spirit of the definition of the NLO K factor as
the ratio of the NLO result to the LO result, one
might start to calculate the K factor for the
differential cross section with the use of the NLO FF's in the
numerator whereas the denominator is evaluated

\noindent
with the LO ones.
However, as the two sets (LO and NLO, respectively) are fitted
independently, they may differ significantly.
In consequence, the K factor will depend 
strongly on the set of fragmentation functions used.
This is
demonstrated in Fig.~\ref{kfac}a where the dashed curve
shows the K factor obtained with our set \cite{binpk} of charged
hadron FF's and where the solid curve has been calculated with
our new set of charged hadron FF's.
Because we used the longitudinal cross section measurement
in our new fit, both the leading and the next--to--leading
results describe the data well, which leads to a K factor close to one,
whereas the K factor for the older set
ranges between 1.45 at small,
and 1.1 at large momentum fraction.
It is thus advisable to keep the FF's at the same order in
both the numerator and the denominator of K,
to assess the size of the NLO corrections to the 
coefficient functions.
This has been done in Fig.~\ref{kfac}b.
The K factor is now insensitive to the set of FF's and is found to
vary little, between 1.6 and 1.8.
Note that the naive results of Fig.~\ref{kfac}a would have been
misleading with respect to the size of the NLO corrections.
Fig.~\ref{kfac}b, on the other hand, confirms the conjecture
prompted by $R_{ee,L}$ that the K factor would turn out to
be large.
\smallskip

\begin{figure}[hht!]
\begin{center} 
\epsfig{file=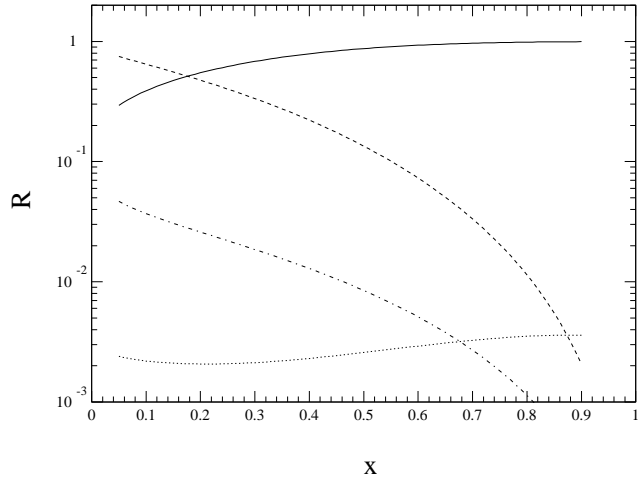,clip=,width=8.8cm}
\end{center} 
\vspace{-1.cm} 
\caption[]{\small  
The individual contributions
to the NLO longitudinal cross section
at the scale $Q=m_Z$, normalized to 1.
The $NS$ quark contribution (solid)
dominates for $x>0.2$, whereas for smaller $x$ the gluon
(dashed) is most important.
Both the $F$ (dotted) and the $PS$ (das--dotted) 
quark contributions are absent at LO.
At NLO, their impact is small and 
the $F$ contribution competes with the negative $PS$ (the
absolute value is plotted) for not too small $x$.
\protect\label{furry}} 
\end{figure}

Secondly, we show the individual contributions to 
the NLO longitudinal cross section, differential in $x$, 
in Fig.~\ref{furry}.
The results are obtained at the $Z$ pole with our new
set of fragmentation functions.
The $NS$ quark combination (solid) dominates for $x>0.2$, 
whereas for smaller $x$ the gluon (dashed) is more important.
The $PS$ quark combination (dash--dotted) and the
contributions $F$ (dotted) from the class $C_F T_R$ 
only appear at NLO.
For not too small $x$, the $F$ competes with the negative $PS$
(the absolute value is plotted).
Whereas the relative weight of the pure singlet rises to
a few percent at low $x$,
the contribution of class $C_F T_R$ is well below 1\% everywhere.
Moreover, the relative importance of $F$ decreases when
moving away from the $Z$ pole due to the decrease of the effective
charge $Q^F(Q^2)$, c.f.\ Fig.~\ref{charge}.
It can thus be neglected.
\smallskip

The longitudinal polarized part contributes only a small
fraction to the unpolarized ($L+T$) cross section,
c.f.\ Fig.~\ref{lrat}.
With respect to measuring the gluon FF, however, it has
the advantage that the gluon appears on an equal
footing with the quarks in (\ref{sigmal}),
whereas its contribution to $\sigma_T$ is subleading.
The gluonic contribution to $\sigma_L$ is also numerically
of the same order as that of the quarks, for not too
large $x$, as can be seen in Fig.~\ref{furry}.
\smallskip

\begin{figure}[hht!]
\begin{center}
\epsfig{file=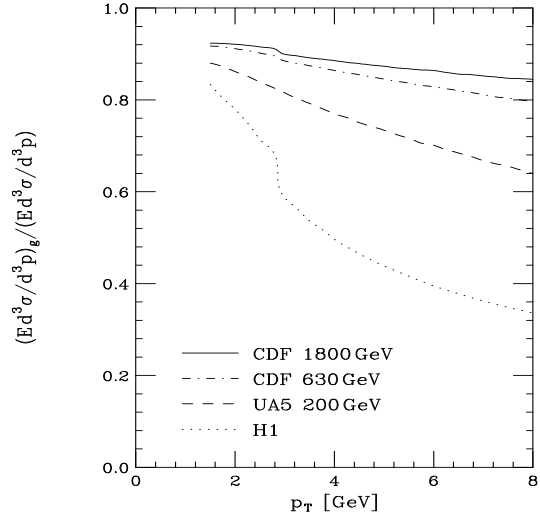, clip=,width=7.cm,height=6.9cm }
\end{center}
\vspace{-0.5cm}
\caption[]{\small 
Fraction of the cross section that is due to gluon fragmentation
(for the set \protect\cite{binpk} of FF's) for various processes and
kinematics.
The results for CDF and UA5 kinematics in $p\overline{p}$ collisions
at 1800, 630, and 200~GeV are shown by the solid,
dash--dotted, and dashed curves, respectively.
The dotted curve gives the fraction for H1 photoproduction kinematics.
The drop at 3~GeV reflects the opening up of the additional
charm channel.
\protect\label{sl0b}}
\end{figure}

The longitudinal cross section, therefore, is a promising
observable for a determination of the gluon FF in \ee annihilation.
The error on
the gluon FF has a huge impact when other processes
are considered.
This is due to large contributions from subprocesses with
initial--state gluons that are usually associated with
final--state gluons.
Fig.~\ref{sl0b} shows the ratio of the cross section, differential
in $p_T$ that is due to gluon fragmentation in 
 $p\overline{p}$ collisions at the TEVATRON and in photoproduction
at HERA.
In particular in $p\overline{p}$ scattering, the
gluon FF dominates even for moderately large transverse momenta $p_T$.
For meaningful predictions of the IPP cross section in these 
processes the gluon must be well determined.
This can be achieved by the use of the longitudinal cross section
in the fit.

\pagebreak
\setlength{\unitlength}{1cm}

\vfill
\clearpage
\section{Fragmentation Functions}
\setcounter{equation}{0}
\bigskip

The investigation of fragmentation functions goes back to
R.D.~Field and R.P.~Feynman who devised the first physical picture
for the fragmentation process \cite{fie1}.
For a long time -- from the early eighties to the early nineties --
only LO fragmentation functions for charged pions, presented
by R.~Baier {\it et al.} \cite{bai1}
and updated by M.~Anselmino {\it et al.} \cite{ans1},
have been available.
In retrospect, it is not easily understood why the
progress in the field of parton density functions during that
time--period has not been accompanied by similar
advances in the field of fragmentation functions, despite
the accumulation of a wealth of data.
Whatever the reasons, the situation has changed for the better
in the past few years by the presentation of 
numerous NLO sets of FF's.
The first work was done by P.~Chiappetta {\it et al.} \cite{chi1}
who extracted a variety of $\pi^0$ sets from
\ee annihilation, fixed--target, and collider data, 
supplemented by data for \ee annihilation obtained from a Monte Carlo
simulation.
We presented NLO sets for charged pions and charged kaons for the
first time \cite{bin1}.
Other, recent, sets are those for $D^{*\pm}$ mesons
obtained from \ee annihilation data by M.~Cacciari and M.~Greco \cite{cac1}
and for photons by L.~Bourhis {\it et al.} \cite{bou1}.
Apart from these sets, and the ones by the author and
collaborators G.~Kramer and B.A.~Kniehl \cite{binpk,bink0,binds}
that are discussed below,
all other sets we are aware of have been fitted to data from
Monte Carlo simulations of \ee annihilation.
Sets for the $\eta$ meson,
for charged pions,
and for charged and neutral kaons have been obtained 
by M.~Greco and collaborators \cite{gre1}.
P.~Nason and B.R.~Webber have concentrated on the sum over charged hadrons,
parametrizing the FF's at the $Z$ pole, and also partly incorporated
experimental data directly \cite{nas1}.
See also \cite{rol1} for a review of recent work on 
fragmentation functions.
\bigskip

Before discussing the new set of charged hadron FF's, as well as
the sets \cite{binpk,bink0,binds} that we published earlier, in subsections
4.1 through 4.4, some general aspects will be covered in this 
introductory section.
Our general policy for both the improved charged hadron fit
performed in this work and for the earlier fits 
is outlined in the following.
Some exceptions that apply to the individual analyses
are specified in the relevant subsections.
\smallskip

As advocated earlier, we use only experimental data from \ee annihilation
to fit the FF's, in order to preserve the most of
the factorization theorem's predictive power.
For its good statistics, data from LEP1 are selected
in all of the analyses presented.
In some of the fits, data at 29~GeV are used in addition to
the 91.2~GeV data to better describe the scaling violation.
The benefits of this addition will be demonstrated in
subsection~5.2.
The hadrons are treated as massless
so that this approach becomes unreliable for $x \to 2m_h/\sqrt{Q^2}$
and a lower limit around $x_{min} \approx 0.1$ is enforced
for the data used in the fits, as well
as for subsequent applications of the FF's in section~5.
Data above $x_{max}=0.8$ are also discarded for poor
statistics and limited reliability of the systematic errors
in the experimental data.
Another problem related to the data is that the phenomenology of
fragmentation given here does not apply to hadrons originating from
weak decays.
Experimentally, such contributions can not be separated from the
strongly produced hadrons at present.
However, the corrections are of order $m^2/Q^2$ and amount only
to about 0.1\% at the $Z$ pole \cite{nas1}.
They are consistently neglected in this work together with all other
possible power--law corrections.
\smallskip

The cross sections have been calculated to NLO only in the
massless theory so all quarks are taken to be massless in this work.
(An exception is the $D^{*\pm}$ analysis where the
$c$ and $b$ quark masses are taken into account also
by a modification of the factorization scheme.)
We adopt the variable flavor number scheme 
for the treatment of the heavy--quark thresholds.
The mass effects of the heavy quarks, $c$ and $b$, would be most
pronounced in the threshold regions around $2m_c$, respectively $2m_b$.
Our massless approach gives reliable results for the $c$ and $b$
fragmentation only sufficiently above their respective threshold.
\smallskip

Novel data from ALEPH \cite{ale0,ale3} and OPAL \cite{opa2} 
at LEP1 that discriminate between
different flavors, enabled us to fit the FF's separately
for each flavor. 
In the analyses presented here, particles and antiparticles are
not discriminated in the final state so that
 $D_{q_i}^h=D_{\overline{q}_i}^h$ for all $q_i$.
The superscript $h^{\pm}$ always refers to the sum of
positively and negatively charged $h$ in our notation.
Some additional constraints from SU(3) symmetry etc.\ are used to
reduce the number of free parameters,
e.g.\ $D_u^{\pi^{\pm}}=D_d^{\pi^{\pm}}$.
\smallskip

The procedure is thus to start at a low scale $\mu_0=\sqrt{2}$~GeV
with $N_F=3$ quark flavors plus the gluon and to parametrize
the FF of parton $a$ to fragment into hadron $h$ in the 
standard form \cite{bai1}
\begin{equation}
D_a^h(x,\mu_0^2)=N_a x^{\alpha_a}(1-x)^{\beta_a}
\STOP
\label{parstand}
\end{equation}
The three quark FF's and the gluon FF are then evolved
to $\mu_c = 2 \, m_c$, where the charm fragmentation function,
parametrized in the same form (\ref{parstand}), is added.
The set of $4+1$ FF's is evolved up to $\mu_b = 2 \, m_b$,
where the bottom quark is added and the now complete
set is finally evolved to 
the scale $M_f$\footnote{In \cite{binpk,bink0}, 
   the heavy quark masses were chosen as
   half that of their lowest bound states, i.e.,
   $2m_c=m_{\eta_c}=2.9788$~GeV and 
   $2m_b=m_{\Upsilon}=9.46037$~GeV according to the
   1994 tables of the Particle Data Group \cite{pdg}.
   For the new fit (as well as in \cite{binds})
   the masses were set to round figures,
   $m_c=1.5$~GeV and $m_b=4.5$~GeV 
   (respectively 5~GeV in \cite{binds}).}.
Although we are primarily interested in NLO sets of FF's,
we simultaneously perform LO fits to check the perturbative
stability.
The evolution and the calculation of the cross sections are
done consistently in NLO (LO) for the
NLO (LO) sets.
Details concerning the evolution are given in
section~2.1 and the numerical techniques pertinent to the solution
of the AP equations are discussed in appendix~C.
\smallskip

A low starting scale, as used in our analyses, has two advantages.
(i) The relatively long evolution to scales where
the FF's are compared to data ensures that those FF's are physical, i.e.\
compatible with the evolution and do not retain gross features of the
functional dependence chosen for the input distributions.
(ii) With due caution, the sets of FF's so obtained may
also be applied at low scales.
This is desirable in particular in $ep$ collisions and has been 
exploited in a number of phenomenological studies of HERA data,
see section~5.
An alternative approach is to use the measured FF's at the $Z$ pole
and evolve them backwards to lower energy scales.
This avenue has been taken in \cite{nas1}.
\smallskip

The evolved set of FF's is convoluted with the partonic
cross sections according to (\ref{sigxpol}) to obtain
the longitudinal or the unpolarized (i.e.\ the sum
of the longitudinal and the transverse) cross sections.
For the leptonic part of the cross section, the interference
of the $Z$ with the photon is fully taken into account
and the electroweak parameters are taken to
be $[\sin^2\theta_W(m_Z)]_{\overline{\rm MS}}=0.2319$,
 $\Gamma_Z=2.4959$~GeV, 
and $m_Z=91.2$~GeV\footnote{
   We used the
   actual parameters from the PDG 
   \cite{pdg}, namely $[\sin^2\theta_W(m_Z)]_{\overline{\rm MS}}=0.2315$,
   $\Gamma_Z=2.491$~GeV, and $m_Z=91.188$~GeV,
   for our new analysis.}
for the numerical evaluation of the cross section.
The strong coupling is consistently evaluated
to two (one) loops for the NLO (LO) fits.
The number of flavors varies from three at low scales to five at
LEP1 energies.
This does not only affect the $\beta$--function of QCD, but also
the value of $\Lambda$, which exhibits discontinuities at
the heavy quark thresholds.
Details on the evaluation of $\Lambda$ and $\alpha_s$ in the
variable flavor number scheme are given in appendix~A.
The scales are identified with the center--of--mass energy of the
hadronic process, $\mu_R=M_f=\sqrt{Q^2}$.
\smallskip

The QCD result is then integrated over the experimental
bins and compared to the data.
The goodness of the fit is measured by the $\chi^2$ per bin,
where for a single data point, 
\begin{equation}
\label{chi}
\chi^2 \equiv \left(\frac{\sigma^{th.}-\sigma^{exp.}}
                  {\Delta\sigma^{exp.}}\right)^2 
\STOP
\end{equation}
The parameters of the fit, i.e.\ the $N_a$, $\alpha_a$ and $\beta_a$,
are optimized through repeated comparison of the evolved FF's to data, 
with the use of {\tt MINUIT} \cite{min}.
\smallskip

The functions so obtained merely give a phenomenological
description of a small subset of \ee data -- those that have been
fitted.
As a check they are compared
to \ee data that have not been used in the fit.
When these data are collected at a different CM energy,
this serves also as a test of scaling violation.
The purpose of the entire procedure is finally to make
predictions with the help of these functions.
To this end, it is desirable to have a parametrization
of the FF's not only as functions of the momentum fraction $x$ but also
of the scale $M_f$.
Such parametrizations have been obtained for the sets
\cite{binpk,bink0} that are discussed in 
subsections~4.1 and 4.3, respectively\footnote{
   A {\tt FORTRAN} routine returning the values of the FF's of
   any parton to charged pions, charged kaons, charged hadrons
   or neutral kaons is available upon request from:
   {\tt binnewie@mail.desy.de}}.
Basically, the functions are parametrized in the form (\ref{parstand}),
where the parameters are now polynomials in
\begin{equation}
\label{sbar}
\overline{s} \equiv \ln\left(\frac{\ln(M_f^2/\Lambda^2)}
{\ln(\mu_a^2/\Lambda^2)}\right) 
\STOP
\end{equation}
Further details of the parametrizations 
are given in the relevant publications.
\smallskip

The following sections will dwell on the
particulars of the individual fits for charged pions, charged kaons,
charged hadrons, neutral kaons, and $D^{*\pm}$, respectively.
In subsection~4.1, the first sets of charged hadron FF's are discussed,
i.e.\ those for charged pions and kaons \cite{binpk} that can,
with some assumptions on the residual contributions to the cross section,
also be applied to indiscriminate charged hadron production.
The treatment of charged hadron FF's will be improved by a novel
fit to both unpolarized and longitudinal cross section data
in the next subsection.
Owing to the inclusion of the 
NLO longitudinal coefficient functions 
discussed in the previous section, this new set will
feature a particularly well constrained gluon FF.
Fragmentation functions for neutral kaons \cite{bink0} are presented
in 4.3.
The discussion of the $D^{*\pm}$ mesons \cite{binds} in
subsection~4.4 introduces some additional concepts in dealing with the
heavy quarks, $c$ and $b$.

\vspace{2cm}
\subsection{Charged Pions and Kaons}
\smallskip

The largest contributions to charged particle production in \ee 
annihilation originate from pions and kaons -- quite naturally we
chose those hadrons for our first NLO fits.
In \cite{bin1}, the FF's were fitted to precise data from TPC \cite{tpc1}
at 29~GeV and successfully described \ee data taken at energies from
5.2~GeV up to 91.2~GeV.
However, we had to assume that the valence--type quarks on one
hand, and the sea--type quarks on the other hand, fragment in the same
way.
When novel data from ALEPH became available 
that discriminate between different quark flavors, this limitation
could be removed.
We were able to obtain a set of FF's \cite{binpk} with well constrained 
individual flavors by fitting to both ALEPH and TPC 
data.
Only this second set will be discussed, as it represents the
updated and improved version of our earlier work.
\smallskip

More specifically, we used data for
charged pion and charged kaon production from ALEPH \cite{ale4}
and TPC \cite{tpc1}.
In addition, the novel data from ALEPH were employed where they
did not discriminate between the charged hadrons but instead
distinguished three cases, namely fragmentation of (i) $u$, $d$, $s$
quarks, (ii) $b$ quarks and (iii) all five quarks \cite{ale0}.
That analysis was done on the basis 
of 40~pb$^{-1}$ of luminosity, taken during 1992 and 1993.
Heavy flavor events were enhanced in the $b$ quark sample with
an impact~parameter~tag and an event~shape~tag.
The overall normalization error was estimated to
amount to 1\%.
\smallskip

In order to determine the flavor differences of the charged
pion and charged kaon FF's in our fit,
the residual contributions to the charged
hadron cross section must be modeled.
Off resonances,
they are mostly due to protons and antiprotons that
are taken to behave like pions with
a relative normalization of
\begin{equation}
\label{ppbarf}
 f(x)= 0.195 -1.35(x-0.35)^2
\end{equation} 
that has been inspired by the differential cross sections for 
inclusive charged pion and proton/antiproton production \cite{ale4}.
Surprisingly, this naive guess turns out to be quite
good not only for the shape of the spectrum but also
for the relative importance of the heavy flavor
contributions.
Very recently, the SLD collaboration presented data
on flavor tagged IPP in $Z$ decays \cite{sld1} that confirm
a close analogy between the heavy quark contributions to
pions and protons for $x$ above 0.1.
Residual contributions from hyperons, charmed mesons etc.\ amount to
less than 10\%, presumably. 
Thus, the $h^{\pm}$ cross section is to a good approximation given by
\begin{equation}
\label{pkhad}
\frac{d\sigma^{h^{\pm}}}{dx}=\left[1+f(x)\right]
      \frac{d\sigma^{\pi^{\pm}}}{dx}
      +\frac{d\sigma^{K^{\pm}}}{dx}
\STOP
\end{equation}
\smallskip

By the use of the novel ALEPH data together with
eq.~(\ref{pkhad}) the $b$ fragmentation could be determined.
This represents a major improvement over our earlier
analysis where we had to assume that the $b$ fragments as the
other sea quarks although it is known to exhibit
a softer $x$--dependence due to its mass.
Data from OPAL \cite{opa2} on $D_G^{h^{\pm}}(x)$,
measured in three jet events, were used in the fit
to address the notorious problem of the gluon FF.
\smallskip

This leaves us with the $u,d,s$, and $c$ quarks.
The bulk of pions produced in strong interactions is due to
 $u$ and $d$ quarks, whereas the $u$ and $s$ quarks are
most prominent in kaon production.
Since we use data for $\pi^{\pm}$ and $K^{\pm}$ together
with the new ALEPH data, this removes most of the
residual freedom for shifting contributions between different flavors.
Only the $c$ quark can not be constrained well by this
procedure as it is neither tagged in \cite{ale0}
nor dominant in either of the 
charged pion or charged kaon data.
\smallskip

The functions were parametrized at the starting scale with the
single, physically motivated constraint that valence--type quarks
should fragment in the same way, i.e.,
\begin{eqnarray}
\label{valence1}
D_u^{\pi^{\pm}} &\n=\n& D_d^{\pi^{\pm}} 
\SPACE {\rm and}
\\
\label{valence2}
D_u^{K^{\pm}} &\n=\n& D_s^{K^{\pm}} 
\STOP
\end{eqnarray}
As we fit data at different CM energies, $\Lambda$ can also be 
determined via scaling violation and is kept as a free parameter.
This gives $2\times5\times3+1=31$ fit parameters.
We exclude data below $x=0.1$ where the formalism
is bound to fail, as well as data above $x=0.8$ that are
plagued by large systematic errors.
\begin{table}[hht!]
\begin{center}
\begin{tabular}{|c|c||c|c|c|}
\hline
set & flavor & $N$ & $\alpha$ & $\beta$ \\
\hline
\hline
$\pi^{\pm}$ & $u$, $d$ 
       & 1.15 (1.09) & -0.74 (-0.85) & 1.43 (1.44) \\
 & $s$ & 4.25 (3.48) & -0.77 (-1.03) & 4.48 (3.90) \\
 & $c$ & 3.99 (4.51) & -0.79 (-0.86) & 4.78 (4.53) \\
 & $b$ & 4.01 (3.60) & -1.03 (-1.13) & 7.86 (7.12) \\
\cline{2-5}
 & $G$ & 5.53 (6.57) & -0.32 (-0.46) & 2.70 (3.01) \\
\hline
\hline
$K^{\pm}$ & $u$, $s$ 
       & 0.31 (0.38) & -0.98 (-1.23) & 0.97 (1.06) \\
 & $d$ & 1.08 (1.12) & -0.82 (-0.92) & 2.55 (2.85) \\
 & $c$ & 0.81 (0.62) & -0.69 (-0.67) & 2.98 (2.48) \\
 & $b$ & 0.61 (0.73) & -0.88 (-0.80) & 2.93 (2.83) \\
\cline{2-5}
 & $G$ & 0.31 (0.37) & -0.17 (-0.21) & 0.89 (3.07) \\
\hline
\end{tabular}
\caption{\small 
Parameters of the charged pion and charged kaon fragmentation functions
obtained in the NLO (LO) fit.
\protect\label{pkpar}}
\end{center}
\end{table}
\smallskip

The values resulting from the fit are given in Table~\ref{pkpar}.
Discussion of the results for $\Lambda$ is
postponed to section~5.1.
Except for $D_G^{K^{\pm}}$, the parameters exhibit good
perturbative stability.
\begin{table}[hht!]
\begin{center}
\begin{tabular}{|c||c|c||c|c||c|c|}
\hline
 $\sqrt{Q^2}$ [GeV] & $\pi^{\pm}$ & ref. & $K^{\pm}$ & ref. 
       & $h^{\pm}$ & ref. \\
\hline
\hline
91.2 & 0.8 (0.9) & \cite{ale4}$^*$ & 0.5 (0.8) & \cite{ale4}$^*$
      & 0.4 (0.5) & \cite{ale0}$^*$ \\
\hline
34.0 -- 35.0 & 1.3 (0.8) & \cite{tas2}   & 0.9 (1.0) & \cite{tas2}
      & 0.4 (0.8) & \cite{cel2}   \\
\hline
29.0 & 1.1 (1.3) & \cite{tpc1}$^*$ & 0.7 (0.8) & \cite{tpc1}$^*$
      & 1.5 (1.4) & \cite{mar5}   \\
\hline
9.98 -- 10.49 & 1.5 (2.0) & \cite{arg2} & 1.1 (1.3) & \cite{arg2}
      & -- & -- \\
\hline
5.2 & 1.4 (0.5) & \cite{das2} & 1.6 (2.1) & \cite{das2}
      & -- & -- \\
\hline
\end{tabular}
\caption{\small
 $\chi^2$ per degree of freedom for the NLO (LO) fits of
charged pions, charged kaons, and the sum over charged hadrons.
The data used in the fits are marked by asterisk.
\protect\label{pkfit}}
\end{center}
\end{table}
\smallskip

\begin{figure}[hht!]
\begin{center} 
\begin{picture}(12.5,8.0)
\put(-1.2,3.4){\rotateleft{$\rm \frac{1}{\sigma_{\rm had}}
      \frac{d\sigma^{\pi^{\pm}}}{dx}$}}
\put(2.8,0.2){$\rm x$}
\put(6.2,3.4){\rotateleft{$\rm \frac{1}{\sigma_{\rm had}}
      \frac{d\sigma^{K^{\pm}}}{dx}$}}
\put(10.3,0.2){$\rm x$}
\put(-0.8,0.7){\makebox{
\epsfig{file=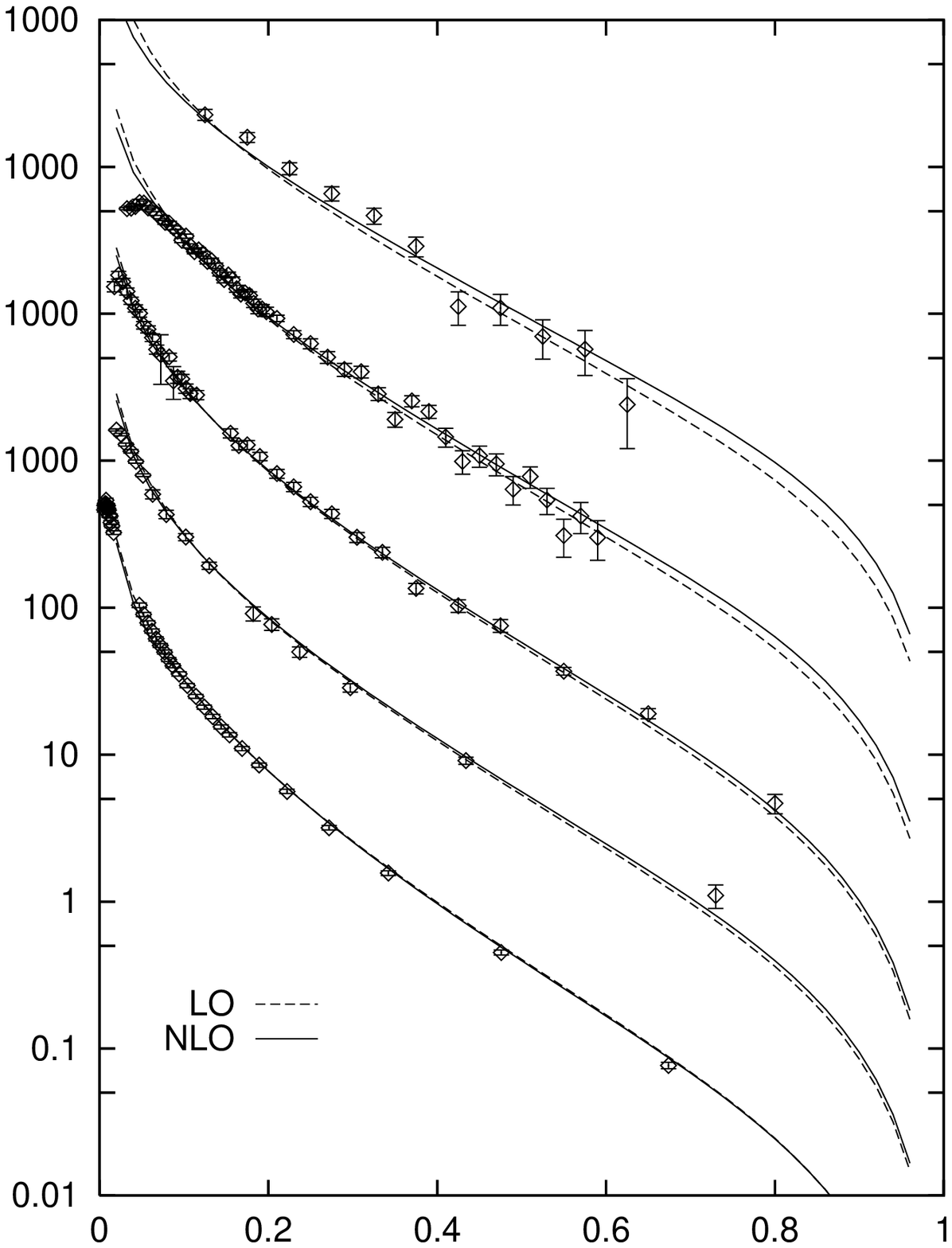,clip=,width=6.3cm}}}
\put(6.8,0.7){\makebox{
\epsfig{file=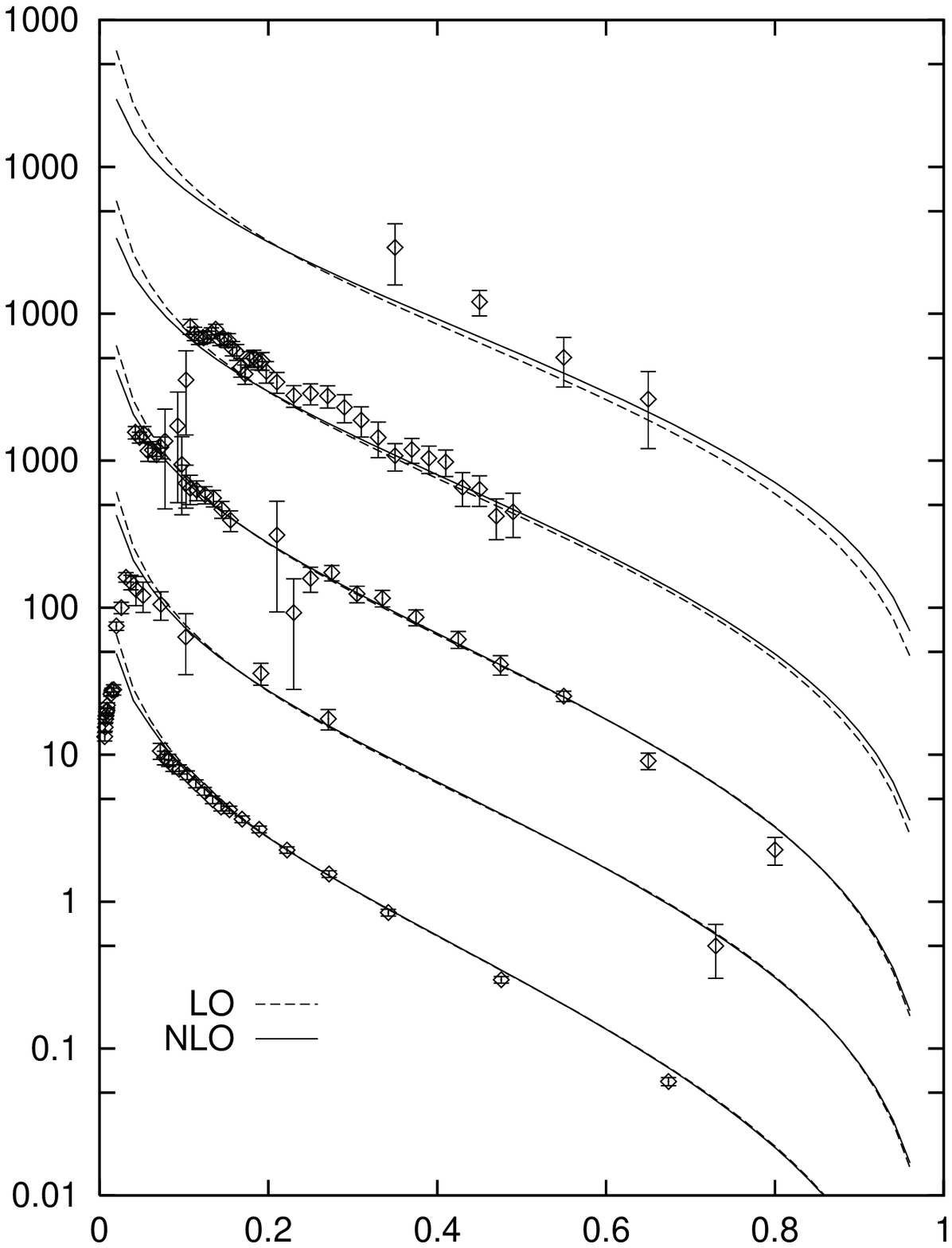,clip=,width=6.3cm}}}
\put(-0.7,0.2){a)}
\put(6.8,0.2){b)}
\end{picture}
\end{center} 
\vspace{-0.4cm} 
\caption[]{\small  
Inclusive production of charged hadrons in
\ee annihilation. 
The plots show the production of
a) 
pions and
b)
kaons.
The data are from top downward
DASP \protect\cite{das2}, ARGUS \protect\cite{arg2}, 
TPC \protect\cite{tpc1}, TASSO \protect\cite{tas2},
and ALEPH \protect\cite{ale4}.
For better separation the data are multiplied by powers of 10.
Solid (dashed) curves give the NLO (LO) results.
\protect\label{pion}} 
\end{figure}

\begin{figure}[hht!]
\begin{center} 
\begin{picture}(12.5,9.5)
\put(1.4,4.8){\rotateleft{$\rm \frac{1}{\sigma_{\rm had}}
      \frac{d\sigma^{h^{\pm}}}{dx}$}}
\put(6.2,0.0){$\rm x$}
\put(2.0,0.7){\makebox{
\epsfig{file=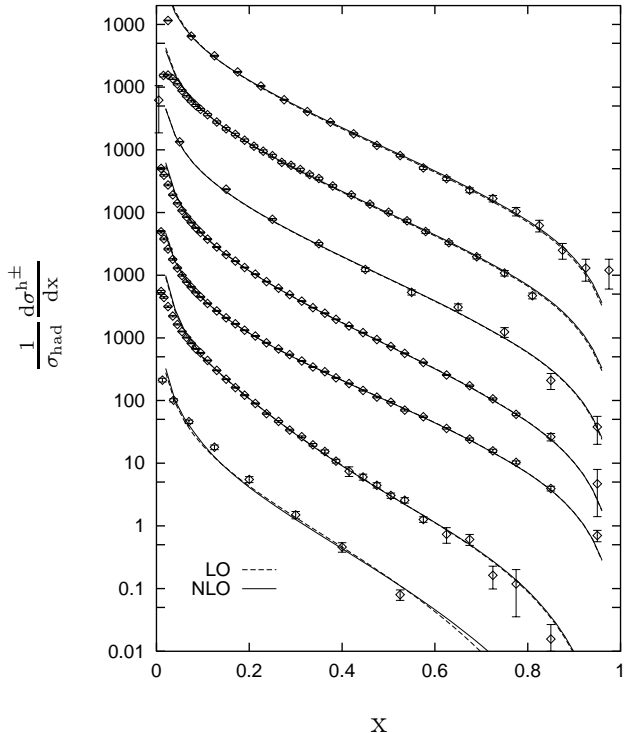 , clip=,width=7.8cm }}}
\end{picture}
\end{center} 
\vspace{-0.3cm} 
\caption[]{\small  
Inclusive production of charged hadrons in
\ee annihilation as in Fig.~\protect\ref{pion}
but for all charged hadrons. 
The data are from top to bottom: MARK~II \protect\cite{mar5},
CELLO \protect\cite{cel2}, AMY \protect\cite{amy1}, 
and ALEPH \protect\cite{ale0}
as well as $u,d,s$--enriched and $b$--enriched 
samples fom ALEPH \cite{ale0}
and $D_G^{h^{\pm}}$ as measured by OPAL \protect\cite{opa2}.
\protect\label{hadron}} 
\end{figure}

The  $\chi^2$ resulting from the fit, as well as those for a selection
of other \ee data, are given 
in Table~\ref{pkfit}\footnote{
   The values for the flavor separated sets of data are not listed,
   they are all less than 1 in NLO with the exemption of the gluon
   data which could not be described well.
   }.
Good agreement with the data over a wide range of scales is achieved.
Most remarkably we find very satisfactory agreement even at
low scales down to 5.2~GeV.
The NLO (LO) cross sections are compared to
 \ee data in Figs.~\ref{pion} and \ref{hadron}.
Figs.~\ref{pion}a, b show the results
for charged pion and charged kaon
production, respectively.
The sum over all charged hadrons is considered in Fig.~\ref{hadron}.
The QCD predictions are in very good agreement with all 
the \ee data, even for $x$ considerably below  $x_{min}=0.1$.
Perturbative stability is evident also by
the near coincidence of the NLO (solid) and LO (dashed) curves.
As a further test, we checked the momentum sum rule;
this has already been discussed in section~2.3.
\smallskip

This concludes the description of our fit \cite{binpk}.
From the good agreement with data at lower CM energies we 
are confident that these FF's describe the \ee annihilation 
data well and may sensibly be applied to other processes in
section~5.
The proton/antiproton final state is not as prominent 
as in \ee annihilation in other processes.
We will thus approximate the subset of all charged hadrons
just by charged pions and charged kaons in our applications
to $ep$, $p\overline{p}$ and $\gamma \gamma$ scattering.

\vspace{2cm}
\subsection{Charged Hadrons -- New Fit}
\smallskip

In the following, the details of the new fit are presented.
The approach is much the same as in the
preceding subsection, i.e.,
we use the data from ALEPH \cite{ale3} 
in the fit\footnote{
   Analyses of the flavor differences
   similar to that of ALEPH \cite{ale3} are currently beeing 
   conducted by the DELPHI and OPAL \cite{opa4} collaborations.
   The longitudinal cross section that is also given in \cite{ale3}
   has also been measured by OPAL \cite{opa3}.
   The data from the ALEPH collaboration have been selected for their 
   better statistics.}.
The data from ALEPH enable a determination of FF's for individual
flavors.
In order to minimize the bias in the fit, we will not distinguish
between pions and kaons but rather deal with the subset of all
charged hadrons.
\smallskip

\begin{figure}[hht!]
\begin{center} 
\epsfig{file=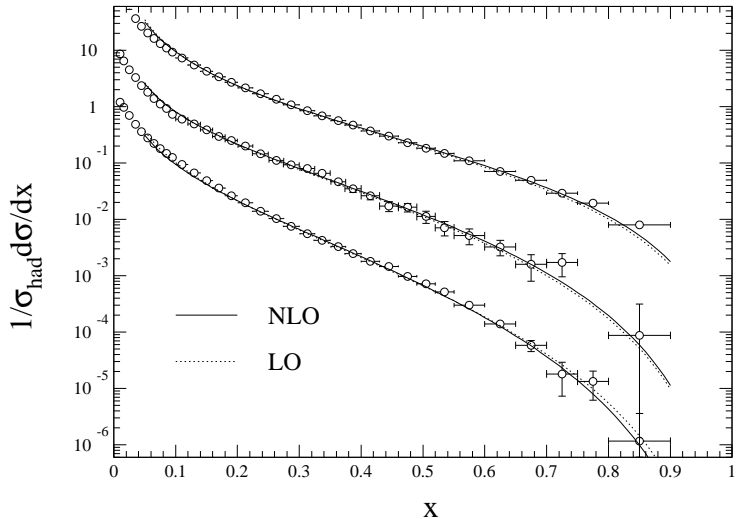,clip=,width=10cm}
\end{center} 
\vspace{-1.2cm} 
\caption[]{\small
Flavor separated data from ALEPH \protect\cite{ale3}
together with the results from the central fit.
From top to bottom,
the differential light quark, charm quark, and
bottom quark distributions are shown,
separated by factors of 10.
\protect\label{flavor}} 
\end{figure}

In contrast to our earlier work, now also a charm--enriched sample
is available in addition to the light--flavor and the bottom--enriched
samples.
We had to assume that the distributions 
are for pure samples in our earlier work.
The values of the mixing matrix have been published in \cite{ale3}
in the meantime.
This enables us to construct pure samples by multiplying the
data by the inverse of the mixing matrix.
Fig.~\ref{flavor} shows the data resulting from this procedure
which have been used for our fit.
From top to bottom, the pure light ($u,d,s$) quark, $c$, and $b$ quark
data are shown, separated by factors of 10.
One point of the charm distribution (bin 0.75 to 0.8) turns
negative in the demixing procedure.
\smallskip

\begin{figure}[hht!]
\begin{center} 
\epsfig{file=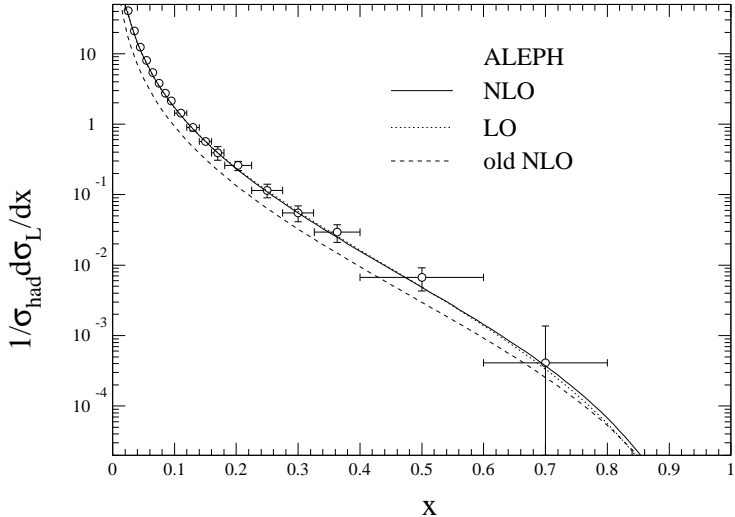,clip=,width=10cm}
\end{center} 
\vspace{-1.cm} 
\caption[]{\small  
The longitudinal cross section in \ee annihilation.
The results from the fit are compared to
the data from ALEPH \protect\cite{ale3}.
Both the NLO (solid) and the LO (dotted) are in
excellent agreement with the measurement, even at very low 
momentum fraction.
The dashed curve shows the result from \protect\cite{binpk} 
at the scale $m_Z$.
\protect\label{long}} 
\end{figure}

In addition to the unpolarized flavor tagged data we also use the
data for the longitudinally polarized cross section \cite{ale3}
that are shown in Fig.~\ref{long}.
In connection with the NLO coefficient functions they help to
constrain the gluon FF.
\smallskip

\begin{table}[hht!]
\begin{center}
\begin{tabular}{|c||r|r|}
\hline
data & $x_{min}$ & $x_{max}$ \\
\hline
$u,d,s$    & 0.14 & 0.80 \\
$c$        & 0.14 & 0.75 \\
$b$        & 0.20 & 0.75 \\
$\sigma_L$ & 0.08 & 0.80 \\
\hline
\end{tabular}
\caption{\small
Cuts on the ALEPH data \protect\cite{ale3} that are used in
the fits.
\protect\label{data}}
\end{center}
\end{table}

Only those bins were used in the fit, for which the correction factors
in the experimental analysis were close to one.
This restriction excluded some high--$x$ bins.
Lower limits were also enforced for theoretical reasons.
For the longitudinal cross section, $x>0.08$ was imposed, whereas
the lower bound was stricter for the flavor separated distributions.
This can be justified because the small $x$ behavior is dominated
by the gluon in any case so that little information is lost
by the exclusion of those points.
The lower and upper bounds for the individual sets of data
are listed in Table~\ref{data}.
\smallskip

The updated ALEPH data enable to fit light quark, $c$ quark, and $b$
quark FF's separately, without any further model assumptions.
While we could reasonably assume that the $u$ and the
$d$ quark behave similar in charged pion fragmentation,
this does not hold for the sum over all charged hadrons.
Therefore, we do not distinguish between the light quarks in this fit.
We use the standard parametrization at the
starting scale so that we have 12 fit parameters.
\smallskip

With the extraction of $\alpha_s$ in $ep$ collisions in mind, which will
be discussed in section~5.2, we will
not fit $\Lambda$ but instead fix the parameter at certain values.
As we do not want to fit $\Lambda$ here, we do not include data
at lower energies in our fit.
We will present separate sets of FF's for 
$\Lambda^{(4),NLO}=150,200,...,400$~MeV.
$\Lambda^{(4),NLO}=350$~MeV corresponds to $\Lambda^{(5),NLO}=229$~MeV
(see appendix~A), which is very close to the value of 227~MeV
we obtained in our earlier fit.
This is our preferred choice, it will be referred to as 
the central value of $\Lambda$.
As always, we will also present a LO fit for the central value.
As the LO does not describe the scaling violation as well we
will only present NLO sets for the other values of $\Lambda$.
For the central choice of $\Lambda$, we also construct 
sets with $\mu_0^2=1$~GeV$^2$
and $\mu_0^2=4$~GeV$^2$, in addition to our standard choice of
$\mu_0^2=2$~GeV$^2$.
\smallskip

\begin{table}[hht!]
\begin{center}
\begin{tabular}{|l||r|r|r|r||r|}
\hline
order & $u,d,s$ & $c$ & $b$ & $\sigma_L$ & average \\
\hline
NLO & 0.91 & 0.34 & 1.63 & 0.07 & 0.78 \\
 LO & 1.00 & 0.35 & 1.15 & 0.09 & 0.69 \\
\hline
\end{tabular}
\caption{\small
$\chi^2$ values for the central sets, fitted 
with $\Lambda^{(4),NLO}=350$~MeV and $\mu_0^2=2$~GeV$^2$.
\protect\label{chx1}}
\end{center}
\end{table}

\begin{table}[hht!]
\begin{center}
\begin{tabular}{|l||r|r|r|r||r|}
\hline
$\mu_0^2$ [GeV$^2$] & $u,d,s$ & $c$ & $b$ & $\sigma_L$ & average \\
\hline
1 & 1.25 & 0.39 & 1.63 & 1.03 & 1.06 \\
4 & 1.06 & 0.35 & 1.99 & 0.43 & 0.98 \\
\hline
\end{tabular}
\caption{\small
$\chi^2$ values for the NLO fits with low and high starting scales,
with $\Lambda^{(4),NLO}=350$~MeV.
\protect\label{chx2}}
\end{center}
\end{table}

\begin{table}[hht!]
\begin{center}
\begin{tabular}{|l||r|r|r|r||r|}
\hline
$\Lambda^{(4)}$ [MeV] & $u,d,s$ & $c$ & $b$ & $\sigma_L$ & average \\
\hline
150 & 0.81 & 0.33 & 1.85 & 0.18 & 0.81 \\
200 & 0.89 & 0.32 & 1.80 & 0.24 & 0.84 \\
250 & 1.36 & 0.40 & 1.64 & 0.64 & 1.03 \\
300 & 0.89 & 0.34 & 1.73 & 0.16 & 0.81 \\
400 & 0.83 & 0.34 & 1.57 & 0.11 & 0.74 \\
\hline
\end{tabular}
\caption{\small
$\chi^2$ values for the NLO fits with the 
various values of $\Lambda$,
and $\mu_0^2=2$~GeV$^2$.
\protect\label{chx3}}
\end{center}
\end{table}

The $\chi^2$ obtained in the fit are given in Tables~\ref{chx1}
through \ref{chx3} for the central values of the
parameters $\Lambda^{(4)}$ and $\mu_0$, for the
high and low starting scales, and for the various $\Lambda$ values,
respectively.
Averaged over all data,
we obtain values around $\chi^2_{d.o.f.}=0.8$
consistently for any choice of $\Lambda$ or $\mu_0$.
\smallskip

The comparison of our central fit results with the ALEPH data \cite{ale3}
have already been shown in Figs.~\ref{flavor} and \ref{long}.
Both the NLO (solid) and the LO (dotted) describe the flavor tagged
data well for $x$ between 0.1 and 0.75.
Our results are slightly low at higher $x$, whereas they overshoot
the data for $x$ below 0.1.
Fig.~\ref{long} demonstrates very good agreement for the
LO as well as NLO longitudinal cross sections, down to very low $x$.
Because the NLO longitudinal coefficient functions were not yet
available then, we combined the NLO FF's with the LO
coefficient functions in \cite{binpk}.
The result is also shown in Fig.~\ref{long} as dashed curve.
As the K factor for $d\sigma/dx$ is large, c.f.\ Fig.~\ref{kfac}b,
the consistent use of the NLO coefficient functions would lead
to considerably improved agreement even without a new fit.
This was demonstrated in \cite{rij2} for our first set \cite{bin1}.
\smallskip

\begin{table}[hht!]
\begin{center}
\small
\begin{tabular}{|c||lll|lll|lll|lll|}
\hline
order & \multicolumn{3}{|c|}{$u,d,s$} &\multicolumn{3}{|c|}{$c$} 
  & \multicolumn{3}{|c|}{$b$} & \multicolumn{3}{|c|}{$G$} \\
 & $N$ & $\alpha$ & $\beta$ & $N$ & $\alpha$ & $\beta$ 
 & $N$ & $\alpha$ & $\beta$ & $N$ & $\alpha$ & $\beta$ \\
\hline
NLO & 2.05 & -0.733 & 1.77 & 4.22 & -0.761 & 3.52 
    & 2.39 & -1.10  & 4.71 & 5.47 & -0.740 & 2.33 \\
 LO & 1.79 & -0.759 & 1.82 & 4.52 & -0.684 & 3.59 
    & 1.81 & -1.26  & 4.35 & 4.30 & -1.17  & 1.25 \\
\hline
\end{tabular}
\normalsize
\caption{\small
The parameters resulting from the central fits in NLO and LO.
\protect\label{chpar}}
\end{center}
\end{table}

\begin{table}[hht!]
\begin{center}
\small
\begin{tabular}{|c||lll|lll|lll|lll|}
\hline
$\Lambda^{(4)}$   & \multicolumn{3}{|c|}{$u,d,s$} 
  & \multicolumn{3}{|c|}{$c$} 
  & \multicolumn{3}{|c|}{$b$} & \multicolumn{3}{|c|}{$G$} \\
$[$MeV$]$  & $N$ & $\alpha$ & $\beta$ & $N$ & $\alpha$ & $\beta$ 
 & $N$ & $\alpha$ & $\beta$ & $N$ & $\alpha$ & $\beta$ \\
\hline
150 & 1.47 & -0.950 & 1.88 & 4.03 & -0.766 & 3.73 
    & 2.29 & -1.07  & 4.72 & 4.97 & -0.935 & 2.33 \\
200 & 1.67 & -0.970 & 1.96 & 3.89 & -0.817 & 3.64 
    & 2.33 & -1.09  & 4.72 & 4.05 & -0.876 & 2.20 \\
250 & 1.88 & -0.693 & 1.81 & 4.38 & -0.740 & 3.78 
    & 2.31 & -1.07  & 4.68 & 4.97 & -0.992 & 2.28 \\
300 & 1.92 & -0.842 & 1.85 & 4.72 & -0.738 & 3.75 
    & 2.48 & -1.07  & 4.77 & 5.07 & -0.729 & 2.30 \\
400 & 1.89 & -0.766 & 1.64 & 3.95 & -0.802 & 3.43 
    & 2.31 & -1.12  & 4.66 & 6.27 & -0.561 & 2.31 \\
\hline
\end{tabular}
\normalsize
\caption{\small
The parameters resulting from the different choices of $\Lambda$ in NLO.
\protect\label{chparp}}
\end{center}
\end{table}

\begin{table}[hht!]
\begin{center}
\small
\begin{tabular}{|c||lll|lll|lll|lll|}
\hline
$\mu_0^2$ 
  & \multicolumn{3}{|c|}{$u,d,s$} &\multicolumn{3}{|c|}{$c$} 
  &\multicolumn{3}{|c|}{$b$} &\multicolumn{3}{|c|}{$G$} \\
$[$GeV$^2]$  & $N$ & $\alpha$ & $\beta$ & $N$ & $\alpha$ & $\beta$ 
 & $N$ & $\alpha$ & $\beta$ & $N$ & $\alpha$ & $\beta$ \\
\hline
1 & 2.59 & -0.638 & 1.70 & 4.11 & -0.839 & 3.47 
   & 2.41 & -1.11  & 4.67 & 5.25 & -0.710 & 2.32 \\
4 & 1.84 & -0.851 & 1.98 & 4.33 & -0.708 & 3.54 
   & 2.35 & -1.05  & 4.64 & 5.17 & -0.683  & 2.24 \\
\hline
\end{tabular}
\normalsize
\caption{\small
The parameters resulting from the fits with low and high starting 
scales $\mu_0^2$.
\protect\label{chparpp}}
\end{center}
\end{table}

Table~\ref{chpar} lists the values of the parameters $N, \alpha$
and $\beta$ that result from the central fits.
For further application, the parameters for the other
values are also given in Tables~\ref{chparp} and \ref{chparpp}.
When comparing the parameters for the different choices
it is apparent that the individual fragmentation functions 
are well constrained at NLO.
In particular for the gluon, the results from the
fits do not depend strongly on the choices of $\Lambda$ or $\mu_0$.
This serves as prove that the gluon is indeed well constrained
by the inclusion of the longitudinal data, at least to
the point that it is not particularly strongly correlated
with the other partons.
As expected from their softer spectra, the values
for the parameter $\beta$ for the heavy $c$ and $b$
are larger than for the light $u,d,s$ quarks.
From around 1.8 for the light quarks, it increases
to 3.7 for the charm and 4.7 for the bottom quark.
\smallskip

\begin{figure}[hht!]
\begin{center} 
\begin{picture}(12.5,5.5)
\put(-1.2,-.3){\makebox{
\epsfig{file=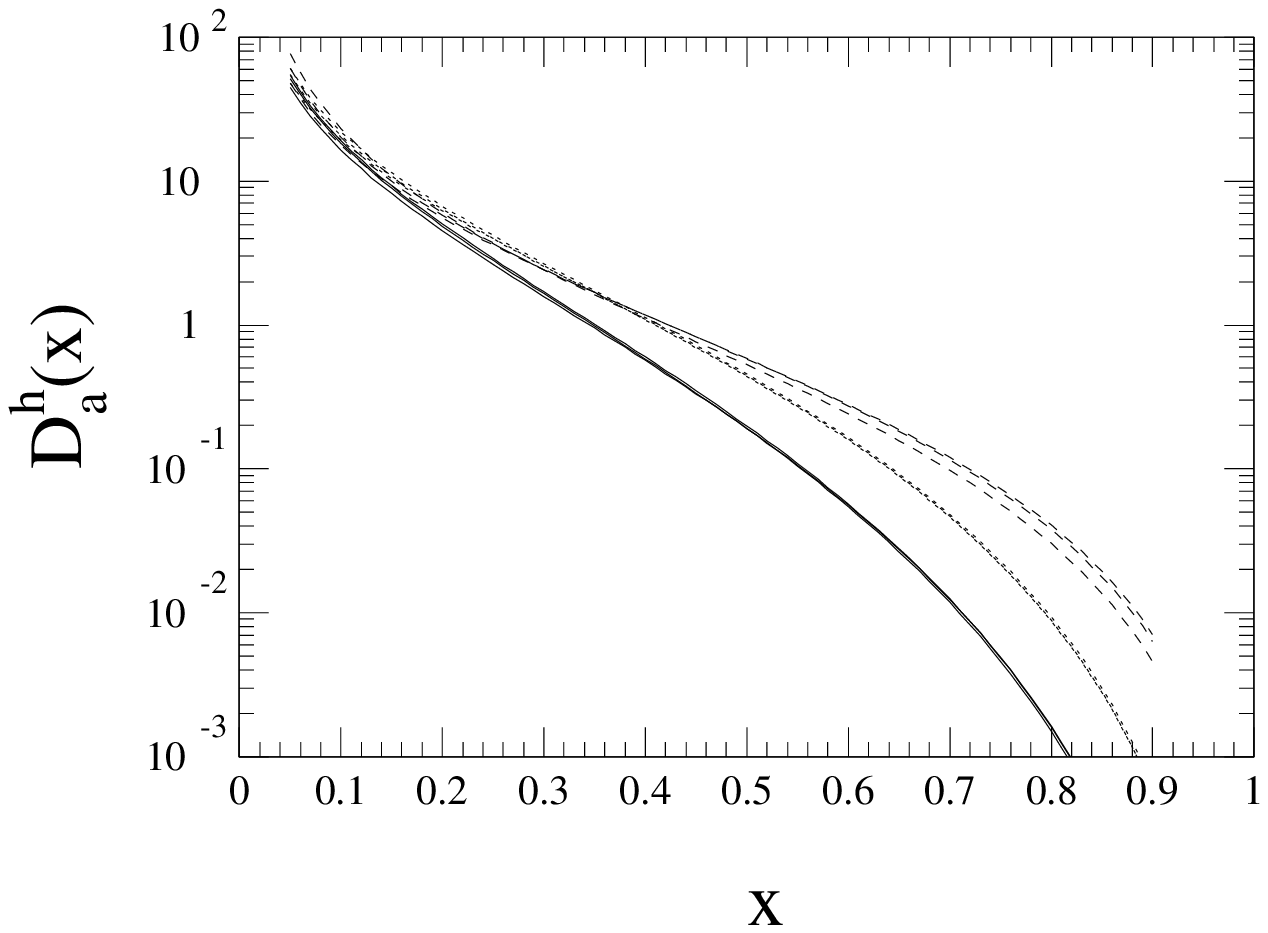,clip=,width=7.3cm}}}
\put(6.4,-.3){\makebox{
\epsfig{file=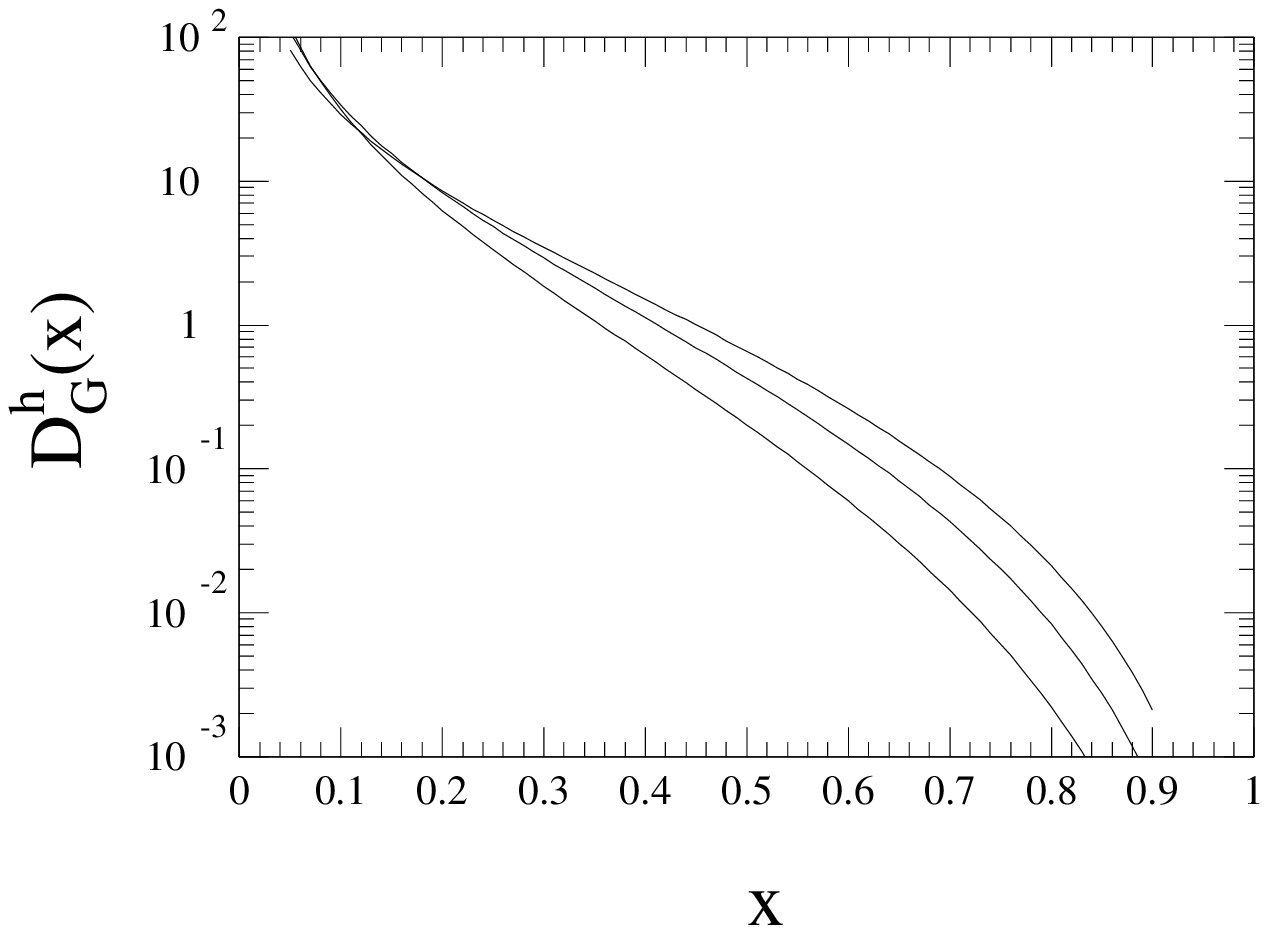,clip=,width=7.3cm}}}
\put(-0.7,0.2){a)}
\put(6.8,0.2){b)}
\end{picture}
\end{center} 
\vspace{-0.4cm} 
\caption[]{\small  
The NLO fragmentation functions of the new $h^{\pm}$
fit.
The spread for different values of the starting scale is shown
at 10~GeV.
a)
The light quark (dashed) charm (dotted) and bottom (solid) FF's.
The three curves for each parton are hardly distinguishable.
b)
The FF of the gluon.
Going upward, the
three solid curves correspond to values of $\mu_0^2$ of
 1, 2, and 4~GeV$^2$, respectively.
\protect\label{spread}} 
\end{figure}

The sets for different choices of $\Lambda$ represent
different physics, namely different strong coupling strength
so that the corresponding FF's are expected to differ.
On the other hand, the starting scale of the evolution, $\mu_0$,
is an unphysical parameter and the results should not depend
on it.
The differences between
fits for various values of $\mu_0$ at the central value of
$\Lambda$ hence give a measure of how well the FF's are 
determined by the fit.
Fig.~\ref{spread}a shows the results from the
fits with $\mu_0^2=1,2$, and 4~GeV$^2$ for the
light quarks (dashed) the charm (dotted) and the bottom 
quark (solid).
In Fig.~\ref{spread}b, the gluon FF's of the same three sets
are plotted as solid lines.
While the quark FF's do not deviate much from each other,
the gluon does exhibit a considerable spread at large $x$.
Only at small $x$ below 0.2, where it 
contributes dominantly to the longitudinal cross section
(cf.\ Fig.~\ref{furry}), is it fixed by the fit.
As the region of relatively small $x$ is the most important, 
both in driving the evolution of the FF's and in the
contribution to the cross section in hadronic or semi--hadronic
scattering,
this nevertheless represents a considerable improvement
over our earlier analyses.
\smallskip

\begin{figure}[hht!]
\begin{center} 
\epsfig{file=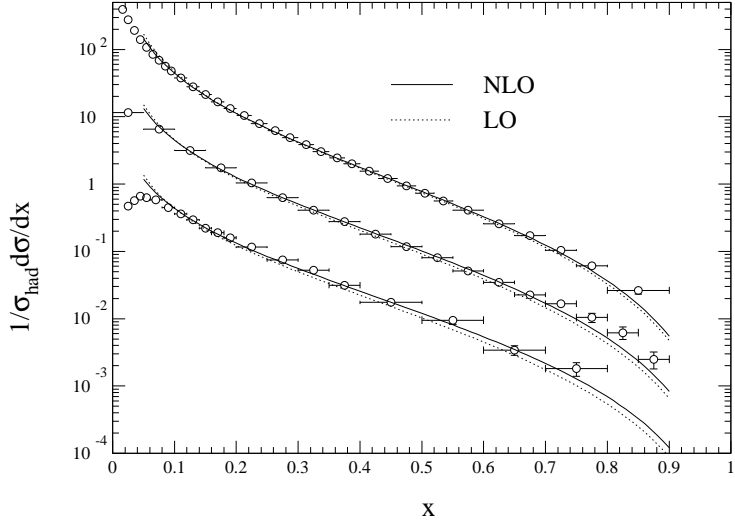,clip=,width=10cm}
\end{center} 
\vspace{-1.2cm} 
\caption[]{\small  
The unpolarized cross section as evaluated with the
new charged hadron FF's is compared to \ee data
at different CM energies.
From top to bottom, the data are from ALEPH 
at 91.2~GeV \protect\cite{ale3}, from
MARK~II at 29~GeV \protect\cite{mar5}, 
and from TASSO at 14~GeV \protect\cite{tas3}.
Our NLO (LO) results are given by the solid (dotted) curves.
For better separation, the data have been divided by powers of 10.
\protect\label{eesig}} 
\end{figure}

Finally, we will compare the predictions of our
new set of FF's to data at lower CM energies.
Fig.~\ref{eesig} shows measurements of the unpolarized
cross section taken by (from top to bottom) ALEPH \cite{ale3},
MARK~II \cite{mar5}, and TASSO \cite{tas3}, at
91.2, 29 and 14~GeV, respectively.
The NLO results (solid) agree well
with the data in the region of reliability between $x=0.1$ and 0.75.
The LO also gives satisfactory agreement.
\smallskip

\begin{figure}[hht!]
\begin{center} 
\epsfig{file=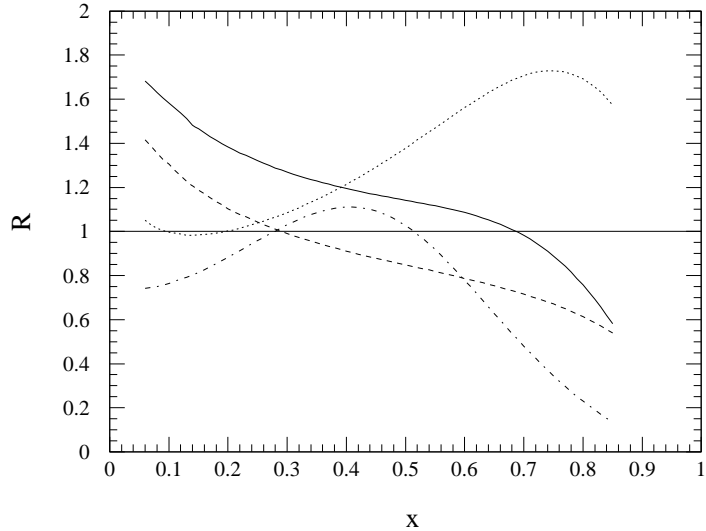,clip=,width=10cm}
\end{center} 
\vspace{-1.2cm} 
\caption[]{\small  
The ratio R of the new set of charged hadron FF's to those
presented in \protect\cite{binpk} at $M_f=10$~GeV.
Shown are the NLO results for the gluon (solid),
the light quarks (dashed), the charm quark (dotted), and the
bottom quark (dash--dotted).
\protect\label{ffcomp}} 
\end{figure} 
\clearpage

We conclude that our new NLO set of charged hadron FF's
features well constrained quark fragmentation functions.
In addition to the bottom quark fragmentation,
the charm quark fragmentation has now also been measured separately.
Moreover, the gluon is better known than in our earlier sets,
at least in the important small--$x$ region.
We will suggest how to use the improved knowledge of the
charged hadron fragmentation functions to measure $\alpha_s$
and to determine the gluon content of the photon in
section~5.
A direct comparison of our new FF's with those of \cite{binpk}
for charged hadrons (pions, kaons, and protons) is shown in
Fig.~\ref{ffcomp}.
We plot the ratio of the NLO results at $M_f=10$~GeV for our new
charged hadron set over the one discussed in the previous subsection.
The additional information on the purity of the samples
influenced the shape of the bottom FF (dash--dotted) at large $x$.
The new fit also profited from new information on the charm fragmentation
not at our disposal in the earlier fit.
Thus, both the marked increase of the charm FF (dotted) and the
decrease of the bottom FF (dash--dotted) at large $x$ 
are signals of improved data.
The light quark fragmentation (dashed) is similar for both sets in the
central region of $x$.
The gluon, however, shows a significant increase at small $x$ compared
to our older set, induced by the improved treatment of the longitudinal
cross section.

\vspace{2cm}
\subsection{Neutral Kaons}
\smallskip

Neutral particles do not allow for a precise determination of 
momentum in conventional experimental setups.
The neutral kaons are studied by their decay products,
experimentally.
The main decay channel for the short--lived
CP--eigenstate is into two charged pions and for a highly energetic
 $K_S^0$ the secondary vertex is located about 3~cm from the primary one.
The longlived eigenstate decays outside the driftchamber.
Thus, it is the $K_S^0$ that is measured, and as it is an equal mixture
of the strangeness eigenstates $K^0$ and $\overline{K}^0$ this corresponds
to measuring the average of 
those two states.
In the following, we shall collectively
use the symbol $K^0$ for the sum of $K_S^0$ and $K_L^0$ 
(or $K^0$ and $\overline{K}^0$).
For not too high momenta, when the localization of the secondary
vertex becomes difficult, the signature is very clear, resulting
in precise measurements even on a large background.
Consequently, spectra of neutral kaons 
in $p\overline{p}$ collisions \cite{ua51,cdf1,ua11}
and in photoproduction \cite{h11,h16} have been measured with small errors.
\smallskip

In order to use these excellent data to test QCD, fragmentation functions
for neutral kaons are required.
This was our motivation for the work presented in \cite{bink0}.
We used \ee data from MARK~II \cite{mar4} and ALEPH \cite{ale2}.
In the spirit of the earlier work on charged pions and kaons 
the standard parametrization (\ref{parstand})
was used for each flavor individually.
Only the valence--type quarks were identified, i.e.\
 $D_d^{K^0}=D_s^{K^0}$.
However, as no experimental information comparable to the one
exploited in subsection~4.1 on flavor differences was available,
the parameters were not well constrained by the fit.
In particular, the gluon
was strongly correlated with the other partons,
which is a notorious problem in fits to \ee data. 
Inspired by isospin symmetry, we made 
the assumption that the gluons behave 
as in charged kaons, i.e.\ the respective FF's were identified at the
starting scale,
\begin{equation}
\label{K0glue}
D_G^{K^0}(x,\mu_0) \equiv D_G^{K^{\pm}}(x,\mu_0)
\STOP
\end{equation}
The $b$~quark is then fixed by the fake scaling effects
that will be discussed in section~5.1.
In essence, the change of shape from 29 to 91.2~GeV, after
subtraction of the scaling violation effect which is 
dominated by the gluon, is attributed to the $b$~quark.
The other sea--type quarks are also 
rather strongly correlated with each other.
 $\Lambda$ was kept fixed at the NLO (LO) value obtained in
\cite{binpk}.
\begin{figure}[hht!]
\begin{center}
\epsfig{file=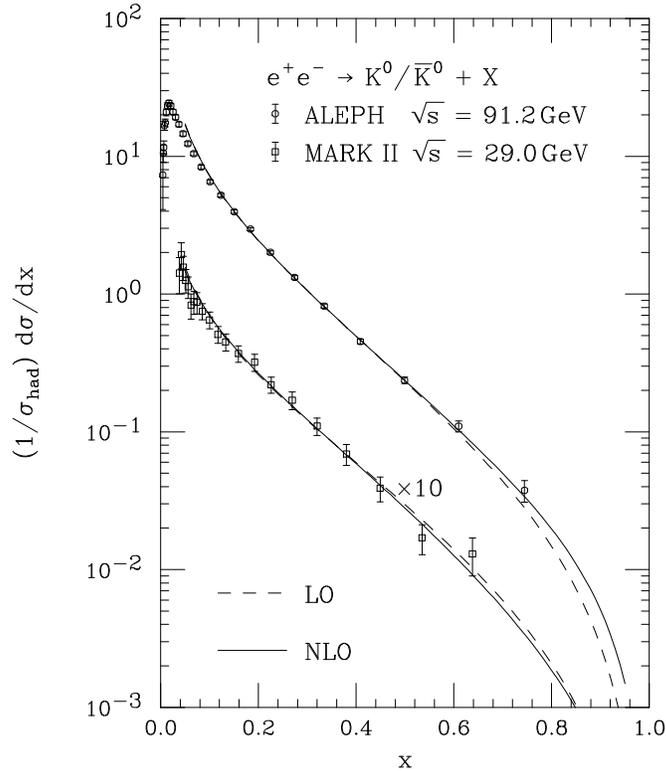,clip=,width=8.7cm}
\end{center}
\vspace{-0.6cm}
\caption[]{\small 
Differential cross sections of inclusive
$K^0 + \overline K^0$ production at LO (dashed lines) 
and NLO (solid lines) as
functions of $x$ at $\protect\sqrt{s}=91.2$ and 29~GeV.
The theoretical calculations are compared with the respective experimental
data by ALEPH \protect\cite{ale2} and MARK~II \protect\cite{mar4}.
For better separation, the distributions at $29$~GeV have been divided by
$10$.
\protect\label{eek0}}
\end{figure}

The quality of the fit to the MARK~II \cite{mar4} and ALEPH \cite{ale2}
data is demonstrated in Fig.~\ref{eek0}.
Both the NLO (solid) and the LO ( dashed) fit the data very well.
The $\chi^2$ resulting from the fit are listed in Table~\ref{k0fit}.
Besides the ones used in the fit, data at lower CM energies
also yield small $\chi^2$ values.
\begin{table}[hht!]
\begin{center}
\begin{tabular}{|c||c|c|c|}
\hline
 $\sqrt{Q^2}$ [GeV] & NLO & LO & ref. \\
\hline
\hline
91.2  & 0.5 & 0.5 & \cite{ale2}$^*$ \\
\hline
35.0  & 0.2 & 0.2 & \cite{cel1} \\
\hline
29.0  & 0.4 & 0.5 & \cite{mar4}$^*$ \\
\hline
10.49 & 1.1 & 1.4 & \cite{cle1} \\
\hline
\end{tabular}
\caption{\small
 $\chi^2$ per degree of freedom for the NLO  and LO fits of
neutral kaons.
The data used in the fits are marked by asterisk.
\protect\label{k0fit}}
\end{center}
\end{table}

\begin{table}[hht!]
\begin{center}
\begin{tabular}{|c|c||c|c|c|}
\hline
set & flavor & $N$ & $\alpha$ & $\beta$ \\
\hline
\hline
$K^0$ & $u$ 
       & 0.53 (0.54) & -0.57 (-0.77) & 1.87 (1.49) \\
 & $d$, $s$ 
       & 1.45 (1.54) & -0.62 (-0.72) & 3.84 (3.70) \\
 & $c$ & 1.70 (1.13) & -0.51 (-0.70) & 3.76 (3.02) \\
 & $b$ & 0.47 (0.64) & -0.66 (-0.63) & 1.49 (1.84) \\
\cline{2-5}
 & $G$ & 0.31 (0.37) & -0.17 (-0.21) & 0.89 (3.07) \\
\hline
\end{tabular}
\caption{\small
The parameters obtained from the NLO (LO) fits for the neutral kaon.
\protect\label{k0par}}
\end{center}
\end{table}

The parameters that result from the fit are given in Table~\ref{k0par}.
The perturbative stability is quite good for the differential cross
sections, although there are some changes in the parameters
of the sea--type quarks from LO to NLO.
The overall agreement with \ee annihilation data is good.
The FF's will be applied to photoproduction and $p\overline{p}$
collisions in subsection~5.1.

\vspace{2cm}
\subsection{$D^{*\pm}$ Mesons}
\smallskip

The H1 \cite{h13} and ZEUS \cite{zeus2} collaborations at HERA
presented new
data on inclusive $D^{*\pm}$ spectra 
that extend up to $p_T=12$~GeV.
In order to obtain definite predictions for $D^{*\pm}$
production cross sections in the massless--charm scheme,
one needs the PDF's of the charm quarks in the proton and
the photon besides the FF's for $D^{*\pm}$ mesons.
In \cite{kni1} it was found that the charm
component in the proton contributes only marginally
to the cross section of inclusive $D^{*\pm}$ photoproduction.
This offers the possibility to use the new data from HERA
together with NLO fragmentation functions for
 $D^{*\pm}$ mesons to obtain information
on the charm distribution in the photon.
This was our motivation in \cite{binds}.
This subsection contains the discussion of
the sets of FF's presented there.
Their application to the determination of the photon PDF's is
postponed to subsection~5.3.
\smallskip

Precise data on inclusive $D^{*\pm}$ production in \ee annihilation
have been presented by the ALEPH \cite{ale1} and
OPAL \cite{opa1} collaborations at LEP1, recently.
The contributions from charm and bottom quarks are disentangled
in those studies and light quarks are negligible in this process.
With the use of these data, we obtained very well constrained sets
of FF's.
In set S we used the standard form (\ref{parstand}) for both
 $c$ and $b$ quarks whereas we used the Peterson distribution 
\cite{pet1},
\begin{equation}
D_a^h(x,\mu_0^2)=N\frac{x(1-x)^2}{[(1-x)^2+\epsilon x]^2}
\label{peterson}
\end{equation}
for the charm quark and the standard form for the bottom quark
in a mixed set (M).
The Peterson distribution 
is particularly suitable to describe FF's that peak at large $x$,
so a good fit could not be obtained when applying it also
to the relatively broad bottom FF.
As the fragmentation of a light quark into a $D^{*\pm}$
is very unlikely, the light quark FF's were set to zero
at the starting scale.
The gluon is generated only through the evolution as well.
Therefore, the evolution starts at $\mu_0=2m_c=3$~GeV with four flavors.
The $b$ quark is added at $2m_b=10$~GeV.
Values for $\Lambda$ were taken from the charged pion and charged
kaon analysis.
We use the $n$--space evolution for set S whereas we found it more 
convenient to evolve set M in $x$--space.
For a comparison of the two techniques see appendix~C.
\smallskip

So far,
the formalism is identical to the one for describing the
fragmentation into light hadrons based on the {\msbar}--factorization scheme. 
To account for the mass of the heavy quarks,
we now adjust the factorization of the
final--state collinear singularities associated with the charm and bottom 
quarks in such a way that it matches the massive calculation. 
This is achieved by a change of the factorization scheme, i.e. 
we substitute
 $P_{a\to Q}^{(0)}(z)\ln(Q^2/M_f^2)\rightarrow
P_{a\to Q}^{(0)}(z)\ln(Q^2/M_f^2)-d_{Qa}(z)$,
in the coefficient functions where the $d_{Qa}$ are given by
\begin{eqnarray}
d_{QQ}(z)
&\n=\n&-P_{Q\to Q}^{(0)}(z)\ln\left(\frac{\mu_0^2}{m_Q^2}\right)
+C_F\left\{-2\delta(1-z)+2\left(\frac{1}{1-z}\right)_+
\right.
\nonumber\\
&&\hspace{2cm}\left.
+4\left[\frac{\ln(1-z)}{1-z}\right]_+
-(1+z)\left[1+2\ln(1-z)\right]\right\}
\COMMA
\label{dQQ}\\
d_{QG}(z)
&\n=\n&-P_{G\to Q}^{(0)}(z)\ln\left(\frac{\mu_0^2}{m_Q^2}\right)
\COMMA
\label{dQG}\\
d_{Qq}(z)
&\n=\n&d_{Q\bar q}(z)=d_{Q\bar Q}(z)=0
\label{dQq}
\COMMA
\end{eqnarray}
with $Q=c,b$ and $q=u,d,s$.
These functions have been extracted from \cite{mel1}
by B.A.~Kniehl, G.~Kramer and M.~Spira in \cite{kni1},
where a detailed discussion of the modified scheme is presented.
The same substitutions must also be performed in any 
massless hard--scattering cross sections that will be
convoluted with these FF's.
When we study $D^{*\pm}$ production in photoproduction in the next
section, this has of course been taken into account appropriately.
\smallskip

\begin{table}[hht!]
\begin{center}
\begin{tabular}{|c|c||c|c|c|c|}
\hline
set & flavor & $N$   & $\alpha$ & $\beta$ & $\epsilon_c$ \\
\hline
\hline
LO S & $c$ & 483 & 8.04 & 3.08 & -- \\
\cline{2-6}
     & $b$ & 204 & 2.96 & 6.05 & -- \\
\hline
NLO S & $c$ & 496 & 8.84 & 2.90 & -- \\
\cline{2-6}
      & $b$ & 173 & 2.81 & 6.20 & -- \\
\hline
\hline
LO M & $c$ & 0.214 & -- & -- & 0.0926 \\
\cline{2-6}
     & $b$ & 197 & 3.04 & 5.92 & -- \\
\hline 
NLO M & $c$ & 0.161 & -- & -- & 0.0674 \\
\cline{2-6}
     & $b$ & 194 & 2.97 & 6.22 & -- \\
\hline 
\end{tabular}
\caption{\small
The fit parameters for the charm and bottom quark FF's of sets S and M
at LO and NLO.
The other FF's are set to zero at the starting scale.
\protect\label{pars}}
\end{center}
\end{table}

The values of $N$, $\alpha$, $\beta$, and $\epsilon_c$ that result from the
LO and NLO fits to these data are displayed in Table~\ref{pars}.
The parameter $\epsilon_c$ appears only in the charm quark FF's of
set M.
Bottom fragmentation is small at low energies and one may
then use $\epsilon_c$ as the single parameter in phenomenological
analyses of $D^{*\pm}$.
This parameter thus supersedes the value by Chrin \cite{chr1}
that was obtained from older data with different methods.
$\epsilon_c$ depends on the scheme of factorization and on the
way perturbative and non--perturbative effects are handled in the
fragmentation functions.
The parameters of the bottom quark FF's of sets S and M are very similar,
i.e., the data fix the charm  and bottom quark FF's independently of
each other.
\smallskip

\begin{figure}[hht!]
\begin{center}
\begin{picture}(12.5,5.5)
\put(-1.7,-.3){\makebox{
\epsfig{file=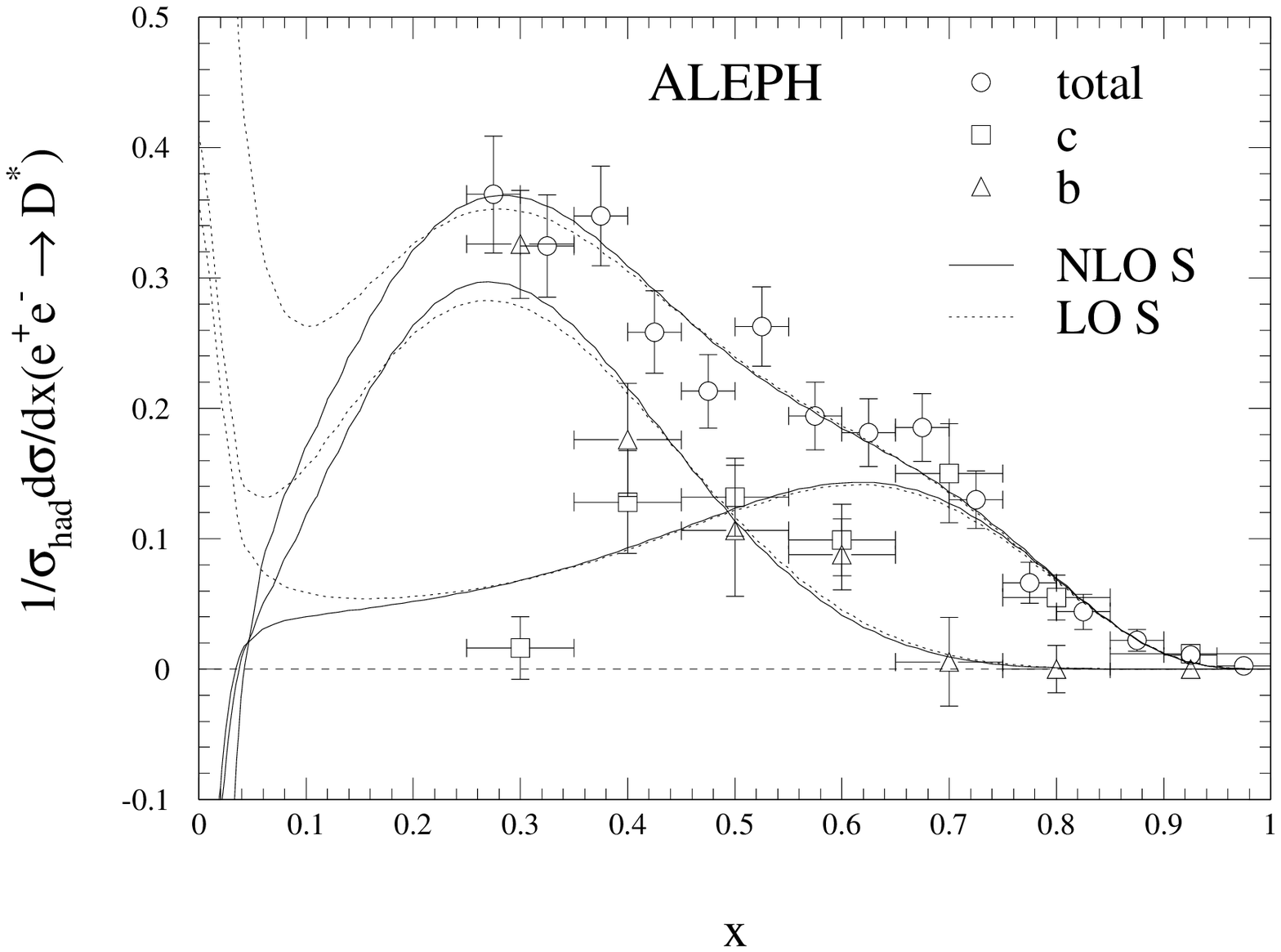,clip=,width=7.3cm}}}
\put(6.4,-.3){\makebox{
\epsfig{file=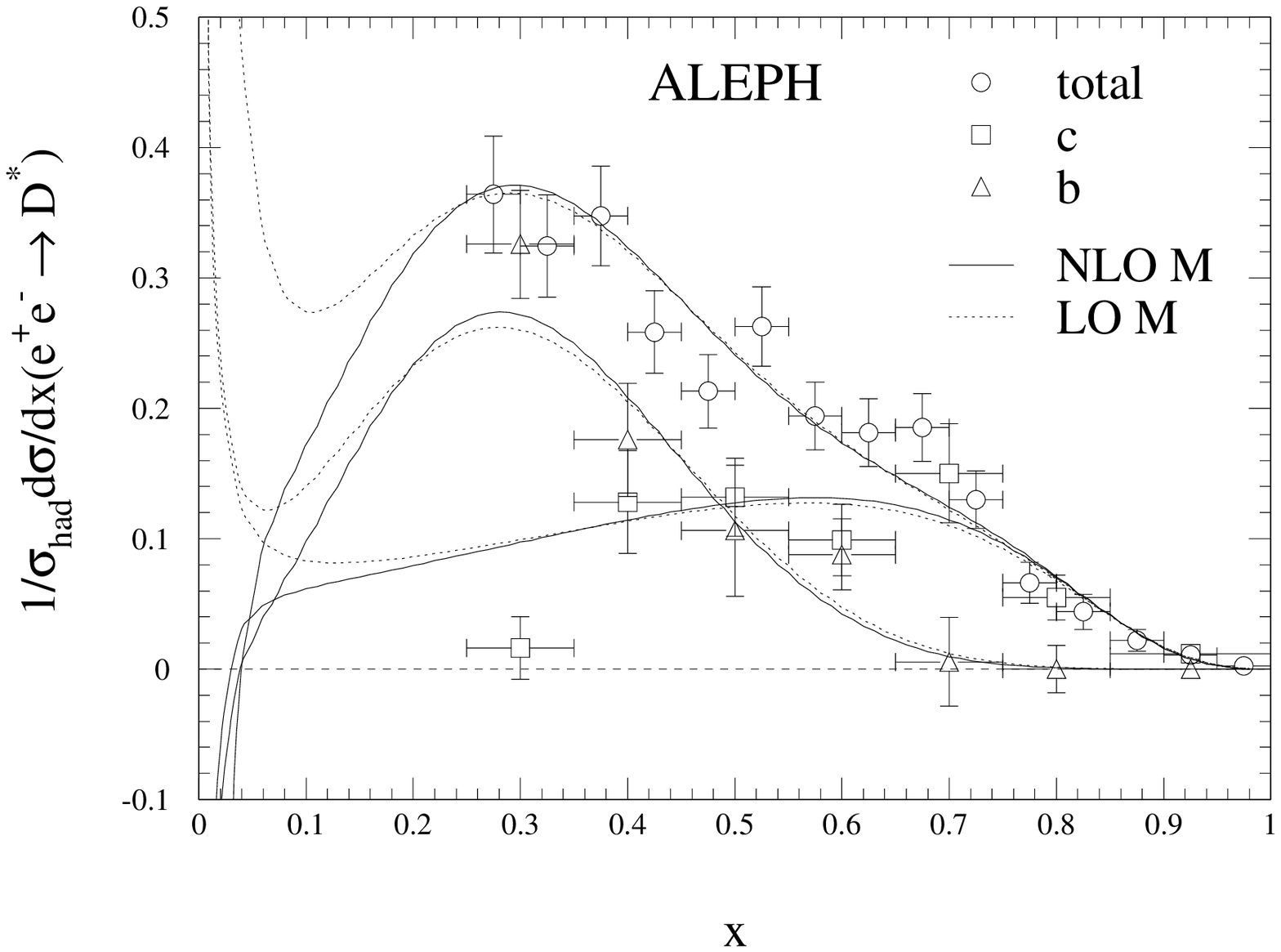,clip=,width=7.3cm}}}
\put(-1.2,0.2){a)}
\put(6.8,0.2){b)}
\end{picture}
\end{center}
\vspace{-0cm}
\caption[]{\small 
Inclusive $D^{*\pm}$ production in \ee
annihilation.
Our predictions for 
a) 
set S and 
b) 
set M are compared
with data from  ALEPH \protect\cite{ale1}.
The three sets of curves and data correspond to the $Z\to c\overline{c}$ 
and $Z\to b\overline{b}$ samples as well as their sum.
The NLO (LO) results are plotted as solid (dotted) lines.
\protect\label{dstaraleph}
}
\end{figure}

\begin{figure}[hht!]
\begin{center}
\begin{picture}(12.5,5.5)
\put(-1.7,-.3){\makebox{
\epsfig{file=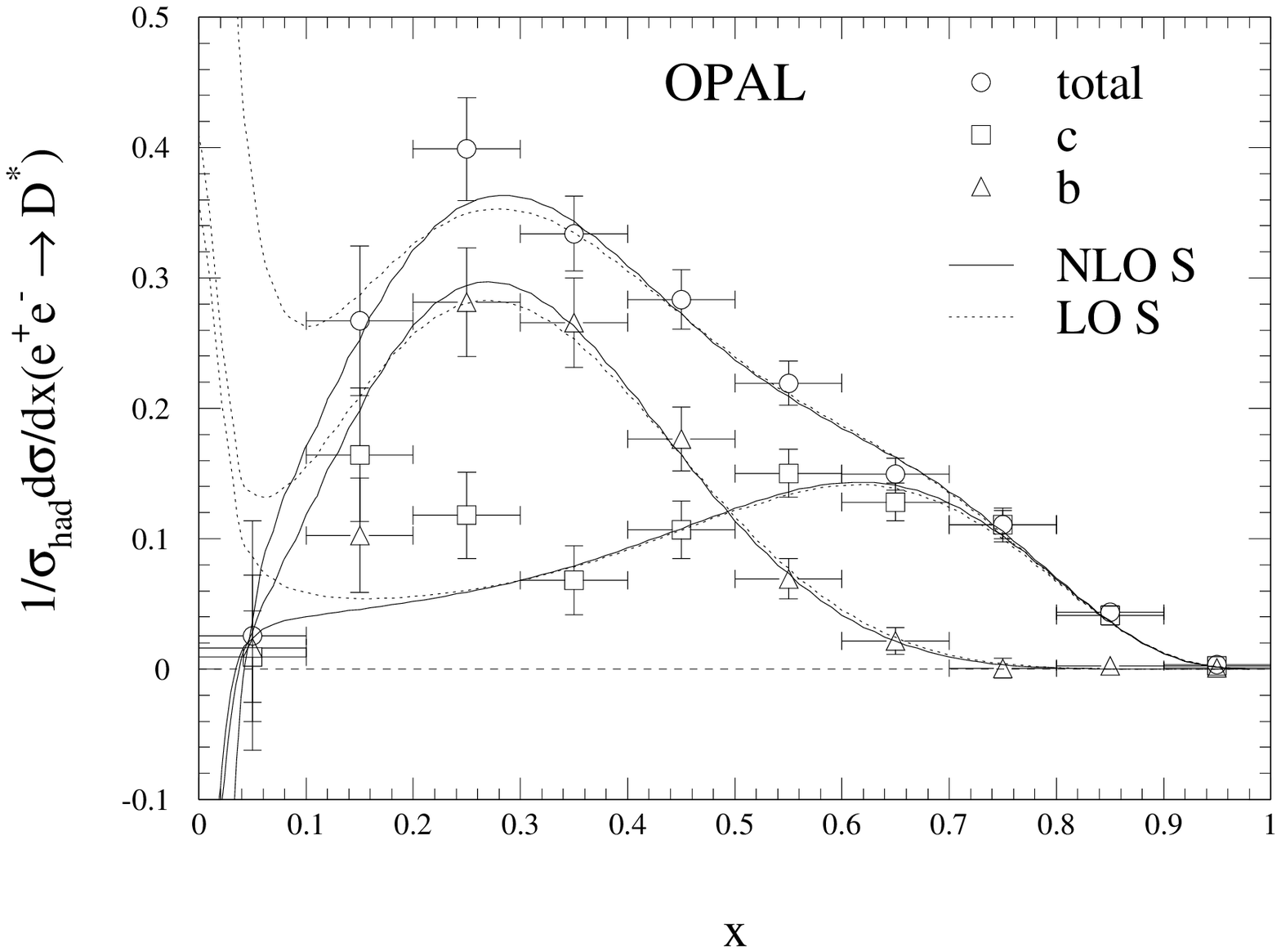,clip=,width=7.3cm}}}
\put(6.4,-.3){\makebox{
\epsfig{file=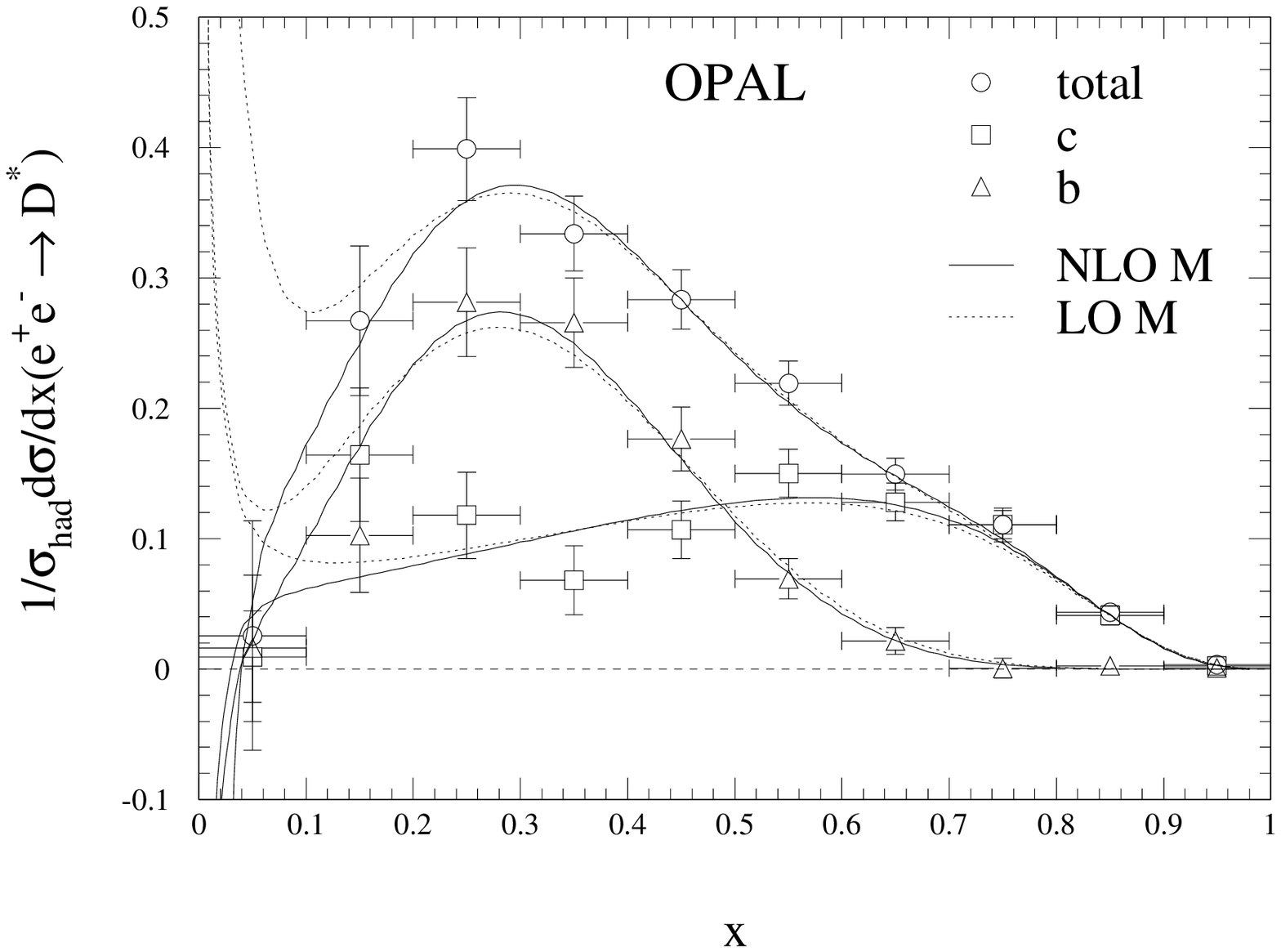,clip=,width=7.3cm}}}
\put(-1.2,0.2){a)}
\put(6.8,0.2){b)}
\end{picture}
\end{center}
\vspace{-0cm}
\caption[]{\small 
As Fig.~\protect\ref{dstaraleph}, but for the 
OPAL \protect\cite{opa1} data.
\protect\label{dstaropal}
}
\end{figure}

The results of our fits S and M are compared to the
ALEPH data \cite{ale1} in Figs.~\ref{dstaraleph}a and b, respectively.
Figs.~\ref{dstaropal}a and b 
shows the analogous comparison for the OPAL data \cite{opa1}.
Except for very small $x$, the NLO (solid) and
LO (dotted) curves are very similar. 
This is also true for the distributions at the starting scale, as may be seen
by comparing the corresponding LO and NLO parameters in Table~\ref{pars}.
Only $\epsilon_c$ of set M changes noticeably from LO to NLO.
For $x \to 0$, we obtain large differences between LO and NLO, indicating
that the perturbative treatment ceases to be valid.
In this limit, also the massless approximation is not good anymore
as pointed out already.
Our results are meaningful only for $x\ge x_{\rm cut}=0.1$ and
the first bin of the OPAL data is excluded from the analysis.
\smallskip

The Figs.~\ref{dstaraleph}a, b 
and \ref{dstaropal}a, b show the
normalized cross section (\ref{sigxpol}), 
which also includes the gluon
contributions in NLO.
These are attributed to the charm and bottom quark
curves, respectively, in proportion to their charge weight.
The contributions from $a=u,\bar u,d,\bar d,s,\bar s$ are 
also included for the curves that give the
total $D^{*\pm}$ cross section.
Numerically, their contribution is less than 2\% of the integrated
cross section.
\smallskip

The $\chi^2$ values obtained with sets S and M at
LO and NLO for the ALEPH and OPAL data, as well as their combination,
are listed in Table~\ref{dsfit}.
Along with the total values, we also give those for the
individual subsets of data.
Set S leads to better $\chi^2$ values than
set M, which is not surprising if one recalls that set S has six degrees of
freedom while set M has only five.
The $\chi^2$ are smaller for the OPAL data than for the ALEPH data,
in general.
\smallskip

\begin{table}[hht!]
\begin{center}
\begin{tabular}{|c||c||c|c|c||c|c|c|}
\hline
set & total & \multicolumn{3}{|c||}{ALEPH}& \multicolumn{3}{|c|}{OPAL}\\
      &       & sum  & $c$  & $b$  & sum  & $c$   & $b$ \\
\hline
\hline
LO S  & 0.81 & 0.73 & 1.29 & 0.65 & 0.48 & 1.02 & 0.82 \\
\hline
NLO S & 0.79 & 0.78 & 1.31 & 0.63 & 0.42 & 1.08 & 0.63 \\
\hline
\hline
LO M  & 1.03 & 1.05 & 2.11 & 0.77 & 0.47 & 1.29 & 0.75 \\
\hline
NLO M & 0.95 & 1.10 & 2.05 & 0.74 & 0.40 & 1.11 & 0.47 \\
\hline
\end{tabular}
\caption{\small
The $\chi^2$ per degree of freedom obtained in the LO and NLO fits 
S and M to the ALEPH \protect\cite{ale1} and OPAL \protect\cite{opa1}
data.
The first bin of the OPAL data has been excluded from the fit.
\protect\label{dsfit}}
\end{center}
\end{table}

The integrals of the charm  and bottom quark FF's into $D^{*\pm}$ mesons over
 $x$ give the branching fractions of these transitions. 
For the reasons given above, we restrict our considerations to the region
 $x_{\rm cut}<x<1$, with $x_{\rm cut}=0.1$, and define
\begin{equation}
\label{dstbranch}
B_Q(M_f)=\int_{x_{\rm cut}}^1dxD_Q(x,M_f^2)
\COMMA
\end{equation}
where $Q=c,b$.
Experimentally, the contribution from the omitted region $0<x<x_{\rm cut}$ is
close to zero with a large error.
When evaluating (\ref{dstbranch}) with our sets S and M,
we find that the values depend only weakly on the scale $M_f$.
The theoretical results for the branching fractions are in good
agreement with the values measured experimentally at LEP1.
For example
in NLO, the sets S and M yield 1.12 and 0.93 respectively
for the ratio of the charm over the bottom branching fractions.
This agrees well with the experimental result
of $B_c(m_Z)/B_b(m_Z)=1.03\pm0.15$ \cite{opa1}.
In a more standard approach, the FF's would have been normalized to
the branching fractions and eq.~(\ref{dstbranch}) would have 
reproduced these experimental results by construction.
In our approach the agreement is nontrivial and indicates the
soundness of the fit procedure.
\smallskip

The first moment of the FF's, normalized to the branching fraction,
gives the mean momentum fraction,
\begin{equation}
\label{dstav}
\langle x\rangle_Q(M_f)=\frac{1}{B_Q(M_f)}\int_{x_{\rm cut}}^1dx\,
xD_Q(x,M_f^2)
\COMMA
\end{equation}
where $Q=c,b$.
The values of $\langle x\rangle_Q(M_f)$ for
$Q=c,b$ evaluated with sets S and M in LO and NLO at $M_f=2m_Q,m_Z$
are collected in Table~\ref{xav}.
The differences between sets S and M and between LO and NLO are small.
The effect of the evolution, however, is significant, e.g.,
$\langle x\rangle_c(M_f)$ decreases from 
about 0.7 at the effective starting scale  to
about 0.5 at $m_Z$.
When these results are compared to the
experimental numbers reported by ALEPH \cite{ale1} and OPAL \cite{opa1},
\begin{eqnarray}
\langle x\rangle_c(m_Z)&\n=\n&0.495{+0.010\atop-0.011}\pm0.007\qquad
({\rm ALEPH}) 
\COMMA
\label{xavALEPH}\\
\langle x\rangle_c(m_Z)&\n=\n&0.515{+0.008\atop-0.005}\pm0.010\qquad
({\rm OPAL}) 
\COMMA
\label{xavOPAL}
\end{eqnarray}
we find good agreement for $\langle x \rangle_c(m_Z)$.
\smallskip

\begin{table}[hht!]
\begin{center}
\begin{tabular}{|c||c|c|c|c|}
\hline
set & $\langle x\rangle_c(2m_c)$ & $\langle x\rangle_c(m_Z)$ &
$\langle x\rangle_b(2m_b)$ & $\langle x\rangle_b(m_Z)$ \\
\hline 
\hline
LO S  & 0.689 & 0.518 & 0.364 & 0.320 \\
\hline 
NLO S & 0.716 & 0.528 & 0.351 & 0.309 \\
\hline
\hline
LO M  & 0.637 & 0.486 & 0.372 & 0.327 \\
\hline
NLO M & 0.664 & 0.495 & 0.359 & 0.315 \\
\hline
\end{tabular}
\caption{\small
Mean momentum fractions of $D^{*\pm}$ mesons produced in
charm and bottom quark fragmentation.
Eq.~(\protect\ref{dstav}) is evaluated at the
respective starting scales and at
$M_f=m_Z$ with sets S and M in both LO and NLO.
\protect\label{xav}}
\end{center}
\end{table}

\begin{figure}[hht!]
\begin{center}
\epsfig{file=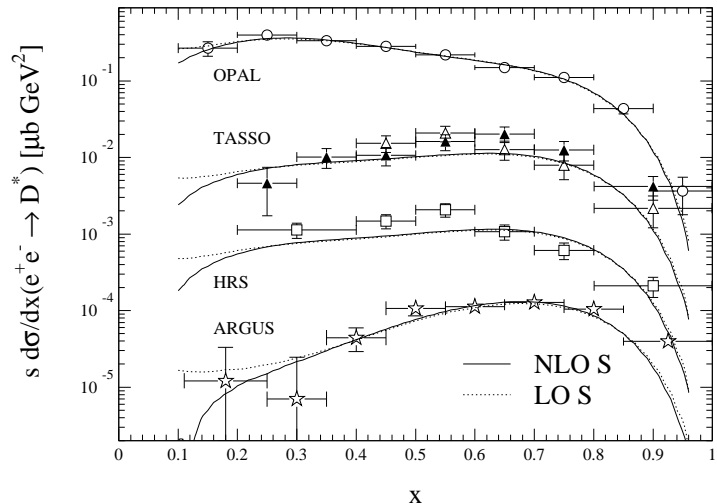 , clip=,width=9.6cm }
\end{center}
\vspace{-0.7cm}
\caption[]{\small
Data for \dstar production in \ee annihilation at a number
of CM energies are compared to the results obtained with the 
standard set of fragmentation functions.
Solid (dotted) curves show the NLO S (LO S) results.
For separation, the data have been rescaled by powers of 10.
The topmost data (circles) are from OPAL \protect\cite{opa1}.
The point at $x=0.05$ is omitted and these data are displayed in the
form $1/\sigma_{\rm had}d\sigma/dx$.
Going down on the plot and in energy, the other data are 
TASSO~1 (open triangles) and TASSO~2 (full triangles) \protect\cite{tas1}
at 34.2~GeV, HRS \protect\cite{hrs1} (squares) at 29~GeV and
ARGUS \protect\cite{arg1} (asterisk) at 10.6~GeV.
\protect\label{dstar}}
\end{figure}

As a final check we compare our FF's to \ee data at lower CM energies.
For the comparison, we selected data from ARGUS \cite{arg1} at
$\sqrt s=10.6$~GeV, from HRS \cite{hrs1} at $\sqrt s=29$~GeV, and from
TASSO \cite{tas1} at $\sqrt s=34.2$~GeV. 
The TASSO collaboration measured two decay channels of the $D^0$ meson, 
namely, $D^0\rightarrow K^-\pi^+\pi^+\pi^-$ (TASSO~1) 
and $D^0 \rightarrow K^-\pi^+$ (TASSO~2).
Fig.~\ref{dstar} shows the ARGUS, HRS and TASSO data on the cross 
section $Q^2d\sigma/dx$ of $e^+ + e^-\to D^{*\pm}+X$, together with
the respective LO and NLO predictions 
based on set S\footnote{For consistency, 
   we corrected the HRS and TASSO data by updating the values of
   relevant branching ratios to the new values from the 1994 tables
   of the Particle Data Group \cite{pdg}, which were also used
   by OPAL \cite{opa1}.
   The ALEPH data used in the fit have also been corrected accordingly.}. 
The theoretical results are calculated with
$N_F=5$ quark flavors (except for the case of ARGUS, where $N_F=4$ is used)
and also the contributions due to gluon and light--flavor fragmentation are
consistently included as in Figs.~\ref{dstaraleph} and \ref{dstaropal}.
For completeness, the OPAL 
data \cite{opa1} on the dimensionless cross section
 $(1/\sigma_{\rm had})d\sigma/dx$ are also included in the plot.
The agreement between the QCD predictions 
and the data is quite satisfactory.
This ensures
that the data indeed exhibit the scaling violation
predicted on the basis of the LO and NLO evolution equations for our FF's.
In fact, the change in the shape of the differential cross section 
with $Q^2$ is mainly due to the bottom quark.
The ARGUS data \cite{arg1} are taken off the resonance, 
at $\sqrt s=10.6$~GeV,
where the bottom quark is not yet active.
The inclusion of the bottom quark contribution leads to a softening of the
distribution, as may be seen in the case of the HRS data \cite{hrs1} at
$\sqrt s=29$~GeV and the TASSO data \cite{tas1} at $\sqrt s=34.2$~GeV.
The evolution from 29~GeV to 34.2~GeV has no discernible effect.
Going to the $Z$ pole at $\sqrt s=91.2$~GeV further increases the
relative importance of the bottom quark fragmentation and leads to an even
softer spectrum.
Except for the problematic region of small $x$, the NLO and LO predictions
are very close to each other.
\smallskip

The successful comparisons in Fig.~\ref{dstar} 
make us confident that our
$D^{*\pm}$ FF's, although constructed at $\sqrt{Q^2}=m_Z$, 
also lead to useful 
descriptions of $D^{*\pm}$ fragmentation at other scales.
In the subsection~5.3, we shall exploit this property together with the
universality of fragmentation to make predictions for inclusive $D^{*\pm}$
photoproduction at HERA.
\smallskip

Recently, Martin {\it et al.} have presented an
alternative method to account for massive quarks 
in parton density functions by modifying the
evolution kernels as well as the coefficient functions \cite{mar3}.
Their approach is well defined also in the problematic threshold
regions and is therefore theoretically very appealing.
An analogous treatment could be applied to the timelike regime
and could represent a considerable improvement upon the
analysis presented here.

\vfill
\clearpage
\section{Applications}
\setcounter{equation}{0}
\bigskip

This section is dedicated to the application of the fragmentation 
functions that have been extracted from \ee data to a variety of 
processes in order to test QCD.
First, universality will be tested in a range of processes
from $ep$ scattering over
$p\overline{p}$ scattering to $\gamma \gamma$ collisions.
Then, we will investigate scaling violation,
not only demonstrating its presence but actually
determining $\Lambda_{\rm  QCD}$
in the $\pi^{\pm}/K^{\pm}$ fit.
We will also discuss a consistent method to extract $\alpha_s$ from
$p_T$ spectra of inclusive particle production in DIS
and $p\overline{p}$ collisions.
Finally, in section~5.3, we will look into the
structure of the photon.
Both its gluon and its charm content can be constrained with
rapidity spectra for inclusive hadron photoproduction at HERA.

\vspace{2cm}
\subsection{Testing Universality}
\smallskip

As a consequence of 
the factorization theorem, the FF's extracted from \ee data
in the previous section can be applied to predict IPP in any
process.
In this section, we will demonstrate this for $ep$ collisions,
 $p\overline{p}$ collisions and $\gamma\gamma$ scattering.
The photoproduction process will be discussed in some
detail as it will also be used to study the PDF's of the photon in
subsection~5.3.
The discussion of $p\overline{p}$ and $\gamma \gamma$ collisions
will be limited to the comparison of data with our predictions.
\smallskip

In electron--proton collisions one usually considers two complementary 
scenarios:
The limites of small virtuality of the photon -- photoproduction --
and of large virtuality -- deep inelastic scattering.
The former contributes the bulk of the $ep$ cross section
so that the experimental measurements benefit from high statistics.
DIS, on the other hand, has the merit to probe the proton on the
smallest scale in position space, giving the most detailed information
about its partonic content.
The intermediate region has not received much attention in the past
but is presently being investigated in a number of new analyses
\cite{sch3,poe2}.
For DIS the situation is somewhat similar insofar as $p_T$
spectra of inclusive particle production have not 
been analyzed to NLO yet, 
this is being changed at the moment \cite{bue2}.
A NLO comparison of the $x_F$ spectra with data from 
H1 \cite{h12a} has been carried out successfully by
D.~Graudenz \cite{gra1} with the use of 
our set of charged particle FF's \cite{binpk}.
\smallskip

\begin{figure}[hht!]
\begin{center}
\epsfig{file=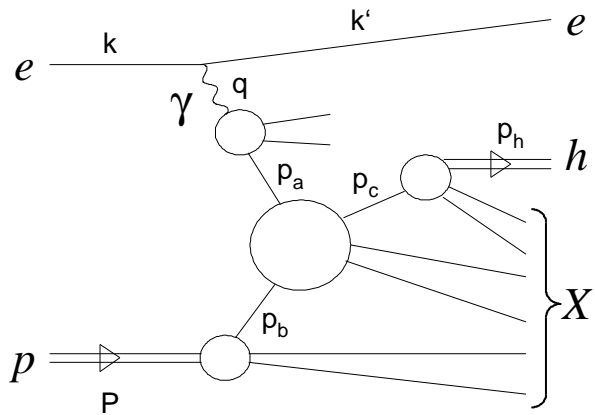 , clip=,width=12cm }
\end{center}
\vspace{-0cm}
\caption[]{\small 
Schematic diagram for IPP in $ep$ scattering. 
\protect\label{photo}}
\end{figure}

The $p_T$ spectra for charged particles have been
measured by the H1 \cite{h12} and the ZEUS \cite{zeus1}
collaborations.
Very recently, the IPP process has been measured in DIS for
the $K^0$ and $D^*$ final states, also \cite{h15,zeus4}.
In this work, however, the DIS process is disregarded in favor of the
photoproduction process,
where excellent
data on inclusive particle production have been 
available for some years.
\smallskip

Before comparing the NLO results on $h^{\pm}$ production
with the data, the relevant kinematics are defined.
The process can be depicted by the schematic diagram in Fig.~\ref{photo},
where the virtuality is $Q^2 \equiv -q^2 \approx 0$ 
and the kinematics of the
final state hadron are determined by its 
transverse momentum $p_T$ and its rapidity $y$.
The CM rapidity, $y_{\rm CM}$, is related to $y_{\rm lab}$ by
\begin{equation}
y_{\rm CM}=y_{\rm lab}-{1\over2}\ln\left(\frac{E_p}{E_e}\right)
\STOP
\end{equation}
In the massless approach, the rapidity is equal to the
pseudorapidity $\eta \equiv -\ln(\tan(\theta/2))$.
As usual, the photon spectrum of the electron/positron is approximated
by the Weiz\-s\"acker--Williams formula \cite{wwa}:
\begin{equation}
f_{\gamma / e} (z)=\frac{\alpha}{2\pi}\left[
  \frac{1+(1-z)^2}{z}\ln \left(\frac{Q_{max}^2}{Q_{min}^2}\right)
 + 2m_e^2z
\left( \frac{1}{Q_{max}^2}- \frac{1}{Q_{min}^2} \right) \right]
\COMMA
\end{equation}
where $z = E_{\gamma}/E_e$.
 $Q_{min}^2 = m_e^2 z^2 / (1-z)$ and
 $Q_{max}^2 = 0.01$~GeV$^2$ ($0.02$~GeV$^2$) for H1 (ZEUS) tagged events
and $Q_{max}^2 = 4$~GeV$^2$ for untagged events (both H1 and ZEUS).
\smallskip

It is well known that photoproduction proceeds via two
distinct mechanisms.
The photon can interact either directly with the partons
originating from the proton (direct photoproduction) 
or via its quark
and gluon content (resolved photoproduction).
The cross section for the resolved process is given by
\begin{eqnarray}
\label{epres}
E_h{d^3\sigma(\gamma p\to h+X)\over d^3p_h} &\n=\n&
\sum_{a,b,c}\int dx_\gamma dx_p{dx_h\over x_h^2}\,
G_a^\gamma(x_\gamma,M_\gamma^2)G_b^p(x_p,M_p^2)D_c^h(x_h,M_f^2)
\nonumber\\
&& \hspace{1cm}
\times
{1\over\pi\hat s}\left[{1\over v}\,{d\sigma_{ab\to c}^0\over dv}
(\hat s,v;\mu_R^2)\delta(1-w)
\right.
\nonumber\\
&& \hspace{1cm}
\left.
+{\alpha_s(\mu_R^2)\over2\pi}K_{ab\to c}
(\hat s,v,w;\mu_R^2,M_\gamma^2,M_p^2,M_f^2)
\right]
\COMMA
\end{eqnarray}
where $d\sigma_{ab\to c}^0/dv$ are the LO hard-scattering cross sections,
and where $v=1+\hat t/\hat s$ and  $w=-\hat u/(\hat s+\hat t)$
are functions of the Mandelstam variables at the parton level.
Those are defined as $\hat s=(p_a+p_b)^2$, $\hat t=(p_a-p_c)^2$, 
and $\hat u=(p_b-p_c)^2$.
The parton momenta are related to the photon, proton, and hadron momenta by
$p_a=x_\gamma q$, $p_b=x_p p_p$, and $p_c=p_h/x_h$.
The indices $a,b,$ and $c$ run over the gluon 
and $N_F$ flavors of quarks and antiquarks.
The $K_{ab\to c}$ functions may be found in Ref.~\cite{ave1} 
for $M_\gamma=M_p$.
The $G_a^{\gamma}$ and $G_b^p$ are the PDF's of the photon and the
proton, respectively.
\smallskip

The NLO cross section of direct photoproduction 
emerges from eq.~(\ref{epres})
by substituting $G_a^\gamma(x_\gamma,M_\gamma^2)=\delta(1-x_\gamma)$,
replacing $d\sigma_{ab\to c}/dv$ and $K_{ab\to c}$ with
$d\sigma_{\gamma b\to c}/dv$ and $K_{\gamma b\to c}$, respectively,
and omitting the sum over $a$.
Analytic expressions for the $K_{\gamma b\to c}$ functions are listed in
Ref.~\cite{gor1}.
\smallskip

When comparing to data, (\ref{epres}) is integrated either over
rapidity or over the transverse momentum $p_T$, leading 
to singly differential $p_T$ or
rapidity spectra, respectively.
The integration boundaries are fixed by the experimental conditions,
details are given in our respective publications. 
Except where stated otherwise, we employ the photon PDF's of 
GRV \cite{glu4}. 
In our recent work on $D^{*\pm}$ production we 
used the set CTEQ4M \cite{cteq1} for the proton PDF's, 
whereas we used CTEQ3M \cite{cteq2} in our earlier analyses.
\smallskip

Very precise data on IPP in photoproduction
have been measured by the two HERA collaborations, ZEUS and H1.
The first data to be published have of course been those for
indiscriminate charged hadron production, as they have the best
statistics \cite{h14,zeus3,hop1}.
Since then, more exclusive analyses have followed suit;
for neutral kaons in \cite{h16} and for $D^{*\pm}$ in
\cite{h13,zeus2}.
\smallskip

\begin{figure}[hht!]
\begin{center}
\epsfig{file=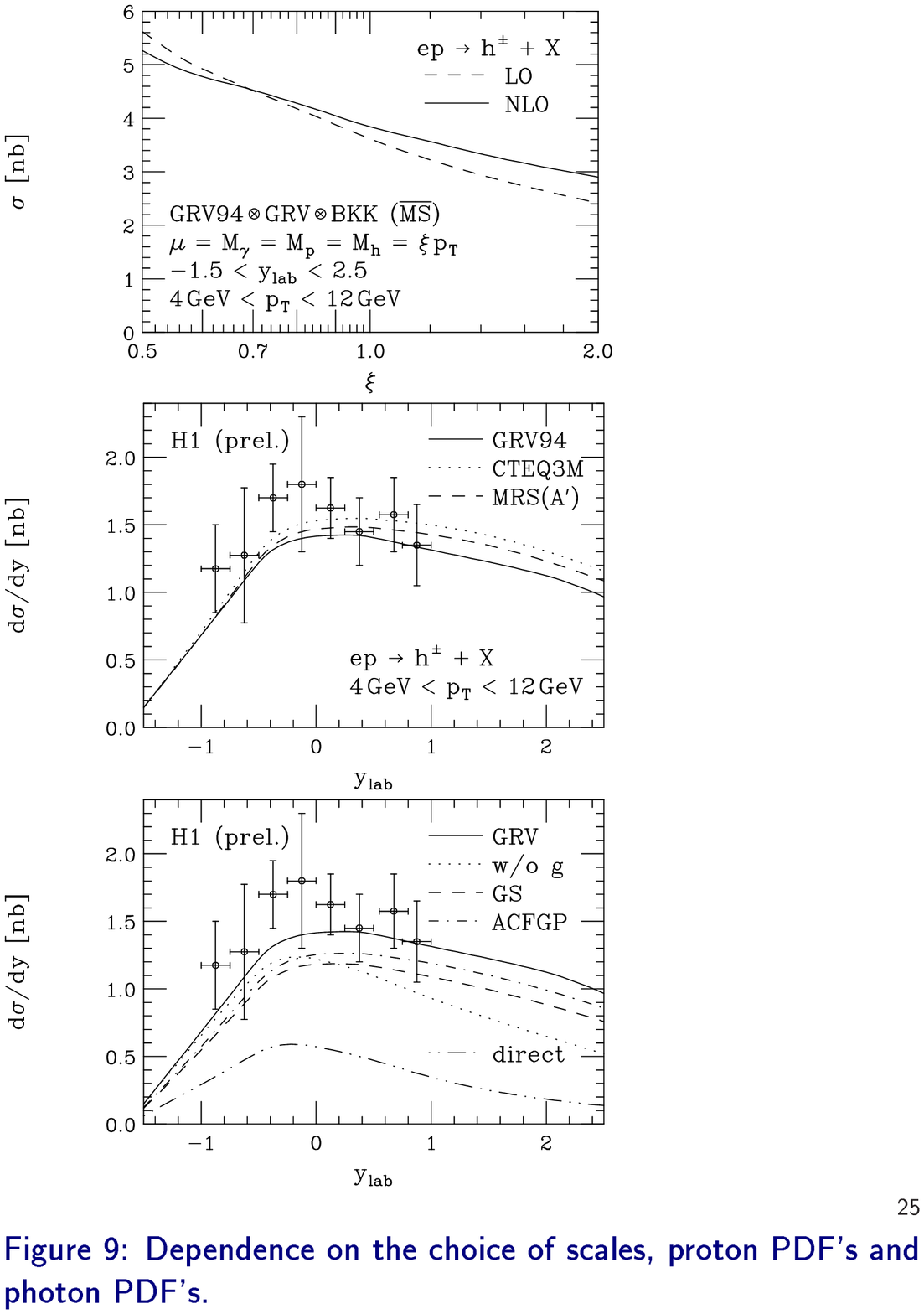,clip=,width=11.5cm }
\end{center}
\vspace{-0.5cm}
\caption[]{\small 
Scale dependence of the photoproduction cross section
\protect\cite{kni4}.
In departure from the usual choice, the proton PDF's
of GRV \cite{glu3} are employed in this plot.
We use the charged hadron FF's of \protect\cite{binpk}.
\protect\label{epscales}}
\end{figure}

\begin{figure}[hht!]
\begin{center}
\epsfig{file=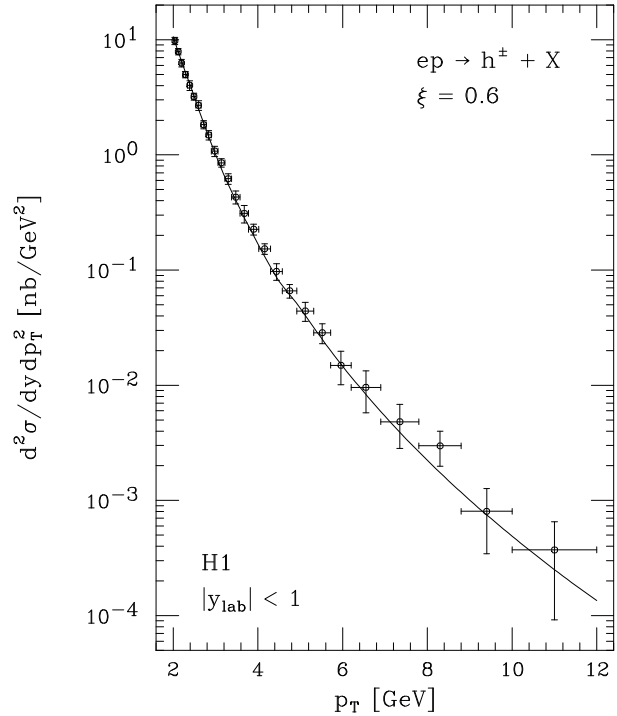,clip=,width=8cm}
\end{center}
\vspace{-0.3cm}
\caption[]{\small 
Photoproduction for $\xi=0.6$ compared to preliminary H1 
data \protect\cite{hop1}.
The curve shows the NLO result for $\xi=0.6$ \protect\cite{kni5}.
The PDF's and FF's are the same as in Fig.~\protect\ref{epscales}.
\protect\label{scal2}}
\end{figure}

The best data on charged hadron production to date are
preliminary ones from H1 \cite{hop1}.
The theoretical predictions possess an inherent ambiguity
due to the unphysical factorization and renormalization scales.
Before comparing to the high precision data, we will
try to reduce this ambiguity by additional conceptual input.
The cross section, when calculated
to all orders, must not depend on the unphysical scales.
The ambiguity introduced by the scales in this sense reflects
our ignorance about unknown higher order terms in the
perturbative expansion.
With the prejudice that higher orders in the expansion should
give only small corrections relative to the NLO, and disregarding
for the moment that the expansion in $\alpha_s$ represents only
an asymptotic expansion, one arrives at the conjecture
that the unknown corrections vanish for those scales where
already the first nonleading corrections (NLO) disappear.
This is known as principle of fastest apparent convergence.
As a first approximation all the scales are identified and
parametrized by
\begin{equation}
\label{xidef}
\mu_R=M_f=M_{\gamma}=M_p=\xi p_T \STOP
\end{equation} 
The cross section integrated over $p_T>4$~GeV 
and $-1.5<y_{\rm lab}<2.5$ is plotted as a function
of the single parameter $\xi$ in Fig.~\ref{epscales}.
The scale dependence is markedly reduced 
from LO (dashed) to NLO (solid).
Following the principle of fastest apparent convergence, we choose
 $\xi=0.6$ for the scales in the comparison with the
 $p_T$ spectrum.
Our curve is plotted together with the H1 preliminary data
in Fig.~\ref{scal2}.
The agreement with the data turns out to be perfect.
This will be exploited further to constrain the
gluon density of the photon in section~5.3.
However,
we want to point out that the principle of
fastest apparent convergence is just one of several
methods for choosing a scale.
It does not
necessarily give the best choice of scales with respect
to the convergence behavior of unknown higher orders.
\smallskip

\begin{figure}[ht!]
\begin{center}
\epsfig{file=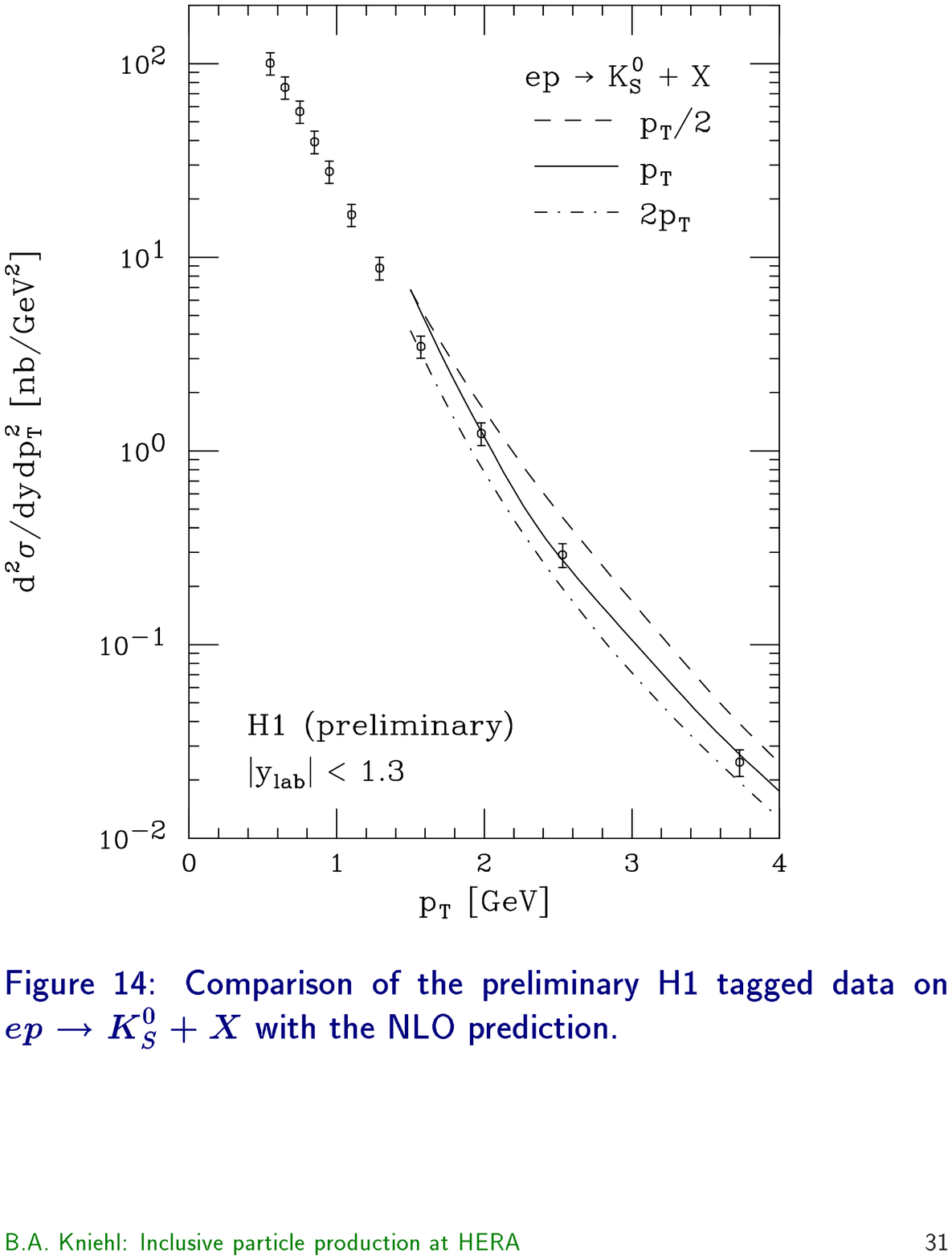,clip=,width=8.cm }
\end{center}
\vspace{-0.5cm}
\caption[]{\small 
Comparison of the preliminary H1 tagged data on
neutral Kaon photoproduction with the NLO prediction
\protect\cite{kni4}.
Solid, dashed, and dash--dotted lines correspond
to the NLO results with $\xi=1/2,1$, and 2, respectively.
\protect\label{epk0}}
\end{figure}

Our NLO predictions for neutral Kaon production in $\gamma p$
collisions are compared to data from H1
\cite{h16} (no longer preliminary) in Fig.~\ref{epk0}.
The solid line gives the result for $\xi=1$ whereas the
dashed (dot--dashed) curves correspond to the choices
 $\xi=1/2$ ($\xi=2$), respectively.
It turns out that the central value agrees best with the data.
Within the theoretical uncertainty simulated by the
three choices of scales, good agreement is found down to
very low scales of $p_T=1.5$~GeV where we would not expect
our results to be reliable.
\smallskip

\begin{figure}[hht!]
\begin{center}
\begin{picture}(12.5,7.5)
\put(-1.2,-.3){\makebox{
\epsfig{file=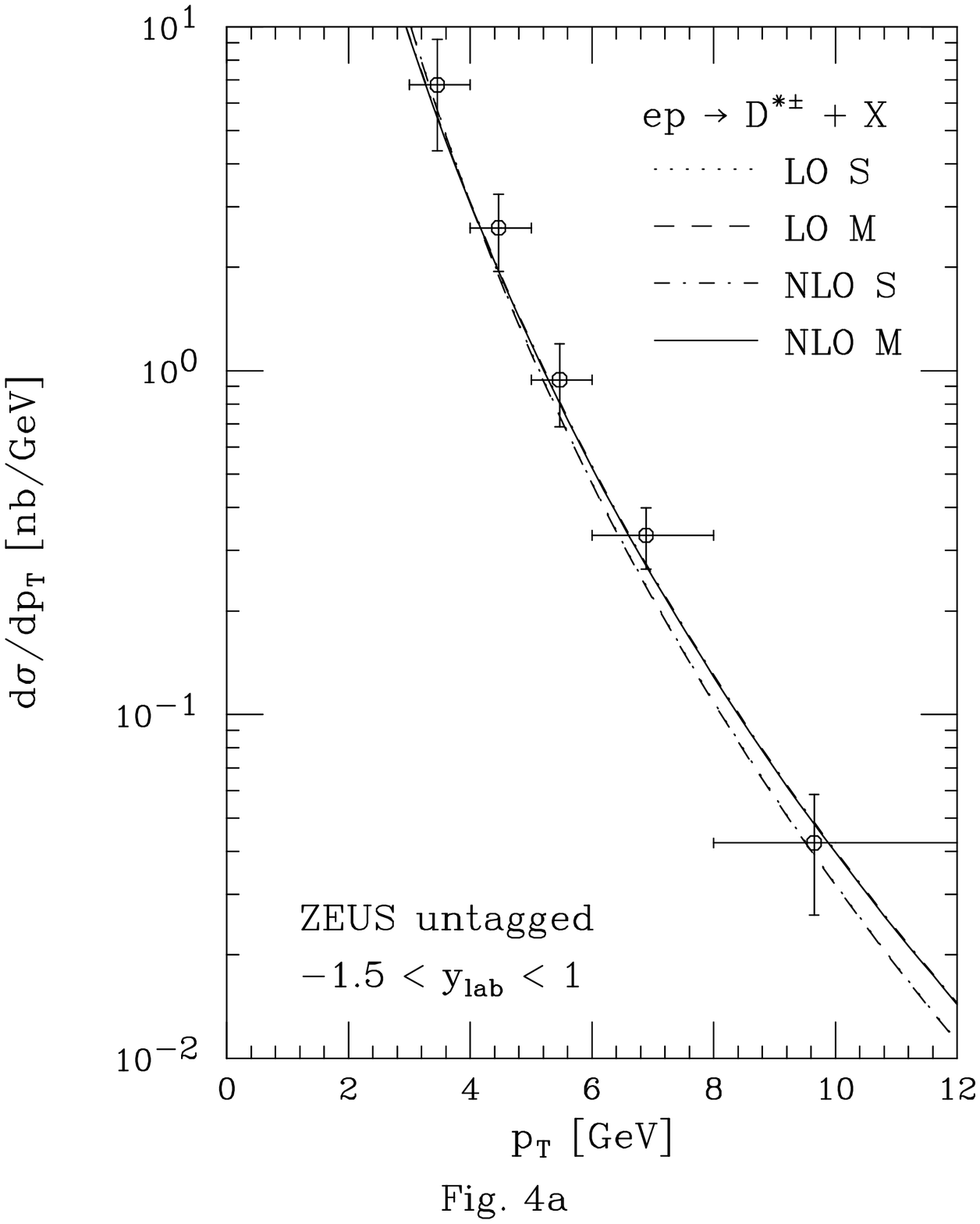,clip=,width=6.6cm}}}
\put(6.4,-.3){\makebox{
\epsfig{file=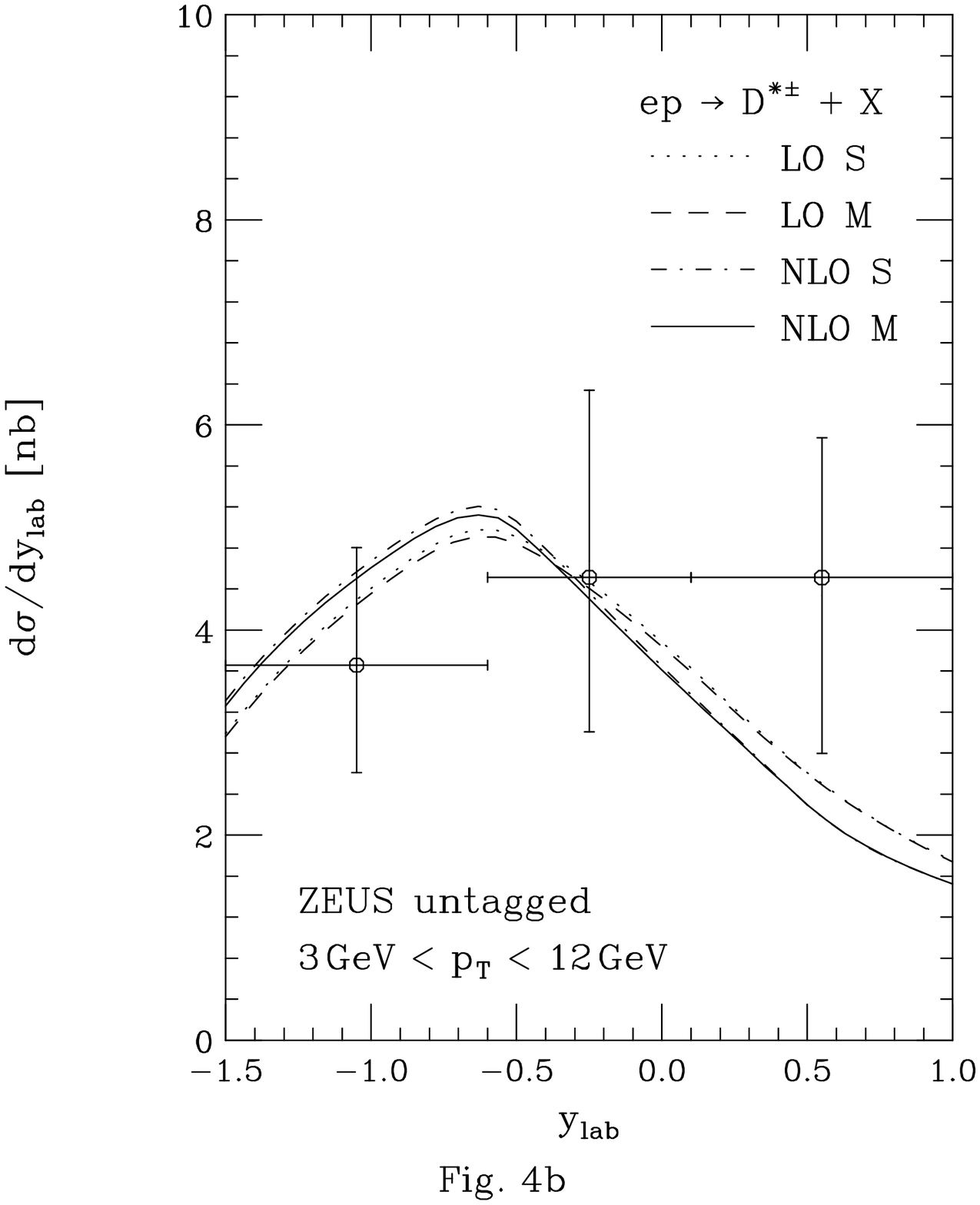,clip=,width=6.6cm}}}
\put(-0.9,0.0){a)}
\put(6.8,0.0){b)}
\end{picture}
\end{center}
\vspace{-0.cm}
\caption[]{\small 
The LO and NLO predictions of inclusive $D^{*\pm}$ photoproduction
in $ep$ collisions based on sets S and M are compared with the ZEUS data
\protect\cite{zeus2}.
We consider 
a) 
the $p_T$ distribution $d\sigma/dp_T$ integrated over
$-1.5<y_{\rm lab}<1$ and 115~GeV${}<W<280$~GeV and
b) 
the $y_{\rm lab}$ distribution $d\sigma/dy_{\rm lab}$ integrated over
3~GeV${}<p_T<12$~GeV and 115~GeV${}<W<280$~GeV.
\protect\label{ep_dstar}}
\end{figure}

The results of our recent work on $D^{*\pm}$ mesons
\cite{binds} are shown
in Fig.~\ref{ep_dstar}a, b.
They are in good agreement with the $p_T$ spectrum measured by
ZEUS \cite{zeus2} down to $p_T=3$~GeV.
Below this scale, the theoretical prediction is zero
as the FF's start their evolution at $2m_c$, effectively.
Integration over $p_T$ yields the $\eta$ spectrum of Fig.~\ref{ep_dstar}b.
Within the large errors, we again agree with the data.
We want to point out that the results of both sets, S and M, are
displayed in Fig.~\ref{ep_dstar}a and b.
However, the LO curves of sets S (dotted) and M (dashed) almost coincide
with each other, as do the NLO curves (dash--dotted and solid, respectively).
In fact, they are absolutely indistinguishable in \ref{ep_dstar}a. 
This serves to prove that our results do not depend on the
functional form of the start parametrizations.
\smallskip

\begin{figure}[hht!]
\begin{center}
\begin{picture}(12.5,7.5)
\put(-1.2,-.3){\makebox{
\epsfig{file=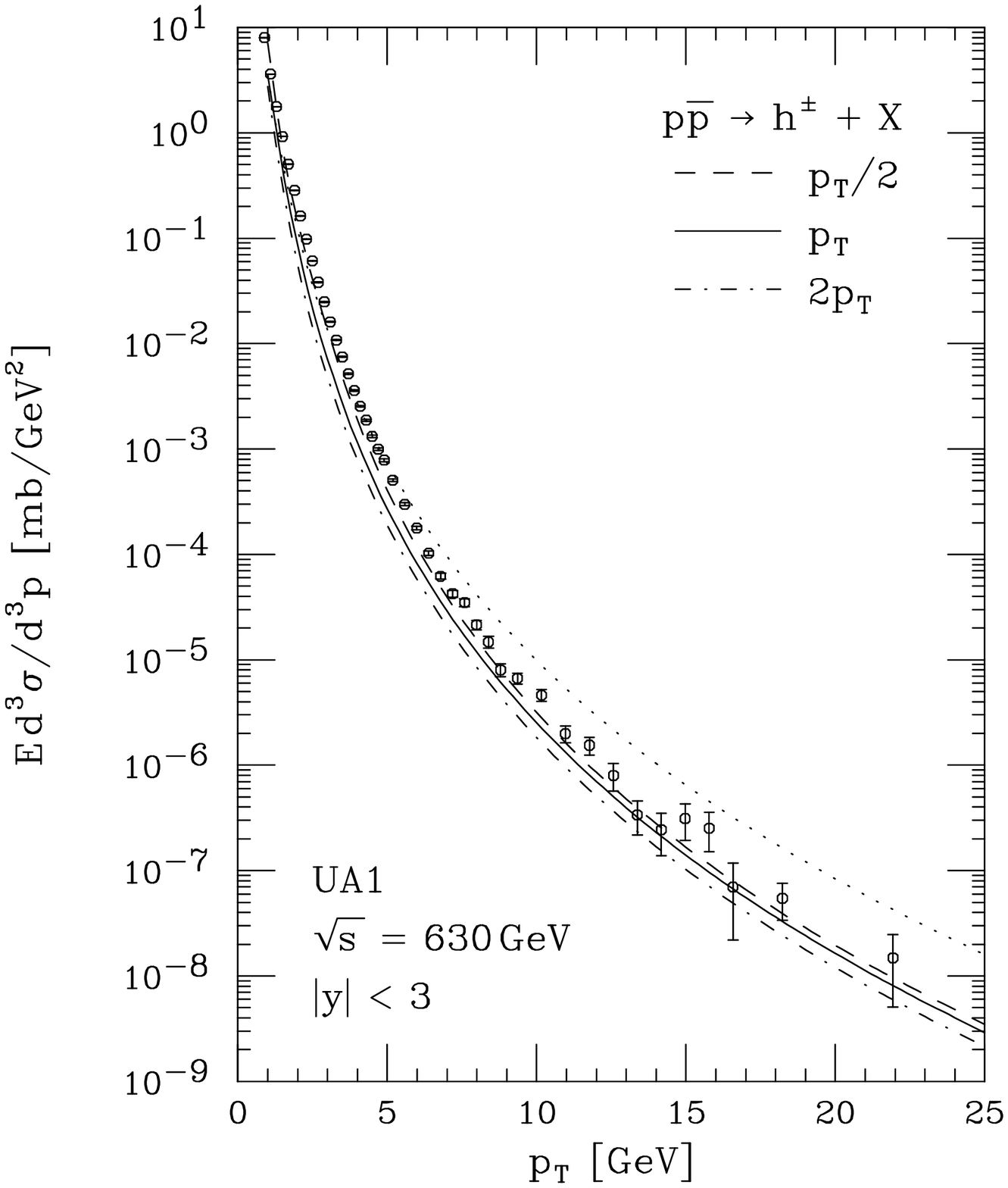,clip=,width=6.6cm}}}
\put(6.4,-.3){\makebox{
\epsfig{file=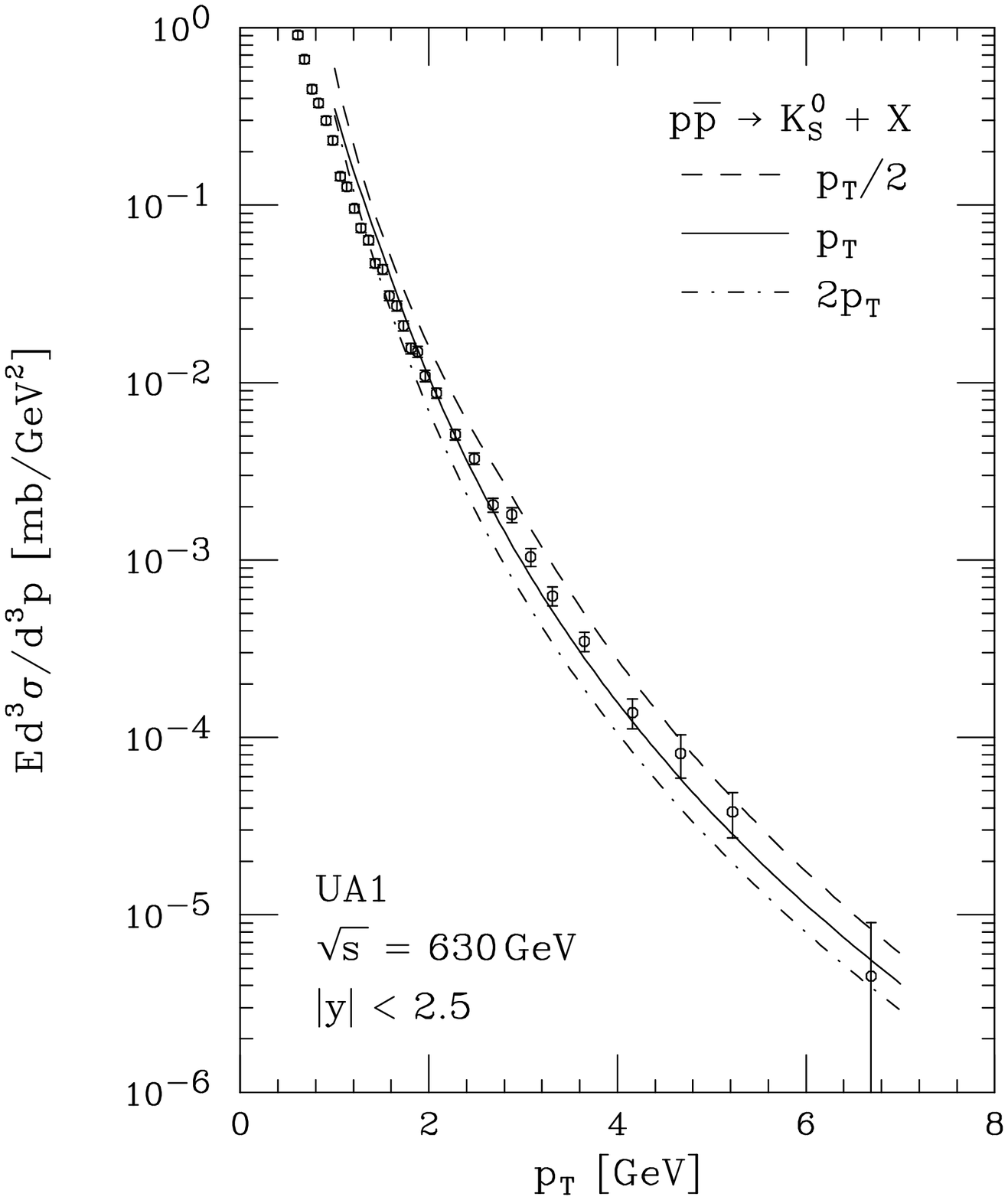,clip=,width=6.6cm}}}
\put(-0.9,-0.2){a)}
\put(6.8,-0.2){b)}
\end{picture}
\end{center}
\vspace{-0.cm}
\caption[]{\small 
Inclusive particle production in
$p \overline{p}$ scattering.
Our NLO QCD results \protect\cite{kni4,kni5}
are compared to data from UA1--MIMI \protect\cite{ua11}
at 630~GeV.
The dashed, solid, and dash--dotted curves correspond to 
scale choices of $\xi=1/2$, 1, and 2, respectively.
a)
Charged hadron production, calculated with our 
set \protect\cite{binpk}.
b)
Neutral kaon production with the FF's of \protect\cite{bink0}.
\protect\label{pphc}}
\end{figure}

We will now turn to $p\overline{p}$ scattering.
It is of particular interest to apply
our FF's to hadronic interactions.
As demonstrated in section~3.4, most of the $p\overline{p}$
cross section comes from the fragmentation of the gluon.
This process may thus be used to test the gluon FF
which is only weakly constrained by our older fits to \ee data.
\smallskip

The IPP cross section has been calculated to NLO in \cite{bor1}
for the first time.
With the use of our older set \cite{bin1}, good agreement
was found with the data from UA1--MIMI,CDF, and 
CDHW \cite{ua11,pp1}.
The NLO predictions that are tested against data here, have been
calculated in \cite{kni5} for the charged particle spectra
and in \cite{bink0} for the neutral kaon spectra.
Fig.~\ref{pphc}a shows inclusive charged particle production
in $p\overline{p}$ collisions.
The NLO results with our set \cite{binpk} are
compared to data from UA1--MIMI \cite{ua11}.
The dashed, solid, and dash--dotted curves
correspond to scales $\xi=1/2,1$, and 2, respectively,
where $\xi$ is defined in (\ref{xidef}).
The lower choice of scales gives satisfactory
agreement with the data over the whole $p_T$ range.
The dotted curve will be explained in the next subsection.
\smallskip

A similar comparison is presented for neutral kaon production
in Fig.~\ref{pphc}b.
The data are from UA1--MIMI \cite{ua11}, also.
Due to the more exclusive final state, the spectrum does not
extend as far as for the charged hadrons.
When comparing with the NLO result for our set of
neutral kaon FF's \cite{bink0} we find again good agreement,
this time for the central choice of scales.
Our results are slightly high
only for $p_T$ below 1.5~GeV, where we do not expect
our formalism to be valid.
\smallskip

\begin{figure}[hht!]
\begin{center}
\epsfig{file=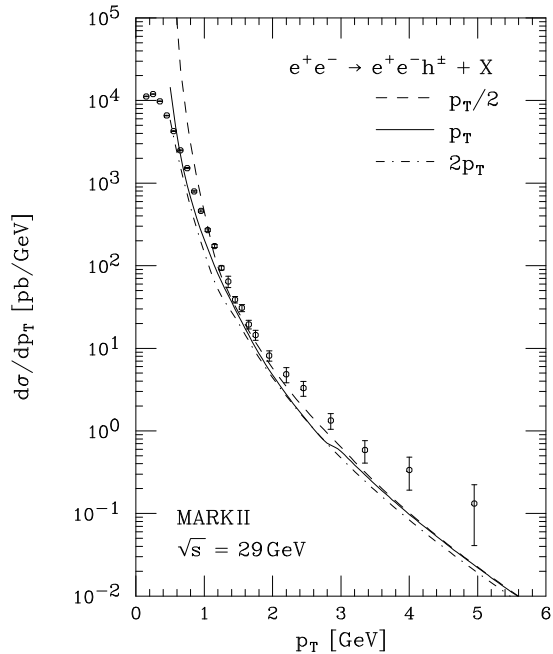, clip=,width=7.2cm}
\end{center}
\vspace{-0.5cm}
\caption[]{\small 
Inclusive 
charged--hadron production in single--tagged $\gamma\gamma$ collisions.
The MARK~II data \protect\cite{mar6} ($\protect\sqrt{S}=29$~GeV)
are compared to the NLO calculations 
of $d\sigma/dp_T$ for scale choices $\xi=1/2$, 1, and 2
\protect\cite{bingg}.
\protect\label{gghcexp}}
\end{figure}

The last process that will be, briefly, discussed
is charged hadron production in $\gamma \gamma$ collisions.
So far, there exist only few data.
However, with the prospect of future measurements at LEP2 and 
at the NLC, this situation is bound to improve.
Here we present the results of \cite{bingg} that draw on the
calculations presented in \cite{gor1}.
The process has been measured by the TASSO collaboration
at PETRA \cite{tas4} at a CM energy of 33.1~GeV,
and by MARK~II at PEP \cite{mar6} at a CM energy of 29~GeV.
We compare the MARK~II data with our results in Fig.~\ref{gghcexp}.
As usual, three curves for different choices of scales
are plotted.
The QCD result is too low, especially at high $p_T$.
The same is observed for the less precise TASSO data (not shown).
This feature has also been observed by other groups, e.g. \cite{gor1},
but is not understood, yet.
\smallskip

\begin{figure}[hht!]
\begin{center}
\begin{picture}(12.5,7.5)
\put(-1.2,-.3){\makebox{
\epsfig{file=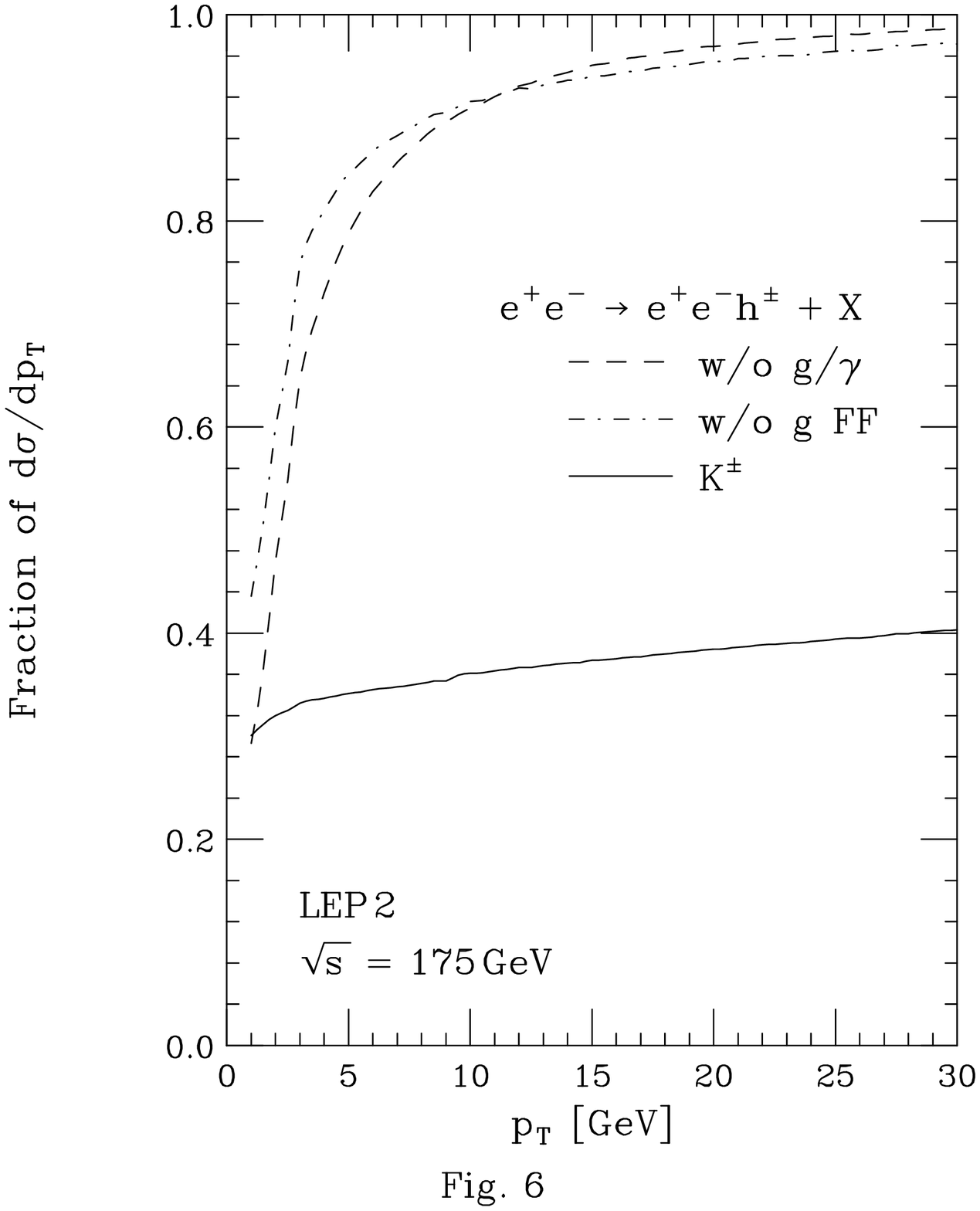,clip=,width=6.6cm}}}
\put(6.4,-.3){\makebox{
\epsfig{file=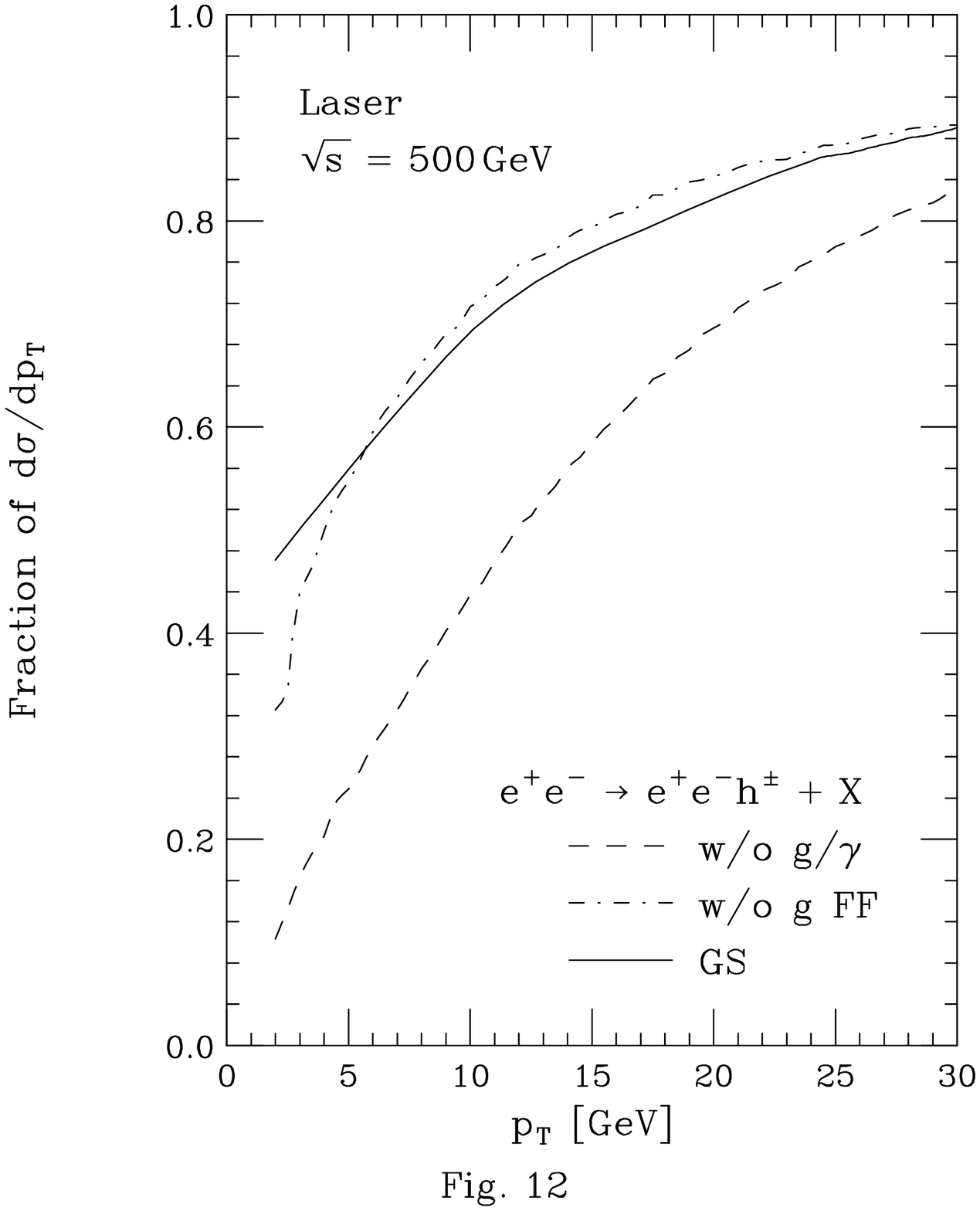,clip=,width=6.6cm}}}
\put(-0.9,0.0){a)}
\put(6.8,0.0){b)}
\end{picture}
\end{center}
\vspace{-0cm}
\caption[]{\small 
Inclusive 
charged--hadron production in single--tagged $\gamma\gamma$ collisions.
a)
The influence of the gluon PDF of the photon,
the gluon FF, and the charged-kaon final states.
Shown are the fractions that remain 
if the gluon is switched off in the photon
PDF's (dashed line) or in the FF's 
(dot--dashed line) as well as the fraction
due to charged-kaon production (solid line).
b)
Influence of the gluon PDF of the photon and
the gluon FF on the total result for the laser spectrum.
Shown are the fractions that remain 
if the gluon is switched off in the photon
PDF's (dashed line) or in the FF's 
(dot--dashed line) as well as the ratio of
the calculation with the GS photon PDF's 
to that with the GRV set (solid line).
\protect\label{gghc}}
\end{figure}

We now move on to make predictions for the future
colliders LEP2 and NLC.
Of the various experimental scenarios for the latter
we will only consider the so--called LASER mode 
where laser light is Compton--backscattered off the $e^+$
and $e^-$ beams.
In Fig.~\ref{gghc} we display the relative importance of the
various contributions to the spectra of
single tagged charged particle production at LEP2 and at the
NLC, respectively.
The curves give the NLO predictions with our set \cite{binpk}
at central scale $\xi=1$.
For LEP2, the direct--direct contribution is dominant for $p_T>3$~GeV
\cite{bingg}.
As a consequence, the impact of the gluon, both 
in the PDF's and in the FF's, on the cross section is small as seen 
in Fig.~\ref{gghc}a.
Switching off either gives the dashed, respectively dot--dashed
lines that rapidly approach 1 with increasing $p_T$.
The fraction due to the kaon is also plotted (solid),
it increases slightly with $p_T$, to about 0.4 at $p_T=30$~GeV.
\smallskip

For the LASER mode of the NLC, on the other hand, the
resolved--resolved contribution dominates up to $p_T=20$~GeV.
This process proceeds mainly via gluon--gluon fusion.
Consequently, the gluon in the photon
is much more important for the NLC.
The dashed line in Fig.~\ref{gghc}b shows the fraction of the
total cross section that remains when the
gluon in the photon is switched off.
The solid line gives the ratio of the NLO results obtained
with the GS photon PDF's \cite{gs1} to those
obtained with the GRV set.
Both curves reveal a significant sensitivity to the photonic PDF's,
in particular of the gluon.
One may hope to exploit it to improve our
knowledge of the gluon density in the photon.
Unfortunately, this sensitivity is accompanied by a sensitivity of
comparable size to the gluon FF, which
constitutes a large part of the cross section (dash--dotted).
Therefore, the latter has to be known well, for a 
meaningful measurement of the gluon in the photon.

\vspace{2cm}
\subsection{Scaling Violation}
\smallskip

The high--precision data acquired at LEP1 are, in union with lower
energy data, an excellent starting point for the determination of
the strong coupling.
In fact, the $\alpha_s$ measurements obtained at LEP1 are among the 
best to date.
Of the numerous analytical methods employed, like event--shape variables
or three jet over two jet rates, we will deal
only with scaling violation of fragmentation functions.
A dedicated analysis along this line 
has been conducted by the ALEPH collaboration 
\cite{ale3}\footnote{A similar analysis has also been 
   conducted by the DELPHI collaboration \cite{del1}.} that 
yields $\alpha_s=0.126\pm 0.009$.
We will take a more qualitative view and restrict ourselves to
the extraction of $\Lambda_{\rm QCD}$ without a quantitative assessment
of the theoretical and experimental errors.
\smallskip

One common problem in the determination of the strong coupling
from IPP in \ee annihilation is fake scaling violation from the $b$ quark
contributions.
The data with the best statistics are from LEP1 so that one usually
uses those together with lower energy data in the $\alpha_s$ analyses.
At the $Z$~pole, the $b$ is strongly enhanced, whereas it is
negligible at lower energies.
Due to its mass, the spectrum of $b$ quark fragmentation is much
softer than that of the light quarks so that an enhancement 
softens the IPP spectrum, mimicking scaling violation.
Fortunately, this contribution to the change of shape is
under good control in our set \cite{binpk} because 
information on fragmentation of the $b$ quark was used in the
fit, as detailed in section~4.1.
\smallskip

We employed the scaling violation effect in
our fit to data from 29 and 91.2~GeV,
to determined $\Lambda_{\rm QCD}^{(5)}$ as one of 31 parameters.
The values obtained from the fit were 0.227 and 0.108 in NLO and LO,
respectively.
This translates into values of the strong coupling at the $Z$ pole of
 $\alpha_s(m_Z^2)= 0.118 (0.122)$ at NLO (LO).
The agreement of both the LO and NLO results with the
value of $0.120 \pm 0.008$, extracted from a global fit to
the observables measured at LEP1 \cite{heb1},
is striking.
\smallskip

In our new fit we chose an alternative approach.
Instead of aiming to determine $\alpha_s$ from \ee
annihilation data, we are interested in the DIS and
photoproduction data.
Compared to {\ee}, the $ep$ process has the
advantage that data at different scales are collected by the
same detector in one measurement.
With other words, the $p_T$ spectra are sensitive to
 $\Lambda_{\rm QCD}$ and can be used to measure $\alpha_s$.
This is demonstrated in Fig.~\ref{pphc}a for $p\overline{p}$ scattering.
The dotted curve results from the NLO calculation with $\xi=1$
when the evolution is switched off in the 
fragmentation functions.
This leads to a significantly harder spectrum and is clearly
inconsistent with the measurement.
Thus the shape of the curve will be affected by a 
variation of $\Lambda$.
In photoproduction, a change of 50~MeV in $\Lambda^{(5)}$
results in a 10\% change of the NLO cross section at high $p_T$.
\smallskip

For \ee annihilation, on the other hand, data at different
experiments have to be combined to measure the
scaling violation.
This introduces large relative normalization errors.
They impair the potential benefit especially of data
at low $Q^2$ which would give a large scaling violation
effect when combined with LEP1 data.
\smallskip

For a consistent fit of $\Lambda$ to DIS or photoproduction
data, $\Lambda$ has to be varied not only in the
coefficient functions, but also in the PDF's and
in the FF's.
PDF sets with variable $\Lambda$ have been presented for the
proton by A.~Vogt \cite{vog2}, some time ago.
No corresponding sets exist for the photon PDF's or for
the FF's.
The former are so poorly known that it does not make sense
at the moment to construct a variable $\Lambda$ set.
The latter are presented in our new $h^{\pm}$ fit,
for the first time.
In the same manner as in \cite{vog2}, 
we kept $\Lambda^{(4)}$ fixed at values of
150, 200, 250, 300, 350 and 400~MeV, respectively,
in our NLO analysis.
\smallskip

\begin{figure}[hht!]
\begin{center}
\epsfig{file=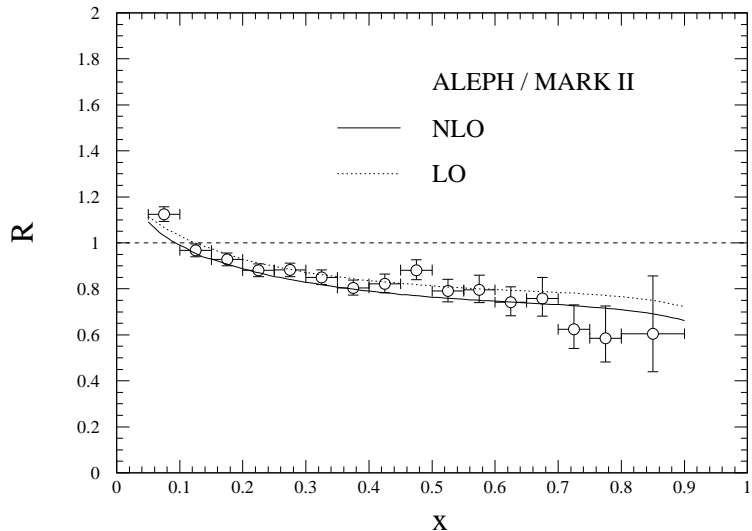,clip=,width=10.2cm}
\end{center}
\vspace{-0.9cm}
\caption[]{\small 
Scaling Violation in \ee annihilation.
The plot shows the ratio R of the differential cross sections for
inclusive charged hadron production at 91.2~GeV to that at 29~GeV.
Data from ALEPH \protect\cite{ale3} and MARK~II 
\protect\cite{mar5} are compared to
the predictions of the new $h^{\pm}$ fragmentation functions.
The solid (dotted) lines correspond to the NLO (LO) results.
\protect\label{scaling}}
\end{figure}

\begin{figure}[hht!]
\begin{center}
\epsfig{file=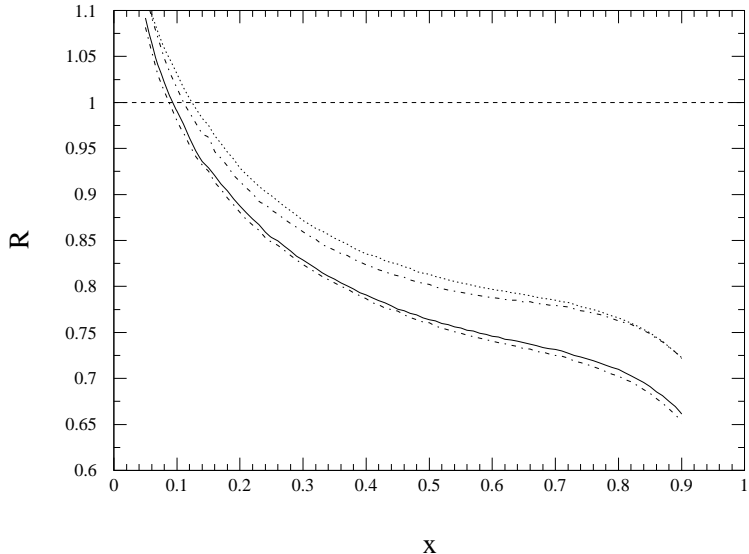,clip=,width=10.2cm}
\end{center}
\vspace{-0.9cm}
\caption[]{\small 
As Fig.~\ref{scaling}, but zoomed and without the data, for clarity.
In addition to the NLO (solid) and LO (dotted) results for our central
choice $\Lambda^{(4)}=350$~MeV,
the range covered by the NLO variable--$\Lambda$ sets is
also shown as dash--dotted curves.
The upper one corresponds to $\Lambda^{(4)}=150$~MeV
and the lower one to $\Lambda^{(4)}=400$~MeV.
\protect\label{scalingl}}
\end{figure}

After completion of the fit, we checked that
the results describe the scaling violation in \ee annihilation
correctly.
The comparison of our central set ($\Lambda^{(4)}=350$~MeV) to
data from ALEPH \cite{ale3} and MARK~II \cite{mar5} at
91.2 and 29~GeV, respectively,
is shown in Fig.~\ref{scaling}.
Both the NLO (solid) and LO (dotted) sets of FF's
lead to good agreement with the scaling violation measurements.
The same holds true for the sets with various $\Lambda$,
as demonstrated in Fig.~\ref{scalingl}.
The results for the individual sets exhibit a steady decrease 
for successively higher values of $\Lambda$,
also for the sets with $\Lambda^{(4)}=200,250,300$~MeV, 
which are not shown.
We conclude that the errors on the scaling violation data obtained
from the combination of MARK~II and ALEPH data are too large
to exclude any of our sets.

\vspace{2cm}
\subsection{The Gluon and Charm Content of the Photon}
\smallskip

Before the advent of HERA, information on the parton content of the
photon came almost exclusively from the measurement of $F_2^{\gamma}$ 
in single tagged $\gamma \gamma$ 
collisions\footnote{In this work we are concerned only with the
   real photon. For a comparison of real and virtual
   photon PDF's, see \cite{sjo1}.}.
Together with the assumption of vector meson dominance (VMD),
the quark PDF's of the photon have been reasonably well determined
from these data.
The gluon, on the other hand, does not contribute to $F_2^{\gamma}$
at LO and is hence poorly known.
Consequently, the quark distributions of the various
parametrizations, e.g. in \cite{glu4,gs1,acfgp},
are similar, whereas the gluon distribution depends 
crucially on the choice of PDF's.
\smallskip

As argued in subsection~5.1,
the IPP process at the NLC in the LASER mode may be used
to measure the gluon content of the photon more precisely. 
However, we will have to wait several years
for the first data from the NLC -- if it will be operated
in the LASER mode at all.
\smallskip

\begin{figure}[hht!]
\begin{center}
\epsfig{file=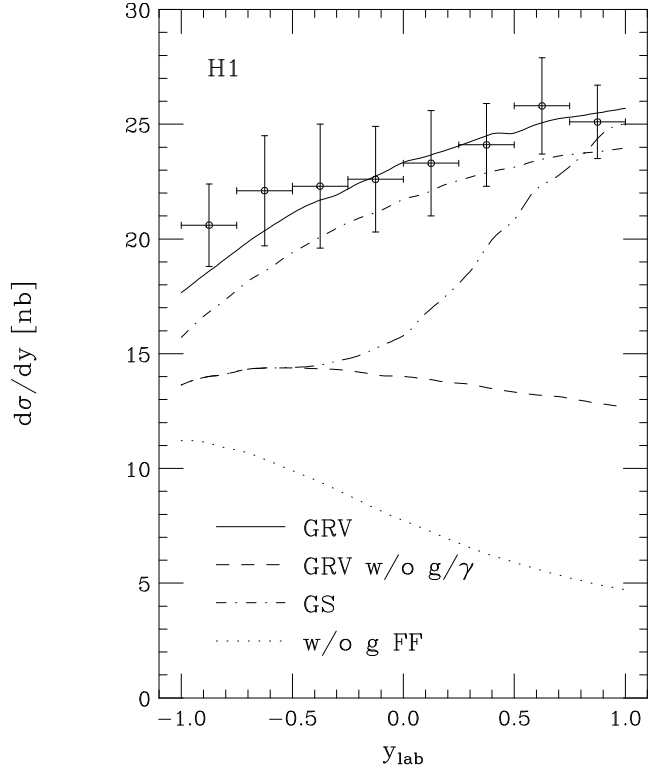,clip=,width=8.5cm }
\end{center}
\vspace{-0.3cm}
\caption[]{\small 
The impact of the gluon in the photon on $\eta$--spectra
of charged particles in photoproduction.
The NLO results \protect\cite{kni5} for $\xi=0.6$
with our set \protect\cite{binpk} 
of charged hadron FF's are compared to preliminary data
from H1 \cite{hop1}.
The cross section has been integrated over $p_T>3$ GeV.
The solid line gives the result with the GRV photon PDF's
\protect\cite{glu4}, whereas the dot--dashed line is
the result obtained with the GS photon PDF's \protect\cite{gs1}.
The dotted (dashed) line results from switching off
the gluon in the fragmentation (in the PDF's).
The dot--dash--dotted curve is discussed in the text. 
\protect\label{eta_charged}}
\end{figure}

In the meantime, HERA is the ideal place to learn more
about the gluon content of the photon.
In particular the rapidity spectra show a significant
sensitivity to the gluon PDF of the photon.
Fig.~\ref{eta_charged} shows the preliminary H1 data 
on inclusive charged hadron production in photoproduction,
integrated over $p_T>3$~GeV, as a function 
of the rapidity \cite{hop1}.
The prediction of our NLO analysis with set \cite{binpk}
of charged hadron FF's is plotted as solid line.
As in the comparison with the $p_T$ spectrum in section 5.1,
we use the GRV \cite{glu4} photon PDF's and the CTEQ3M \cite{cteq2}
proton PDF's and choose
the optimized scale $\xi=0.6$ in our calculation.
The agreement with the data is near perfect over the whole 
measured region of $-1 < y_{\rm lab}<1$.
When using the GS \cite{gs1} PDF's instead (dot--dashed)
the theoretical prediction decreases only slightly and is still 
consistent with the data.
However, both sets are constructed with similar prejudice concerning
the VMD input \cite{sjo1} and it would therefore be misleading to
take this small difference as a measure of the sensitivity
of the observable to the gluon density in the photon.
The sensitivity is revealed much more clearly by switching off
the gluon in the photon, which gives the dashed curve.
It is clearly inconsistent with the data in both normalization
and shape.
Moreover, the rapidity spectrum is sensitive to the gluon
density at both small and large $x_{\gamma}$, 
where $x_{\gamma}$ is the fraction of the photons momentum that
is carried by the gluon.
A toy distribution that is similar to that of GRV around 
 $x_{\gamma}=0.1$ but zero else results in the dot--dash--dotted
curve.
This curve approaches the GRV result in the forward region 
(positive $y_{\rm lab}$) and
is identical to the no gluon result in the backward region.
\smallskip

We conclude that the gluon distribution can be probed
by IPP in photoproduction from $x_{\gamma} \approx 0.05$
to $x_{\gamma}$ considerably above 0.2.
However, the rapidity spectrum is even more sensitive
to the gluon FF.
The result of turning off the gluon fragmentation (in the cross section)
is plotted as dotted curve.
Obviously, the precision of any future measurement of the gluon 
content of the photon in this process depends heavily
on the uncertainty of the gluon FF.
This holds in particular in the forward region where the
gluon fragmentation contributes up to 80\% of the cross section.
Our improved charged hadron set should be sufficiently precise
to permit an analysis along these lines.
\smallskip

\begin{figure}[hht!]
\begin{center}
\begin{picture}(12.5,8.2)
\put(-1.2,-.3){\makebox{
\epsfig{file=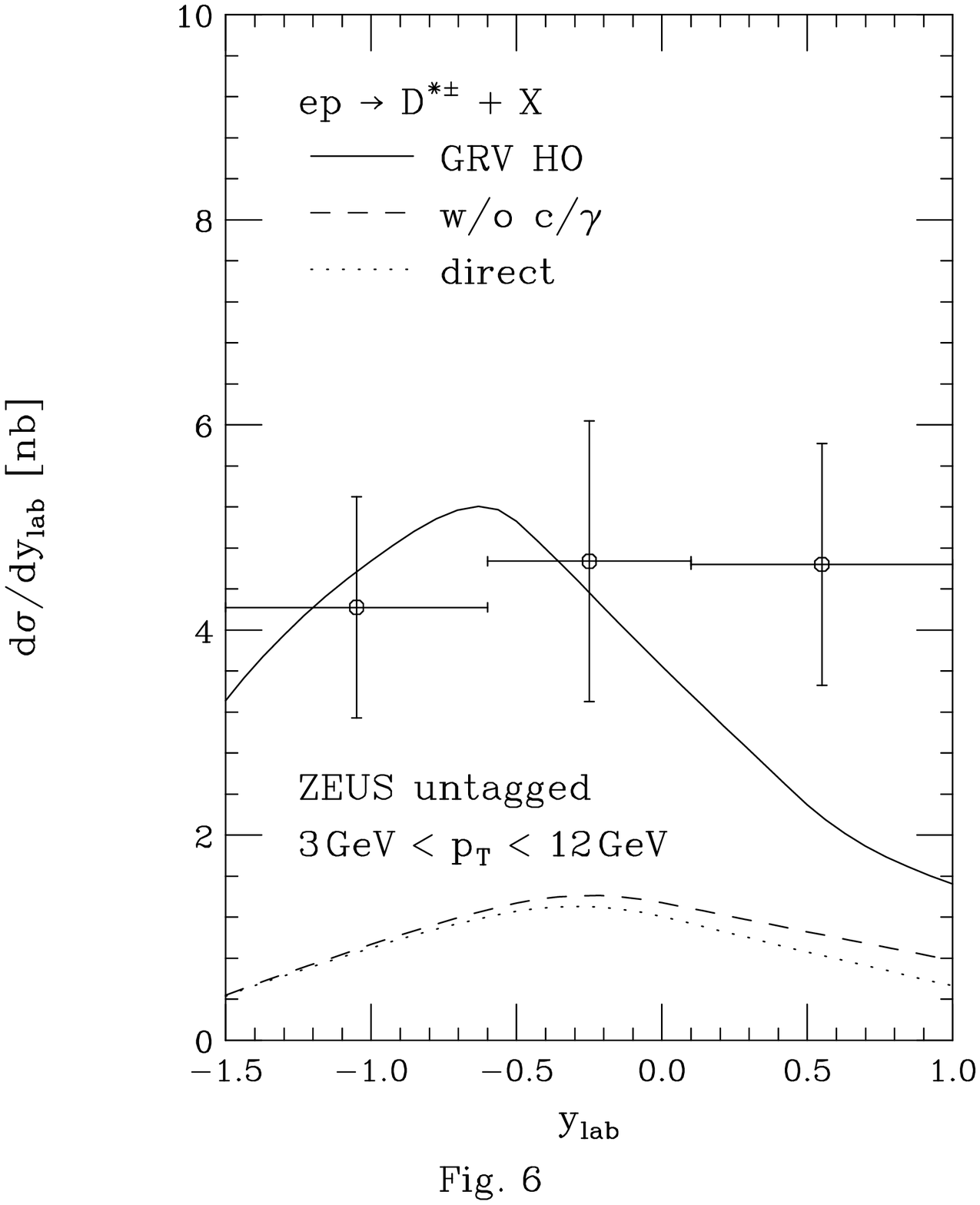,clip=,width=7cm}}}
\put(6.4,-.3){\makebox{
\epsfig{file=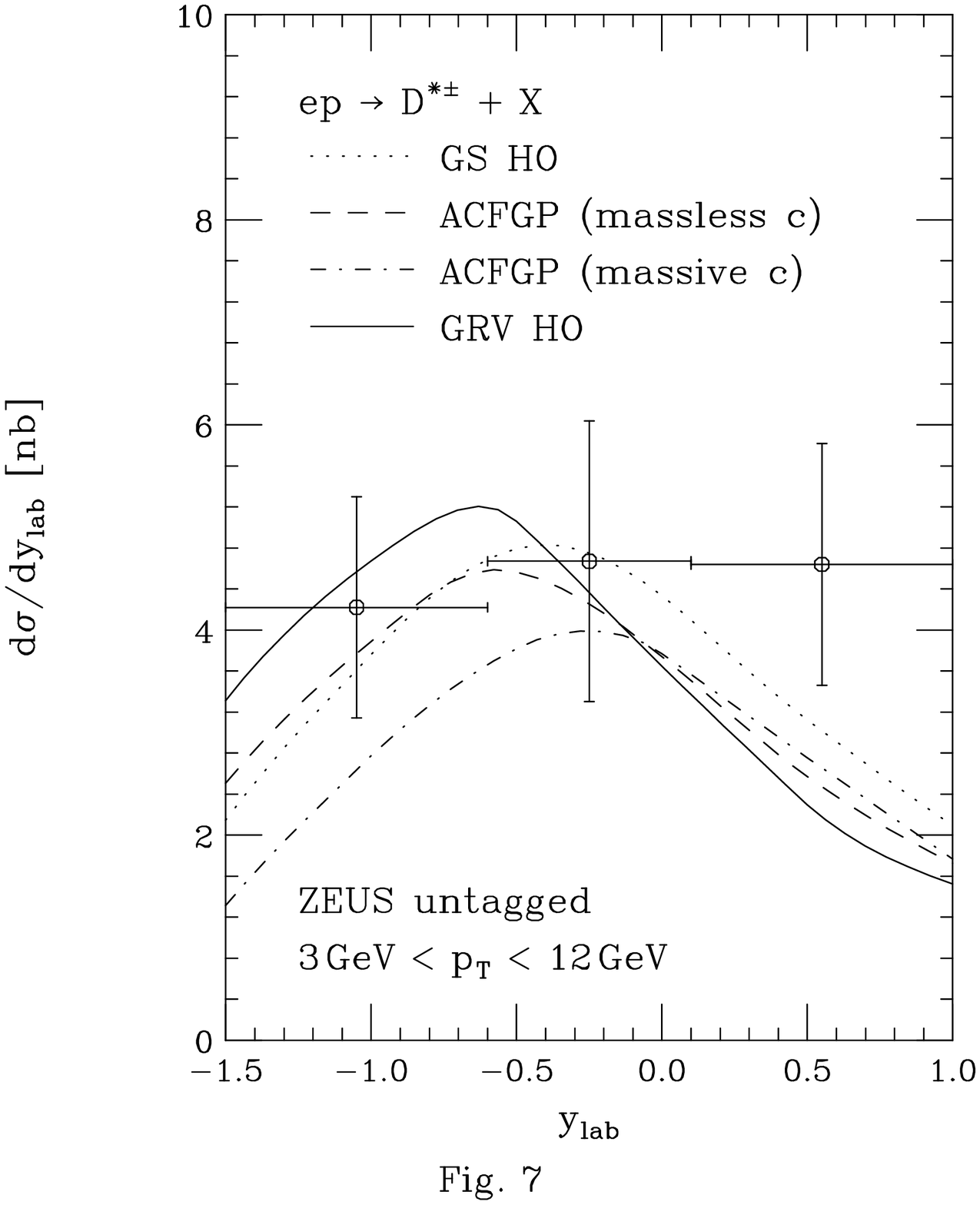,clip=,width=7cm}}}
\put(-0.9,0.0){a)}
\put(6.8,0.0){b)}
\end{picture}
\end{center}
\vspace{-0cm}
\caption[]{\small 
The impact of the charm quark in the photon on $\eta$--spectra
of $D^{*\pm}$ production in photoproduction.
Our NLO predictions are compared to data from 
ZEUS \protect\cite{zeus2}.
The solid line, calculated with the GRV photon PDF's 
\protect\cite{glu4} gives reasonable agreement with the data.
a)
The result with no charm in the photon and the direct contribution
are also given.
b)
Sensitivity of the cross section to the differences between
the photon PDF's.
The NLO results are shown for the sets of GRV \protect\cite{glu4},
GS \protect\cite{gs1}, and ACFGP \protect\cite{acfgp}. 
\protect\label{eta_dstar}}
\end{figure}

The charm distribution in the photon is not well
constrained by the $F_2^{\gamma}$ data, either.
At HERA, it can be probed by inclusive production of
charmed mesons in photoproduction.
Precise data have become available
for the production of $D^{*\pm}$ mesons, recently \cite{h13,zeus2}.
The FF's are well known as has been demonstrated in section~5.1.
The main limitation for the measurement is the experimental error.
Fig.~\ref{eta_dstar} shows the comparison of our NLO predictions
with data from ZEUS \cite{zeus2} for the
rapidity spectra of the cross section integrated over $p_T>3$~GeV.
The solid curve in Fig.~\ref{eta_dstar}a represents our
standard choice of the GRV set of photon PDF's,
the CTEQ4M \cite{cteq1} proton PDF's, and scale $\xi=1$.
For this choice, we find good agreement with the measurement in the central
and backward region, whereas it is about $1.8 \sigma$ low
for the bin in the forward region.
The dotted curve represents the direct contribution.
It is far below the data.
When only the charm distribution in the photon is switched off,
the dashed curve results.
It almost coincides with the dotted curve, except for the
forward region, where it is slightly higher.
Clearly, the dominant resolved part of the cross section is almost
entirely due to the charm density.
Therefore, a measurement of the rapidity spectrum of
$D^{*\pm}$ corresponds, after subtraction of the well
determined direct part, directly to the measurement of the
charm density in the photon.
\smallskip

Fig.~\ref{eta_dstar}b shows the NLO predictions of various
PDF sets, compared to the same data.
In addition to GRV, the set of GS \cite{gs1} is plotted as dotted
line.
ACFGP \cite{acfgp} presents two sets, with a massless charm (dashed)
and with a massive one (dash--dotted).
GS is slightly favored over GRV by the data, whereas the massive
set of ACFGP compares least favorable to the data.
Unfortunately, the statistics are not sufficient at the
moment, to exclude any of the above sets.

\vfill
\clearpage
\section{Summary and Outlook}
\setcounter{equation}{0}

In this thesis, we set out to extract fragmentation
functions for charged hadrons to NLO accuracy from
\ee annihilation data solely.
Former fits revealed that it is not possible to unambiguously
determine the gluon FF with data on unpolarized
cross sections alone \cite{binpk,bink0,binds,bin1}.
We thus used data on the longitudinal polarized cross section
to constrain the gluon fragmentation.
\smallskip

To this end, the NLO longitudinal coefficient functions
are required.
We completed the calculation of P.J.~Rijken and W.L.~van~Neerven \cite{rij1}
by the color class $C_F T_R$ and checked the pure singlet
and the class $N_F T_R C_F$ of the nonsinglet.
These coefficient functions were then used together
with those for the transverse cross section, to fit FF's
to ALEPH \cite{ale3} data on the longitudinal and the unpolarized
cross section.
In the unpolarized case, ALEPH distinguished three different
data samples for (i) mainly $u,d,s$ quarks (ii) mainly $c$ 
fragmentation and (iii) a $b$ quark enriched sample.
Our analysis yields well constrained light quark,
 $c$, and $b$ quark fragmentation functions
for charged hadrons.
Owing to the inclusion of the NLO longitudinal
coefficient functions in the fit, the gluon
is also well constrained in the important small $x$ region.
However, it is not constrained well for intermediate and
large $x$.
As the gluon fragmentation is effectively probed at relatively
small $x$ in many processes, this analysis nevertheless
represents a significant improvement
over our earlier $h^{\pm}$ sets.
\smallskip

In particular, the new sets could be used to measure $\alpha_s$ via
inclusive particle production in photoproduction, DIS or
 $p\overline{p}$ scattering, where the gluon dominates 
the cross section.
For a consistent analysis along those lines, the
value of $\Lambda$ must be varied not only in the formula
for the strong coupling, but also in the PDF's and the FF's.
It is thus necessary to perform separate fits for a range
of specified $\Lambda$ values.
For the PDF's of the proton this has been done some time ago by
A.~Vogt \cite{vog2}.
In this thesis,
we provide an analogous set of functions
for the fragmentation process, for the first time.
\smallskip

Our fragmentation functions may also be put to good use
in constraining the charm and the gluon densities of the
resolved photon from rapidity spectra as measured at HERA.
With the precision that is expected from 
the luminosity accumulated in the past years,
this is arguably the best place to learn about the
charm PDF.
Finally, these FF's will give particularly precise
predictions for cross sections at future experiments
like LEP2 and NLC.
First data from LEP1.5 \cite{leh1} 
are already awaiting comparison with theory.
All of this has hardly been tackled, yet, promising
fascinating insights to be gained in future work.
\smallskip

Another application of fragmentation functions is
to study
scaling violation in \ee annihilation.
In \cite{binpk} we fitted
$\Lambda$, along with the other parameters,
to data at 29~GeV and at 91.2~GeV.
We obtained very sensible values for $\alpha_s(m_Z)$
of 0.118 (0.122) in NLO (LO), however,
we could not assess their uncertainty by this simple approach.
A dedicated study with complete error analysis can give
competitive errors for the value of the strong coupling,
as demonstrated, e.g., by ALEPH \cite{ale3}.
\smallskip

Universality, a key prediction of the factorization
theorem, can also be tested by applying FF's that were
fitted to \ee annihilation data to other processes.
This has also been demonstrated successfully for
$ep$--, $p\overline{p}$--, and $\gamma\gamma$--scattering,
and is briefly reviewed in this work.
With the improved set of FF's, it will be possible to reduce  
the theoretical uncertainties considerably. 
\smallskip

\vfill
\clearpage
%%%%%%%%%%%%%%%%%%%%%%%%%%%%%%%%%%%%%%%%%%%%%%%%%%%%%%%%%%%%%%%%%%%%%%%%%%%
\vspace{2cm}
{\Large\bf Acknowledgments}
\bigskip

It is a pleasure to thank
Professor Dr.~G.~Kramer for his support during the preparation of this 
dissertation.
I also want to thank Professor Dr.~G.~Kramer and 
PD Dr.~B.A.~Kniehl for the good 
collaboration over the past years on much of the work presented here.
\smallskip

I enjoyed a very fruitful cooperation with K.~Johannsen and 
S.~Aid from H1 on the subject of neutral Kaon production.
M.~Erdmann and W.~Hoprich, also from H1, participated
prominently in the analysis of the parton content of the photon.
\smallskip

I am indebted to B.~P\"otter and to M.~Klasen
for a careful reading of the manuscript.
Special thanks are due to all my colleagues at the institute
for providing a very pleasant working environment.
\smallskip

This project could not have been tackled without the access
to the DESY computer network or without the support by the
group R2, notably by E.--L.~Bohnen and K.~Wipf.
\smallskip

Financial support was provided by the Bundesministerium f\"ur Bildung und
Wissenschaft, Forschung und Technologie (BMBF), Bonn, Germany under contract
05 7HH92P(0), 
and by the European Union within the program {\it Human Capital
and Mobility} through the network {\it Physics at High Energy Colliders}
under contract CHRX--CT93--0357 (DG12 COMA). 
\smallskip

\vfill
\clearpage
\begin{appendix}
\renewcommand{\theequation}{\Alph{section}.\arabic{equation}}
\section{The Strong Coupling and $\Lambda_{\rm QCD}$}
\setcounter{equation}{0}
\bigskip

The coupling constant of QCD satisfies the renormalization group equation
\begin{equation}
\label{rg}
\mu^2\frac{d\alpha_s}{d\mu^2}=\beta(\alpha_s)
\STOP
\end{equation}
This differential equation determines the scale dependence, hence
$\alpha_s$ is known up to a constant.
To NLO, the first two terms in the expansion are kept,
\begin{equation}
\label{betafunc}
\beta (\alpha_s)=
-\beta_0\frac{\alpha_s^2}{4\pi}
-\beta_1\frac{\alpha_s^3}{(4\pi)^2}
+{\cal O}(\alpha_s^4) \, , \,\,\,\,
\beta_0=\frac{33-2N_F}{3}\,, \,\,
\beta_1=\frac{306-38N_F}{3} 
\COMMA
\end{equation}
and to this order, the solution of eq.~(\ref{rg}) can be expanded to 
\begin{equation}
\label{twoloop}
\frac{\alpha_s(\mu^2)}{2\pi} =
                  \frac{2}{\beta_0 \ln{(\mu^2/\Lambda^2)}}
\left[1-\frac{\beta_1}{\beta_0^2}\frac{\ln{\ln{(\mu^2/\Lambda^2)}}}
{\ln{(\mu^2/\Lambda^2)}} \right]
\end{equation}
up to terms of order $[\ln(\mu^2/\Lambda^2)]^{-3}$ that
can not be consistently included at NLO.
(Other approximations are in use also, introducing 
numerically small ambiguities in the interpretation of
 $\alpha_s$ and $\Lambda$ --- see \cite{ell3})
To LO, the logarithmically suppressed term in the square 
brackets is absent.
The parameter $\Lambda$ is the one constant, mentioned above that
fixes the boundary conditions of (\ref{rg}). 
It can not be calculated from perturbative QCD but must be 
determined from experiment.
\smallskip

In this work, $\Lambda$ is used as a phenomenological parameter to fix
the value of the strong coupling at some point.
For ease of notation this parameter is at places given without 
indices.
However it does depend on a number of details in the calculation
of the cross section.
Great care should be taken when comparing $\Lambda$ values of different
analyses.
First, this QCD parameter depends on the renormalization scheme used.
In this work the {\msbar}--scheme \cite{msbar} is used throughout.
For two distinct schemes, say scheme $A$ and scheme $B$, the
coupling constant is identical to LO but differs by a finite renormalization
at NLO \cite{ell3},
\begin{equation}
\label{alpharen}
\frac{\alpha_s^B}{2\pi} = \frac{\alpha_s^A}{2\pi} 
   \left[1+c_1\frac{\alpha_s^A}{2\pi}+{\cal O}
   \left(\left(\frac{\alpha_s}{2\pi}\right)^2\right)\right]
\STOP
\end{equation} 
The first two coefficients of the QCD $\beta$--function are unchanged
by such a transformation so that $\beta_0, \beta_1$ and $c_1$
determine any scheme transformation (to NLO). 
In fact, only the leading coefficient of the $\beta$ function 
is needed for the transformation of $\Lambda$ \cite{ell3},
\begin{equation}
\label{lambdaren}
\Lambda^B = \Lambda^A \exp \left( \frac{c_1}{\beta_0} \right)
\STOP
\end{equation} 
Second, as $\beta$ depends on the number of flavors, 
it is clear that $\Lambda$ does, too.
When an effective theory is applied as done in this work, 
all the quark masses are taken to be massless in the calculation of the
hard processes and 
the masses of the heavy quarks ($c$ and $b$) are incorporated
by the introduction of thresholds, only.
This leads to a variable number of flavors,
depending on the scale, and consequently the $\Lambda$ parameter
must be adjusted accordingly.
The matching at the flavor--thresholds must be done in such a way
that the strong coupling is continuous.
The respective equations are rather intricate, for practical
purposes it is sufficient to use the approximations presented by
Marciano \cite{mar2},
\begin{eqnarray}
\label{lambdamatching}
\Lambda_{\overline{\rm MS}}^{(4)} &\n \approx \n&
  \Lambda_{\overline{\rm MS}}^{(5)}\left(
    \frac{\mu_5}{\Lambda_{\overline{\rm MS}}^{(5)}}
    \right)^{\frac{2}{25}}\left[
    2\ln\left(\frac{\mu_5}{\Lambda_{\overline{\rm MS}}^{(5)}}
      \right)\right]^{\frac{963}{14375}}
\nonumber 
\SPACE {\rm and}
\\
\Lambda_{\overline{\rm MS}}^{(3)} &\n \approx \n&
  \Lambda_{\overline{\rm MS}}^{(4)}\left(
   \frac{\mu_4}{\Lambda_{\overline{\rm MS}}^{(4)}}
    \right)^{\frac{2}{27}}\left[
    2\ln\left(\frac{\mu_4}{\Lambda_{\overline{\rm MS}}^{(4)}}
      \right)\right]^{\frac{107}{2025}}
\COMMA
\end{eqnarray}
where the parameters $\mu_4$ and $\mu_5$ specify the threshold scales
for the transitions of 3 to 4, respectively 4 to 5 flavors in the VFNS.
In this work, the thresholds are set at twice the heavy quark masses:
 $\mu_4=2m_c\approx 3$~GeV and $\mu_5=2m_b\approx 10$~GeV.
Equations~(\ref{lambdamatching}) apply to the 
next--to--leading $\Lambda$, in LO
the factor in square brackets is absent and the relations are then exact.
\begin{figure}[hht!]
\begin{center}
\epsfig{file=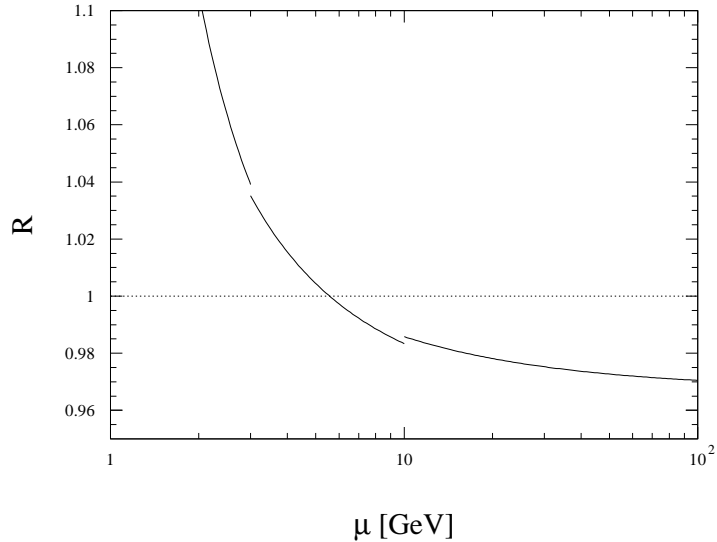 , clip=,width=9.5cm }
\end{center}
\vspace{-0.4cm}
\caption[]{\small 
The strong coupling with the Marciano scheme for 
$\Lambda$--matching \protect\cite{mar2}.
Because of the limited accuracy of
(\protect\ref{lambdamatching}), the ratio of the two--loop to
the one--loop $\alpha_s$ exhibits small discontinuities
at the thresholds.
\protect\label{matching}}
\end{figure}

The accuracy of the $\Lambda$ matching prescriptions (\ref{lambdamatching})
is demonstrated in Fig.~\ref{matching}.
The quantity plotted is the ratio of the two--loop $\alpha_s$ 
(eq.~(\ref{twoloop})) over the one--loop $\alpha_s$ as a function of the
scale $\mu$.
As the matching prescription is exact in LO, the discontinuities at the
thresholds are due to the error of the approximation for the NLO matching
(\ref{lambdamatching}).
The error is small compared to the effect of the higher order term 
in (\ref{twoloop}).
\smallskip

\begin{table}[hht!]
\begin{center}
\begin{tabular}{|r|r||r|r||r|}
\hline
 $\Lambda^{(4),NLO}$ & $\Lambda^{(4),LO}$ & $\Lambda^{(5),NLO}$ & 
 $\Lambda^{(5),LO}$ & $\alpha_s(\Lambda,m_Z)$ \\
\hline 
\hline
150 &  52 &  90 &  33.5 & 0.1036 \\
200 &  71 & 123 &  46.5 & 0.1081 \\
250 &  90 & 158 &  60.0 & 0.1119 \\
300 & 109 & 193 &  74.0 & 0.1152 \\
350 & 129 & 229 &  89.0 & 0.1182 \\
400 & 149 & 265 & 104.0 & 0.1209 \\
\hline
\end{tabular}
\caption{\small
Matching of $\Lambda^{(4)}$ to $\Lambda^{(5)}$.
The $\Lambda$ values are given in units of MeV.
\protect\label{lambda5to4}}
\end{center}
\end{table}

In photoproduction and in other processes with relatively low
hard scales, only four flavors are active over much of the
kinematical region and one commonly cites the $\Lambda$ value for
four flavors.
In our new charged hadron fit we cite the $\Lambda^{(4)}$
values.
They are chosen as in \cite{vog2} so that our set of FF's can be
easily matched with the respective PDF's for a
determination of $\alpha_s$.
Table~\ref{lambda5to4} lists the corresponding $\Lambda^{(5)}$
values according to Marcianos matching prescription adopted here.
The LO values are fixed by the requirement that the
physical $\alpha_s(\Lambda^{(5)},m_Z)$
should not depend on the order of perturbation theory.

\vfill
\clearpage
\section{Phase Space Integrals}
\setcounter{equation}{0}
\bigskip

In order to perform the integration over four of the five degrees of 
freedom of the four--particle phase space, we regard the process
in a specific system \cite{kra2}.
This is the CM system of partons 1 and 2, where the coordinate system
has been rotated in such a way that of the five invariants in
the irreducible set, $y_{12}$, $y_{123}$, $x_3$, $y_{234}$, 
and $y_{23}$, only $y_{23}$ depends on the angle $\phi$.
Also, only $y_{234}$ and $y_{23}$ depend on the second angle, $\theta$.
In this system, the angular integrations are relatively easy.
With the definition of the integration measures
\begin{eqnarray}
{\rm \bf I}_{\phi} \left[ f(\phi) \right] 
      &\n \equiv \n& \int_0^{\pi} \frac{d\phi}{N_{\phi}} 
                    \sin^{-2\ve} \phi \, f(\phi)\\
\nonumber\\
{\rm \bf I}_v \left[ f(v) \right]
      &\n \equiv \n& \int_0^1 \frac{dv}{N_v} 
                    v^{-\ve} (1-v)^{-\ve} \, f(v) 
\label{measures}
\end{eqnarray}
the phase space integrals required for the angular integrations 
of $\PS^{(4)}$
are given by
\begin{eqnarray}
\label{pint}
{\rm \bf I}_{\phi} \left[1\right] &\n \equiv \n& 1 \hspace{.7cm} 
, \hspace{.3cm} {\rm i.e.} \hspace{.4cm}
   N_{\phi} = 2^{2\ve}\pi\frac{\Gamma(1-2\ve)}
                                      {\Gamma^2(1-\ve)} \label{anorm1} \\
{\rm \bf I}_{\phi} \left[ \cos^{2k+1}\phi \right] 
                  &\n=\n& 0 \hspace{.7cm} , \hspace{.3cm} k \, \in \, \N_0 \\
{\rm \bf I}_{\phi} \left[ \cos^2\phi \right] 
                  &\n=\n& \frac{1}{2} + \frac{1}{2}\ve 
                        +{\cal O}(\ve^2) \\
{\rm \bf I}_{\phi} \left[ \cos^4\phi \right] 
                  &\n=\n& \frac{3}{8} + \frac{9}{16}\ve 
                        +{\cal O}(\ve^2) 
\end{eqnarray}
for the integration over the polar angle $\phi $.
The normalization of eq.~(\ref{anorm1}) is determined from
\begin{equation}
\label{intbeta}
\int_0^1 dz \, z^a \, (1-z)^b = B(1+a, 1+b) 
    = \frac{\Gamma(1+a)\Gamma(1+b)}{\Gamma(2+a+b)}
\STOP
\end{equation}
\smallskip

The results for the azimuthal integration are (with the
substitution $v={\scriptstyle \frac{1}{2}}(1-\cos\theta)$)
\begin{eqnarray}
\label{vint}
{\rm \bf I}_v \left[ 1 \right] &\n \equiv \n& 1  \hspace{.7cm} 
, \hspace{.3cm} {\rm i.e.} \hspace{.4cm}
   N_v = {\rm B}(1-\ve,1-\ve) \label{anorm2} \\
{\rm \bf I}_v \left[ v^k \right] &\n=\n& \frac{{\rm B}(1+k-\ve,1-\ve)}
    {{\rm B}(1-\ve,1-\ve)} \\
{\rm \bf I}_v \left[ \frac{1}{v+D} \right] 
       &\n=\n& \ln\left(\frac{D+1}{D}\right) 
       -\ve \left[\li\left(\frac{D+1}{D}\right) 
                  - \li\left(\frac{D}{D+1}\right) \right] 
                        +{\cal O}(\ve^2) \\
{\rm \bf I}_v \left[ \frac{1}{(v+D)^2} \right]
       &\n=\n& \frac{1}{D(1+D)}
       + \ve (2D+1)\ln\left(\frac{D+1}{D}\right)
                        +{\cal O}(\ve^2) 
\COMMA
\end{eqnarray}
where the expansion of Euler's Beta function in (\ref{anorm2}) can be
performed with the use of
\begin{eqnarray}
\Gamma(1+a) &\n\equiv \n& a \, \Gamma(a) \, \, , \hspace{.7cm}
\Gamma(1) \equiv 1
\label{gamdef}
\\
\Gamma(1+\ve) &\n=\n& 1 -\gamma_{\tt E} \ve + \frac{1}{2} 
       \left(\gamma_{\tt E}^2+\zeta(2) \right) \ve^2 
       +{\cal O}\left(\ve^3\right)
\hspace{1mm} ; \hspace{3.5mm} \gamma_{\tt E}\approx 0.577\, 215 \STOP
\label{gamexp}
\end{eqnarray}
The remaining azimuthal integrals are of the form
\begin{equation}
{\rm \bf I}_v \left[ \frac{v^{k+1}}{(v+D)^{l+1}} \right] 
   \hspace{.7cm} , \hspace{.3cm} {\rm with} 
     \hspace{0.5cm}  k,l \, \in \, \N_0
\COMMA
\end{equation}
and are solved by recursive application of the decomposition
\begin{equation}
\frac{v^{k+1}}{(v+D)^{l+1}} = 
\frac{v^k}{(v+D)^l} - D \frac{v^k}{(v+D)^{l+1}}
\STOP
\end{equation}
\smallskip
The above angular integrals are in agreement with \cite{bee1} 
and have also been checked numerically.
\smallskip

The integrals required for the energy integrations are too numerous
to list them here.
For the regular terms, the limit $\ve \to 0$ can be taken in the
integration measures (\ref{meas1}) and (\ref{meas2}) and one
has then to deal with standard definite integrals, merely.
Those can be found in \cite{integrals}.
In the singular terms (\ref{energysing}) the poles are
extracted with the substitution
\begin{equation}
\label{pole}
z^{-1-\ve}=-\frac{1}{\ve}\delta(z)+\left(\frac{1}{z}\right)_+ 
   +{\cal O}(\ve)
\STOP
\end{equation}
Due to the functional form of $g(z,x,r)$ in (\ref{energysing}),
the plus distribution gives no 
contribution to the integral and up to ${\cal O}(\ve^0)$ the
singular term can be replaced by $-1/\ve$.
However, the second energy integration must keep track of the
terms proportional to $\ve$ that, together with the pole, give
rise to finite terms.
For this purpose, the expressions in the second integral which is
free of singularities are expanded in powers of $\ve$ according to
(\ref{logexp}).
This, again, leads to standard integrals.
\smallskip

Bookkeeping is facilitated by employing an algebraic computer
program.
In this work, {\tt MAPLE} \cite{map} has been used for the four integrations
over phase space.
\smallskip

\vfill
\clearpage
\section{Numerical Techniques for Evolution}
\setcounter{equation}{0}
\bigskip

In this appendix, the different methods for the numerical solution
of the Altarelli--Parisi evolution equations are discussed.
Before turning to the details of the computer codes,
the Mellin transform will be introduced as the basic tool
for the more elegant of the two computer codes -- the $n$--space code.
\smallskip

The Mellin transform technique, however,  
is not just a convenient way to solve the AP
evolution equations.
The essential feature of the Mellin transformation
\begin{equation}
\label{mtrans}
f^{[M]}(n) \equiv \int_0^1 dx x^{n-1} f(x)
\end{equation}
is that it renders convolutions to products,
\begin{equation}
\label{mconv}
(f\otimes g)^{[M]}(n)= f^{[M]}(n) \cdot g^{[M]}(n)
\STOP
\end{equation}
This can be employed for many calculational purposes, see \cite{gra2}.
Here, it will be exploited to reduce the
system of differential equations describing the
evolution of FF's 
to a system of linear algebraic equations.
This system is solved analytically \cite{fur1} and the
solution is then transformed back to $x$--space via
the inverse Mellin transform
\begin{equation}
\label{mback}
f(x) = \frac{1}{2\pi i} \int_{
C} dn x^{-n} f^{[M]}(n)
\STOP
\end{equation}
The Mellin technique has the additional advantage that one keeps
exactly those terms that have to be kept in the order in perturbation
theory considered, whereas in $x$--space evolution one can not help
but sum up part of the higher orders. 
The $x$--space evolution is in that sense somewhat inconsistent.
\smallskip

A drawback of the Mellin technique is, however that the input
for the evolution must be known analytically in order to make use
of the techniques advantages. 
Even if it is known analytically, it is not always possible to
perform the Mellin transform analytically which then
spoils the techniques main numerical advantages.
For details of the numeric in $n$--space, see \cite{vog1,glu1}.
\smallskip

We will closely follow \cite{fur1}
in the ensuing sketch of the analytical solution in $n$--space.
With the introduction of the new variable
\begin{equation}
t= \frac{2}{\beta_0}\ln{\left(\frac{\alpha_s(\mu_0^2)}
       {\alpha_s(\mu^2)}\right)} 
\COMMA
\end{equation}
the set of Mellin--transformed
AP~equations (\ref{APplus}) -- (\ref{APS2}) can be written as
\begin{eqnarray*}
\partial_t M_{(-),i}&\n=\n&  \gamma_{(-)}M_{(-),i}
\\
\partial_t M_{(+),i}&\n=\n&   \gamma_{(+)}M_{(+),i}
\\
\partial_t M_{\Sigma}&\n=\n& \gamma_{\Sigma}M_{\Sigma} 
       +2 N_F \gamma_{q\to G}M_G
\\
\partial_t M_G &\n=\n&  \frac{1}{2 N_F}\gamma_{G\to q}M_{\Sigma} 
     +\gamma_{G\to G}M_G
\STOP
\end{eqnarray*}
The convolutions have been rendered to products of the
new functions $M\equiv D^{[M]}$ and $\gamma\equiv P^{[M]}$.
For the standard parametrization (\ref{parstand}), the moments
of the FF's are Beta--functions (\ref{intbeta}).
The timelike anomalous dimensions $\gamma$ can be found in
\cite{glu2}.
The introduction of matrix notation 
\begin{equation}
\gamma\equiv
\left(  \begin{array}{cc}
            \gamma_{\Sigma} & \gamma_{q\to G} \\
            \gamma_{G\to q} & \gamma_{G\to G} \\
         \end{array} \right)  \,\,\,\,\, ,\,\,\,
M\equiv
\left(  \begin{array}{c}
            \frac{1}{2N_F} M_{\Sigma}  \\
            M_{G}       \\
         \end{array} \right) 
\nonumber
\end{equation}
further simplifies above equations ($M_{(-)}$ is disposable as it 
does not enter the cross sections (\ref{sigxpol})).
\begin{eqnarray}
\partial_t M_{(+),i}&\n=\n& \gamma_{(+)}\,M_{(+),i}
\\
\partial_t M&\n=\n& \gamma\,M
\STOP
\end{eqnarray}
In the following, matrix notation is implicit where appropriate.
The solution of the evolution equations can be expressed
by the action of evolution operators $E$
\begin{equation}
M(t,n)=E\,M(0,n)
\COMMA
\end{equation}
where for the non--singlet
\begin{equation}
E_{(+)}(t,n)=\left[1+\frac{\alpha_s(t)-\alpha_s(0)}{2\pi}\left(
            \frac{\beta_1}{\beta_0^2}\gamma_{(+)}^{(0)}(n)
            -\frac{2}{\beta_0}\gamma_{(+)}^{(1)}(n)\right)\right]
            e^{\gamma_{(+)}^{(0)}(n)t}
\end{equation}
and for the singlet
\begin{eqnarray}
E(t,n)&\n=\n&\left[e_1+\frac{\alpha_s(t)-\alpha_s(0)}{2\pi}\left(
            -\frac{2}{\beta_0}e_1Re_1+
            \frac{e_2Re_1}{\lambda_1-\lambda_2-
                  {\scriptstyle \frac{1}{2}}\beta_0}  \right)\right]
            e^{\lambda_1t}   
\nonumber  
\\
      & & +\left[e_2+\frac{\alpha_s(t)-\alpha_s(0)}{2\pi}\left(
            -\frac{2}{\beta_0}e_2Re_2+
            \frac{e_1Re_2}{\lambda_2-\lambda_1-
                  {\scriptstyle \frac{1}{2}}\beta_0}  \right)\right]
            e^{\lambda_2t}
\STOP
\end{eqnarray}
The Matrix $R$ is defined as 
\begin{equation}
R=\gamma^{(1)}-\frac{\beta_1}{2\beta_0}\gamma^{(0)}
\STOP
\end{equation}
The $\lambda_i$ are the eigenvalues of the matrix $\gamma^{(0)}$,
\begin{equation}
\lambda_{1,2}=\frac{1}{2}\left[
         \gamma_{q\to q}^{V(0)}+\gamma_{G\to G}^{(0)}\pm
         \sqrt{(\gamma_{q\to q}^{V(0)}-\gamma_{G\to G}^{(0)})^2
          +4\gamma_{G\to q}^{(0)}\gamma_{q\to G}^{(0)}}\right] 
\COMMA
\end{equation}
and the matrices
\begin{equation}
e_{1,2}\equiv\frac{1}{\lambda_{1,2}-\lambda_{2,1}}
   \left[\gamma^{(0)}-\lambda_{2,1}{\bf 1}\right]
\end{equation}
are projectors, with the properties 
\begin{equation}
e_{1,2}^2=e_{1,2} \,\,\, , \,\,\, e_1e_2=0 \,\,\, , \,\,\,
e_1+e_2={\bf 1} \,\,\, , \,\,\, \gamma^{(0)}=\lambda_1e_1+\lambda_2e_2 
\STOP
\end{equation}
Here, ${\bf 1}$ stands for the one in the matrix--space.
The evolved FF's in $x$--space are then obtained via numerical
back--transformation of the $M(t,n)$, according to (\ref{mback}).
\smallskip

In the spacelike case, a number of computer codes for the
numerical solution of the evolution equations have become available 
by now \cite{pcode}.
The theoretical uncertainties and the differences between the
various codes are discussed in \cite{blu1}.
In the timelike case, progress has been quite recent.
Nason, e.g., wrote the possibly first evolution program for the
timelike region \cite{nas1}. 
Spira and Kniehl completed computer
codes in the recent past \cite{ucode}, in addition to the 
codes by the author that have been employed in 
\cite{binpk,bink0,binds,bin1,bin0}.
Our codes for both the $n$--space and the $x$--space will be briefly
discussed below.
\smallskip

\begin{figure}[hht!]
\begin{center}
\epsfig{file=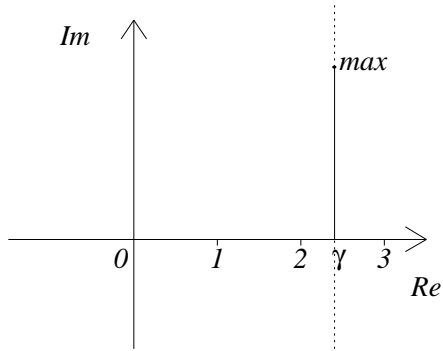,clip=,width=10cm }
\end{center}
\vspace{-0cm}
\caption[]{\small 
Contour for the Mellin back--transformation.
\protect\label{contour}}
\end{figure}

\noindent
\underline{(i) $n$--space:}
In $n$--space, the AP--equations are solved analytically, as outlined
in this section.
An adaptive routine is used for the back--transformation 
into $x$--space (\ref{mback}).
The contour is indicated in Fig.~\ref{contour}
as the dotted line.
It runs parallel to the imaginary axis, through $(\gamma,0)$,
and is closed at infinity.
The real part of the kernel in (\ref{mback}) is symmetric to the real
axis so that it is sufficient to integrate over half the contour.
For large imaginary parts of $n$, the kernel rapidly approaches
zero so that the integration can be terminated at rather small values
of $|{\rm max}|={\cal O}(20)$.
Both $|{\rm max}|$ and $\gamma$ are chosen adaptively.
The precision can thus be set to any desired value,
usually it is of the order of 0.5\%.
For the calculation of the complex logarithmic gamma functions
the {\tt NAG FORTRAN} library \cite{nag} is used.
The code has been checked numerically with the program by Spira \cite{kni3}.
\smallskip

\noindent
\underline{(ii) $x$--space:}
In this brute--force approach, the AP--equations are solved
by iterative application of their integral form (\ref{ap})
where the convolution $P\otimes D$ is evaluated with linear
interpolation between $n={\cal O}(200)$ regularly spaced values
and (\ref{ap}) is iterated of the order of 50 times.
The accuracy of this routine is of the order of 1\% for $x<0.8$.
We find perfect agreement with the program by Kniehl \cite{kni3}.
\smallskip

For both the $x$-- and the $n$-- space evolution, the
accuracy deteriorates as $x\to1$.
As this is true also for the data (because $\sigma \to 0$, too,
and hence also the statistics),
the impact of these numerical errors is small in most analyses.
If desired, arbitrary precision can of course be obtained for any
value of the momentum fraction $x$, albeit at high CPU time cost.
\smallskip

\begin{figure}[hht!]
\begin{center}
\epsfig{file=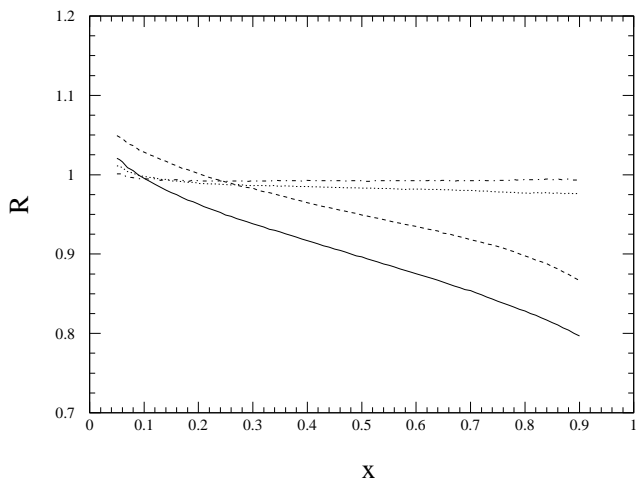,clip=,width=8.8cm }
\end{center}
\vspace{-1.2cm}
\caption[]{\small 
Ambiguities of the solution of the AP equation.
The curves show the ratios of the results of the
 $n$--space evolution over those for the $x$--space
evolution.
Our new NLO sets of FF were evolved from $\mu_0$ to $m_Z$.
The solid, dashed, dotted and dash-dotted lines correspond
to the fragmentation functions of the gluon, the light quarks,
the $c$ and the $b$, respectively.
\protect\label{xvsn}}
\end{figure}

A fundamental difference between the $x$--space
and the $n$--space techniques turns up in the way
the perturbative expansion is truncated.
Whereas one consistently keeps terms of ${\cal O}(\alpha_s^2)$
(${\cal O}(\alpha_s)$) only in the NLO (LO) $n$--space evolution,
the repeated evaluation of (\ref{ap}) in the $x$--space
approach amounts to a partial resummation
of higher order terms.
However, as the deviation is numerically rather minor
it may be disregarded for most practical purposes.
Fig.~\ref{xvsn} shows the ratio of the FF's
obtained at $m_Z$ when using the $n$--space technique over
those obtained in the $x$--space.
In both cases, our new NLO set for charged hadrons was used as
input at $\mu_0$.
The largest deviations from 1 show up in the fragmentation of the
gluon (solid line).
The light quarks (dashed) are also affected slightly 
whereas the charm and bottom FF's are insensitive to the
choice of technique.
In general, the effect is largest for the partons that dominate 
the evolution.
(See \cite{blu1} for a detailed discussion
of the ambiguities in the closely related spacelike case.)

\vfill
\clearpage
\section{Coefficient Functions}
\setcounter{equation}{0}
\bigskip

The order $\alpha_s^0$ coefficient functions in 
eqns.~(\ref{sigTNSren}) -- (\ref{sigTGren}) are given by
\begin{eqnarray}
\COFm_{T,q}^{NS,(0)}(z)&\n=\n& \delta(1-z)
\label{CTqNS0}
\\
\nonumber\\
\COFm_{k,a}^{\lb,(0)}(z)&\n=\n& 0 \hspace{0.6cm} {\rm else.}
\end{eqnarray}
Only at ${\cal O}(\alpha_s)$ do the coefficient
functions differ from the naive parton model expectation
\cite{alt1},
\begin{eqnarray}
\COFm_{T,q}^{NS,(1)}(z)&\n=\n&
C_F\left[\frac{3}{2}(1-z)
+4\frac{\ln(z)}{1-z}-(1+z)\left(\ln(1-z)+2\ln(z)\right)
\right.\nonumber\\
& &{}+\left.
2\left(\frac{\ln{(1-z)}}{1-z}\right)_+
-\frac{3}{2}\left(\frac{1}{1-z}\right)_+
+\left(4\zeta(2)-\frac{9}{2}\right)\delta(1-z)\right]
\label{CTqNS1}
\\
\nonumber\\
\COFm_{T,q}^{PS,(1)}(z)&\n=\n& \COFm_{T,q}^{F,(1)}(z) = 0
\\
\nonumber\\
\COFm_{T,G}^{(1)}(z)&\n=\n&
C_F\left[ \frac{1+(1-z)^2}{z}\left(\ln{(1-z)}+2\ln(z)\right) 
-2\frac{1-z}{z} \right]
\label{CTG1}
\\
\nonumber\\
\COFm_{L,q}^{NS,(1)}(z)&\n=\n& C_F \left[ 1 \right]
\label{CLqNS1}
\\
\nonumber\\
\COFm_{L,q}^{PS,(1)}(z)&\n=\n&\COFm_{L,q}^{F,(1)}(z) = 0
\\
\nonumber\\
\COFm_{L,G}^{(1)}(z)&\n=\n& C_F \left[ 4 \frac{1-z}{z} \right]
\label{CLG1}
\STOP
\end{eqnarray}

For the factorization, also the $a$ functions proportional to $\ve$
as introduced in (\ref{baresigNS}), (\ref{baresigG}) are presented,
\begin{eqnarray}
\label{afctnq}
a_{L,q}^{(1)}(z)&\n=\n&C_F \left[1-2\ln(z)-\ln(1-z) \right]
\COMMA
\\
a_{L,G}^{(1)}(z)&\n=\n&C_F \left[4\frac{1-z}{z}\left(2-2\ln(z)-\ln(1-z)
    \right) \right]
\STOP
\label{afctnG}
\end{eqnarray}
\smallskip

The $\COFm_T^{(2)}$ are not required in this
work, see \cite{rij2} for the expressions.
The same applies to the $\COFm_A^{(1)}$ and $\COFm_A^{(2)}$ that can
be found in \cite{rij3}.
All of the $\COFm_L^{(2)}$ are collected in the following
equations for convenience.
The color class $C_F T_R$ of the longitudinal cross section,
specified by the coefficient function $\COFm_{L,q}^{F}$ 
is presented here for the first time.
Class $N_F T_R C_F$ of $\COFm_{L,q}^{NS}$ as well as $\COFm_{L,q}^{PS}$
have been checked and the result of \cite{rij1}, where
all the other classes are taken from, has been confirmed.
\small
\begin{eqnarray}
\COFm_{L,q}^{NS,(2)}(z) &\n=\n&
C_F^2\left\{4S_{1,2}(1-z)+8S_{1,2}(-z) -12\lithree(-z)
+8\ln(1+z)\li(-z)
\right.
\nonumber\\ 
&& \left. \hspace{1.0cm}
+4\ln(z)\li(-z)+4\zeta(2)\ln(1+z) -4\zeta(2)\ln(1-z)
\right.
\nonumber\\ 
&& \left. \hspace{1.0cm}
+4\ln(z)\ln^2(1+z) -2\ln^2(z)\ln(1+z)-4\zeta(3)
\right.
\nonumber\\ 
&& \left. \hspace{1.0cm}
+\left(-4+\frac{12}{5}\frac{1}{z^2}-8z-\frac{8}{5}z^3\right)
\left(\li(-z)+\ln(z)\ln(1+z)\right)
\right.
\nonumber\\ 
&& \left. \hspace{1.0cm}
-3\li(1-z)+\ln(z)\ln(1-z)+\left(4-8z-\frac{8}{5}z^3\right)\zeta(2)
\right.
\nonumber\\ 
&& \left. \hspace{1.0cm}
+\ln^2(1-z)+\left(-\frac{3}{2}+4z+\frac{4}{5}z^3\right)\ln^2(z)
+\left(\frac{7}{2}+z\right)\ln(1-z)
\right.
\nonumber\\ 
&& \left. \hspace{1.0cm}
+\left(\frac{17}{10}-\frac{12}{5}\frac{1}{z}+\frac{6}{5}z
+\frac{8}{5}z^2\right)\ln(z)-\frac{147}{20}+\frac{12}{5}\frac{1}{z}
-\frac{9}{10}z+\frac{8}{5}z^2
\right\}
\nonumber\\ 
&&  \hspace{0cm}
+N_C C_F \left\{-2S_{1,2}(1-z)-4S_{1,2}(-z)+6\lithree(-z)
\right.
\nonumber\\ 
&& \left. \hspace{1.0cm}
-4\ln(1+z)\li(-z) -2\zeta(2)\ln(1+z)+2\zeta(2)\ln(1-z)
\right.
\nonumber\\ 
&& \left. \hspace{1.0cm}
-2\ln(z)\li(-z)-2\ln(z)\ln^2(1+z)+\ln^2(z)\ln(1+z)+2\zeta(3)
\right.
\nonumber\\ 
&& \left. \hspace{1.0cm}
+\left(2-\frac{6}{5}\frac{1}{z^2}+4z+\frac{4}{5}z^3\right)
\left(\li(-z)+\ln(z)\ln(1+z)\right)
\right.\nonumber\\ 
&& \left. \hspace{1.0cm}
+\zeta(2)\left(4z+\frac{4}{5}z^3\right)
-\left(2z+\frac{2}{5}z^3\right)\ln^2(z) -\frac{23}{6}\ln(1-z)
\right.
\nonumber\\ 
&& \left. \hspace{1.0cm}
+\left(-\frac{73}{30}+\frac{6}{5}\frac{1}{z}+\frac{2}{5}z
-\frac{4}{5}z^2\right)\ln(z)
+\frac{1729}{180}-\frac{6}{5}\frac{1}{z}-\frac{49}{30}z
-\frac{4}{5}z^2
\right\}
\nonumber \\
&&  \hspace{0cm}
+N_F T_R C_F \left\{\frac{2}{3}(\ln(1-z)+\ln(z))
-\frac{25}{9}+\frac{2}{3}z\right\}
\label{CLqNS2}
\\
\nonumber \\
\COFm_{L,q}^{PS,(2)}(z) &\n=\n&
N_F T_R C_F \left\{
4\li(1-z)+4\ln(z)\ln(1-z)+6\ln^2(z)
-\frac{28}{3}-4\frac{1}{z}+\frac{52}{3}z-4z^2
\right.
\nonumber\\ 
&& \left. \hspace{.6cm}
+\left(\frac{8}{3}\frac{1}{z}-4z+\frac{4}{3}z^2\right)\ln(1-z)
+\left(-8+\frac{16}{3}\frac{1}{z}-8z+\frac{4}{3}z^2\right)\ln(z)
\right\}
\label{CLqPS2}
\\
\nonumber \\
\COFm_{L,q}^{F,(2)}(z) &\n=\n&
C_F T_R \left\{
\left(\li(1-z)-\zeta(2)\right)\left(-4+16\frac{1}{z}-16\frac{1}{z^2}\right)
-\frac{8}{5}z^3\zeta(2)
\right.
\nonumber\\ 
&& \left. \hspace{1.0cm}
+\li(z)\left(\frac{2}{3}+\frac{4}{5}\frac{1}{z^2}
\right)
+\li(-z)\left(\frac{8}{3}-\frac{32}{15}z^3+\frac{16}{5}\frac{1}{z^2}\right)
\right.
\nonumber\\ 
&& \left. \hspace{1.0cm}
+\ln(1-z)\left(-2+6\frac{1}{z}-4\frac{1}{z^2}\right)
+\ln(z)\left(-\frac{16}{15}z-\frac{22}{5}+\frac{44}{5}\frac{1}{z}
\right. \right. 
\nonumber\\ 
&& \left. \hspace{1.0cm}
+\frac{8}{3}\ln(1+z)
+\frac{32}{15}z^2-\frac{32}{15}z^3\ln(1+z)
+\frac{16}{5}\frac{1}{z^2}\ln(1+z)+\frac{16}{5}z^3\ln(z)\right)
\nonumber\\ 
&& \left. \hspace{1.0cm}
-\frac{64}{5}\frac{1}{z}+\frac{56}{5}-\frac{8}{15}z+\frac{32}{15}z^2
\right\}
\label{CLqF2}
\\
\nonumber \\
\COFm_{L,G}^{(2)}(z) &\n=\n&
C_F^2\left\{
\left(-\frac{8}{3}+\frac{16}{5}\frac{1}{z^2}+\frac{8}{15}z^3\right)
\left(\li(-z)+\ln(z)\ln(1+z)\right)
\right.
\nonumber\\ 
&& \left. \hspace{1.0cm}
+4\li(1-z)+4\ln(z)\ln(1-z)+\frac{8}{15}\zeta(2)z^3
\right.
\nonumber\\ 
&& \left. \hspace{1.0cm}
+\left(6-\frac{4}{15}z^3\right)\ln^2(z)
+\left(-6+8\frac{1}{z}-2z\right)\ln(1-z)
\right.
\nonumber\\ 
&& \left. \hspace{1.0cm}
+\left(-\frac{2}{5}+\frac{24}{5}\frac{1}{z}-\frac{56}{15}z
-\frac{8}{15}z^2\right)\ln(z)+\frac{6}{5}-\frac{24}{5}\frac{1}{z}
+\frac{62}{15}z-\frac{8}{15}z^2
\right\}
\nonumber\\ 
&& \hspace{0cm}
+N_C C_F \left\{
\left(8+8\frac{1}{z}\right)\left(\li(-z)+\ln(z)\ln(1+z)\right)
-16\frac{1}{z}\li(1-z)
\right.
\nonumber\\ 
&& \left. \hspace{1.0cm}
-16\ln(z)\ln(1-z)+\zeta(2)\left(-16+24\frac{1}{z}\right)
+\left(-4+4\frac{1}{z}\right)\ln^2(1-z)
\right.
\nonumber\\ 
&& \left. \hspace{1.0cm}
-\left(12+16\frac{1}{z}\right)
\ln^2(z)+\left(36-\frac{116}{3}\frac{1}{z}+4z-\frac{4}{3}z^2\right)
\ln(1-z)
\right.
\nonumber\\ 
&& \left. \hspace{1.0cm}
+\left(28-\frac{88}{3}\frac{1}{z}+8z-\frac{4}{3}z^2\right)\ln(z)
-\frac{80}{3}+\frac{112}{3}\frac{1}{z}-\frac{40}{3}z
+\frac{8}{3}z^2
\right\}
\label{CLG2}
\end{eqnarray}
\normalsize
\smallskip

The plus--distributions that appear in the coefficient functions and
in the splitting functions can be defined via the convolution
with a regular, so--called test--, function.
\begin{equation}
\int\limits_0^1\!\!dz\,\left[f(z)\right]_+ \, g(z)
    \equiv\int\limits_0^1\!\!dz\,f(z)\, \left(g(z)-g(1)\right)
\label{plus}
\end{equation}
The subtraction is to be taken at the locus of the singularity,
which in this work is 1, mostly.
The convolution of a plus--distribution with a regular function is
then given by adding the missing contribution to eq.~(\ref{plus}),
\begin{equation}
\label{plusconv}
\left(f\right)_+ \otimes g =
\int_x^1 \frac{dz}{z} f\left(\frac{x}{z}\right) \left[
z-\frac{x}{z}g(x) \right] 
-g(x)\int_0^x dz f(z)
\STOP
\end{equation}
\smallskip

Nielsen's generalized polylogarithms \cite{nie1} are defined as
\begin{eqnarray}
\li(z) &\n\equiv \n&   -\int_0^z dy \frac{\ln(1-y)}{y}
\label{Li2def}
\\
S_{1,2}(z) &\n\equiv \n& \frac{1}{2}\int_0^z dy \frac{\ln^2(1-y)}{y}
\label{S12def}
\\
\lithree (z) &\n\equiv \n& \int_0^z dy \frac{\li(y)}{y}
\label{Li3def}
\STOP
\end{eqnarray}
Riemann's zeta function has the values
\begin{eqnarray}
\zeta(2) &\n=\n& \frac{\pi^2}{6} \approx 1.644\, 934
\label{zeta2}
\\
\zeta(3) &\n\approx \n& 1.202\, 057
\label{zeta3}
\STOP
\end{eqnarray}

\vfill
\end{appendix}
%%%%%%%%%%%%%%%%%%%%%%%%%%%%%%%%%%%%%%%%%%%%%%%%%%%%%%%%%%%%%%%%%%%%%%%%%%%

\end{document}